\pdfoutput=1
\title{A Quadratic Manifold for Model Order Reduction of Nonlinear Structural Dynamics}
\author{
	Shobhit Jain\footnote{Institute for Mechanical Systems, ETH Z{\"u}rich, Leonhardstra{\ss}e 21, 8092 Z{\"u}rich, Switzerland}, Paolo Tiso$ {}^{*} $, Johannes B. Rutzmoser$ {}^{\dagger} $, Daniel J. Rixen\footnote{Chair of Applied Mechanics, Technical University of Munich, Boltzmannstra{\ss}e 15, D - 85748 Garching, Germany}  \\
}
\date{\today}
\documentclass[10pt]{article}
\usepackage{lineno,hyperref}
\usepackage[margin=2.5cm]{geometry}

\usepackage{booktabs}
\usepackage{amsmath}
\usepackage{amssymb}
\usepackage{siunitx}
\usepackage{subcaption}
\usepackage{float}
\usepackage{placeins}
\usepackage{parskip}
\usepackage{graphicx}
\usepackage{cleveref}
\usepackage{amssymb}
\usepackage{amsmath,amsfonts}
\usepackage{amsbsy}
\usepackage{moreverb}
\usepackage{mathrsfs}
\usepackage{enumitem}
\usepackage{xcolor}
\usepackage{framed}
\usepackage{footmisc}
\usepackage{amsthm}
\usepackage{stackengine}
\setstackgap{S}{1pt}
\newcommand{\parderv}[2]{\frac{\partial #1}{\partial #2}}
\newcommand{\pardervd}[2]{\dfrac{\partial #1}{\partial #2}}
\usepackage{stackengine}
\usepackage{algpseudocode,algorithm,algorithmicx}

\algrenewcommand\algorithmicrequire{\textbf{Input:}}
\algrenewcommand\algorithmicensure{\textbf{Output:}}
\algdef{SE}[DOWHILE]{Do}{doWhile}{\algorithmicdo}[1]{\algorithmicwhile\ #1}%

\setstackgap{S}{1pt}

\newtheorem{thm}{Theorem}
\newtheorem{rmk}{Remark}

\newcommand{\review}[1]{#1}
\definecolor{reviewblue}{RGB}{0,101,189}

\begin{document}
\maketitle

	\begin{abstract}
		This paper describes the use of a quadratic manifold for the model order reduction of structural dynamics problems featuring geometric nonlinearities. The manifold is tangent to a subspace spanned by the most relevant vibration modes, and its curvature is provided by modal derivatives obtained by sensitivity analysis of the eigenvalue problem, or its static approximation, along the vibration modes. The construction of the quadratic manifold requires minimal computational effort once the vibration modes are known. The reduced order model is then obtained by Galerkin projection, where the configuration-dependent tangent space of the manifold is used to project the discretized equations of motion.
	\end{abstract}




\section{Introduction}

The use of large Finite Element (FE) models for nonlinear structural analysis is becoming a pressing need in several industrial fields, as for instance the mechanical, aerospace and biomedical. Nowadays, it is relatively easy to generate large models that account for extremely detailed geometric features and material distribution. However, such models are often of prohibitive size and routine simulations to explore different load scenarios, geometric layouts and material choice are severely limited.
Among other nonlinear effects, geometric nonlinearities mainly characterize thin-walled structural components that are typically employed when high stiffness-to-weight and strength-to-weight ratios must be achieved. The redirection of stresses due to non-infinitesimal deflections causes peculiar behaviors as bending and torsion-stretching coupling, buckling, snap-through and mode jumping \cite{Riks19971}. In this context, Reduced Order Models (ROMs) are paramount to enable sound design and optimization
activities. In a broad sense, ROMs are low order realizations of the original model, often referred as High Fidelity Model (HFM). This reduction is achieved through a projection of the full model onto a Reduced Order Basis (ROB) which spans the subspace in which the solution is assumed to lie.


\review{An established method to obtain accurate ROMs by Galerkin projection is the Proper Orthogonal Decomposition \cite{Han01012002,Amsallem:2010aa}, where the reduction basis is constructed using the solution snapshots of the HFM.} Albeit optimal in a sense, it bears the drawback of requiring the full solution. Nonetheless, it is meaningfully applied in the so-called \emph{many-queries} scenarios, for which the cost of the full training simulations is justified. In a preliminary design context, however, the resources required for such an approach might not be available. In this case, it is desirable to build a ROM \emph{not} with the reliance on full simulations, but rather using certain intrinsic characteristics of the underlying physical system, which are usually available at a very small fraction of the computational cost associated to such full simulation(s).

Modal truncation and superposition is a standard practice for linear structural dynamics, as it enables the decoupling of the linear governing equation to readily \review{assess} the dynamic response. However, a reduction based solely on Vibration Modes (VMs) would perform poorly in the presence of geometric nonlinearities, as they typically do not capture the relevant bending/torsion-stretching coupling. This would require the inclusion of in-plane displacement dominated fields in the basis. An appealing enrichment to a ROB of few VMs is constituted by the Modal Derivatives (MDs), which were originally proposed  in \cite{IdelCard2}.  These are computed by differentiating the eigenvalue problem associated to small, undamped vibrations with respect to the modal amplitudes. A static version of their construction (i.e. neglecting the inertial terms) enjoys computational advantages: the MDs thus obtained, are the solutions of a set of linear systems where the coefficient matrix is factorized only once and the right hand sides are symmetric functions of the VMs.

In a reduced basis approach, the MDs could be appended to a ROB constituted by the dominant VMs. This approach leads to very accurate results \cite{SLAATS19951155,TensorRef}. Unfortunately, the number of MDs that can be generated, grows quadratically with the size of the VMs basis used to generate them, thereby severely hampering the efficiency of the method. However, the MDs are in fact the curvature of a quadratic manifold arising from the Taylor expansion of the physical displacement into the direction of the dominant vibration modes. As such, the modal amplitudes associated to the MDs are enslaved, in a quadratic fashion, to those of the VMs, and hence do not require independent reduced unknowns for their description. This approach can often be supported by a sound theoretical justification in examples which are characterized by a special dichotomy in time scales, and corresponds to neglecting the inertial forces associated to the fast dynamics of the system at hand \cite{Rutzmoser2014}. More specifically, the static MDs provide a second order approximation to the underlying critical manifold in such examples \cite{slowfast}. This leads to the notion of the solution lying on a \emph{quadratic} manifold, parameterized by the amplitudes of the dominant VMs. This idea already appeared in a static context when evaluating the initial post-buckling response of thin-walled structures using a perturbation approach \cite{PaoloThesis,Menken1997473}, and, with a very similar framework, in the computation of the backbone curves for nonlinear harmonic responses \cite{wedel1991vibration}. 

\review{The use and efficacy of such a Quadratic Manifold to construct a ROM for dynamic applications remains unexplored and is the focus of this work. In this work we propose a unified approach to construct a ROM using a quadratic manifold comprised of VMs and MDs. The classical notion of the Galerkin projection is extended here to projection on a tangent, configuration-dependent space, which is variationally consistent with the nonlinear mapping between modal and full DOFs. Further, we test this approach on a simple, illustrative example as well as a realistic, industrial structure and compare it with established reduction techniques.}

It is well known that once a ROB has been constructed, significant speed-ups could be obtained by equipping the ROM with one of the many available \emph{hyper-reduction} techniques, \cite{hyperreduction,DEIM,UDEIM,ECSW1,ECSW2} which aim at scaling the cost of evaluation of the reduced nonlinear term down to the order of the number of reduced variables, and not that of the original HFM. Regardless of the specific method adopted, the accuracy of any ROM is determined by the choice of the associated reduction subpace. To this effect, this paper focuses only on the reduction subspace and its generalization to a curved manifold, and speed-up will not be discussed here.

This paper is organized as follows. The generalization of the Galerkin projection onto a nonlinear manifold is sketched in Section \ref{Chap:MOR}. The construction of a MD-based linear manifold is discussed in Section \ref{Sec:LM}. The quadratic manifold is then introduced in Section \ref{Sec:QM}. Numerical results are presented and discussed in \ref{chap:Results}, and finally, the conclusions are given in Section \ref{chap:conlusions}. The appendix describes the comparison of th


\section{Model Order Reduction}
\label{Chap:MOR}
\review{The dynamical response of a structure to externally applied loading is obtained by solving an Initial Value Problem (IVP). This IVP is characterized by a system of second-order Ordinary Differential Equations (ODEs) usually resulting from the FE discretization of the governing Partial Differential Equations (PDEs), and can be written in the following form}:
\begin{equation}
\label{eqn:govEqn}
\begin{gathered}
\mathbf{M}\ddot{\mathbf{u}}(t) + \mathbf{C}\dot{\mathbf{u}}(t) + \mathbf{f}(\mathbf{u}(t)) = \mathbf{g}(t)   \\
\mathbf{u}(t_0) = \mathbf{u}_0 \\
\dot{\mathbf{u}}(t_0) = \mathbf{v}_0,
\end{gathered}
\end{equation}
where the solution $ \mathbf{u}(t)\in \mathbb{R}^n $ is a high dimensional generalized displacement vector with the initial conditions $ \mathbf{u}_0 $ for displacements and $ \mathbf{v}_0 $ for velocities given as inputs at initial time $ t_0 $, $ \mathbf{M}\in \mathbb{R}^{n\times n} $ is the mass matrix, $ \mathbf{C} \in \mathbb{R}^{n\times n}$ is the damping matrix, $ \mathbf{f}(\mathbf{u}) : \mathbb{R}^n \mapsto \mathbb{R}^n$ is the nonlinear internal force and $ \mathbf{g}(t) \in \mathbb{R}^n $ is the time dependent external load vector. These ODEs are further discretized in time using a suitable time integration scheme, resulting in a high-dimensional, fully discrete, nonlinear system of algebraic equations, to be iteratively solved at each time step with a Newton method (for example). \review{This full solution bears a prohibitive computational cost even for a single-query scenario, not to mention the case when the time integration needs to be performed several times, e.g., to explore different operational scenarios.}

Fortunately, in structural dynamics applications, a relatively small number of "modal" coordinates are expected to govern the system response. This is to say that, in general, the solution may be assumed to evolve on a low dimensional manifold in $ \mathbb{R}^n $. In other words, we seek a mapping $ \boldsymbol{\Gamma}: \mathbb{R}^n \mapsto \mathbb{R}^m $ with $ m \ll n $ such that
\begin{equation}\label{eq:map}
\mathbf{u}(t) \approx \boldsymbol{\Gamma}(\mathbf{q}(t)),
\end{equation}
where $ \mathbf{\Gamma} $ is a general nonlinear mapping and $ \mathbf{q} \in \mathbb{R}^m $ is the reduced vector of unknowns. The semi-discrete equations for dynamic equilibrium in \eqref{eqn:govEqn} can be written in the following variational or weak form (time dependency is omitted for clarity purposes):

\begin{equation}
\label{eqn:varEqm}
\begin{gathered}
\left[ \mathbf{M}\ddot{\mathbf{u}} + \mathbf{C}\dot{\mathbf{u}} + \mathbf{f}(\mathbf{u}) \right] \cdot \delta\mathbf{u} = \mathbf{g}  \cdot \delta\mathbf{u},
\end{gathered}
\end{equation}
where $ \delta\mathbf{u} $ is an admissible variation in the solution $ \mathbf{u} $. By introducing the lower dimensional approximation \eqref{eq:map} into \eqref{eqn:varEqm}, we obtain
\begin{equation}
\label{eqn:varRedEqm}
\left[ \mathbf{M}\ddot{\boldsymbol{\Gamma}}(\mathbf{q}) + \mathbf{C}\dot{\boldsymbol{\Gamma}}(\mathbf{q}) + \mathbf{f}(\boldsymbol{\Gamma}(\mathbf{q})) \right] \cdot \delta\boldsymbol{\Gamma}(\mathbf{q}) = \mathbf{g}  \cdot  \delta\boldsymbol{\Gamma}(\mathbf{q}).
\end{equation}
Observing that the variation $ \delta\boldsymbol{\Gamma}(\mathbf{q}) $ is given by $ \parderv{\boldsymbol{\Gamma}(\mathbf{q})}{\mathbf{q}} \delta\mathbf{q} $, and $\delta\mathbf{q}$ being  arbitrary,  we finally obtain
\begin{equation}
\label{eqn:RedEqm}
\mathbf{P_{\Gamma}}^T\left[ \mathbf{M}\ddot{\boldsymbol{\Gamma}}(\mathbf{q}) + \mathbf{C}\dot{\boldsymbol{\Gamma}}(\mathbf{q}) + \mathbf{f}(\boldsymbol{\Gamma}(\mathbf{q})) \right]  = \mathbf{P_{\Gamma}}^T \mathbf{g},
\end{equation}
where $  \mathbf{P_{\Gamma}} $ denotes the tangent subspace $ \parderv{\boldsymbol{\Gamma}(\mathbf{q})}{\mathbf{q}} $.

If the mapping function is chosen to be linear such that $ \boldsymbol{\Gamma}(\mathbf{q}):= \mathbf{Vq} $ (where $ \mathbf{V}\in \mathbb{R}^{n\times m} $ is typically a basis spanning some lower dimensional subspace in $ \mathbb{R}^n $ in which the solutions is assumed to live), the above treatment leads to the \textit{Bubnov-Galerkin} or simply the  \textit{Galerkin Projection}. The reduced ODEs can then be simplified as
\begin{equation}
\label{eqn:GalerkinProj}
\underbrace{\mathbf{V}^T\mathbf{MV}}_{\tilde{\mathbf{M}}}\ddot{\mathbf{q}}(t) + \underbrace{\mathbf{V}^T\mathbf{CV}}_{\tilde{\mathbf{C}}}\dot{\mathbf{q}}(t) + \underbrace{\mathbf{V}^T\mathbf{f}(\mathbf{Vq}(t)) }_{\tilde{\mathbf{f}}(\mathbf{q}(t))} = \mathbf{V}^T\mathbf{g}(t),   \\
\end{equation}
where $ \tilde{\mathbf{M}} $, $ \tilde{\mathbf{C}} \in \mathbb{R}^{m\times m}$  are the reduced mass and damping matrices, respectively. For a linear system, one would have $ \mathbf{f(u)} = \mathbf{Ku} $ ($ \mathbf{K} \in \mathbb{R}^{n\times n}$ being the linear stiffness matrix), and a reduced stiffness matrix $ \tilde{\mathbf{K}} = \mathbf{V}^T\mathbf{K}\mathbf{V} \in \mathbb{R}^{m\times m} $ is also obtained.

The choice of projection basis $ \mathbf{V} $ (or the mapping $ \boldsymbol{\Gamma}(\mathbf{q}) $) is critical in determining the accuracy of the reduced solution. The size of the basis (or the reduced number of unknowns) is important in determining the speed-up in computation time. In further sections, we consider the candidates for such linear and nonlinear mappings.

\section{Linear Manifold}
\label{Sec:LM}

The existence of an invariant subspace (or manifold) is a key requirement for reduction of system \eqref{eqn:govEqn}, as described above. Upon reduction over an invariant \textit{linear} subspace, we refer to the reduced solution to lie on a \textit{Linear Manifold}. Finding a suitable invariant subspace is by no means trivial, if at all possible. In the linear mappings context, the Proper Orthogonal Decomposition (POD) is a remarkably versatile and robust method. However, one of its drawbacks is the need for training snapshots of solution vectors which are obtained from a full nonlinear run. Typically, a reduction basis constructed in such a manner is suitable only for the trained set(s) of loading(s). One natural question that arises then, is that if it is possible to obtain a Reduced Order Model (ROM) without the need of a full nonlinear run.

In structural dynamics, linear modal analysis is a powerful and insightful tool for preliminary analysis of any system on a \emph{linear} level. The use of a truncated set of Vibration Modes (VMs) certainly goes a long way towards reduction of a linear system without the need of a full solution. The concept of reduction using a linear basis of vibration modes can be extended to nonlinear systems by the use of \textit{Modal Derivatives}. We briefly review these concepts in the following sections.

\subsection{Vibration Modes}

The system in \eqref{eqn:govEqn} can be linearized around its static equilibrium position. Without the loss of generality, we assume the equilibrium configuration is $ \mathbf{u}_{eq} = \mathbf{0}$ to obtain
%
\begin{align}
\label{Eqn:linEOM}
\mathbf{M}\ddot{\mathbf{u}} + \mathbf{C}\dot{\mathbf{u}} + \mathbf{K}|_{eq}\mathbf{u}= \mathbf{g}(t),
\end{align}
where
\begin{equation}
\mathbf{K}|_{eq}=\left.\frac{\partial\mathbf{f(u)}}{\partial\mathbf{u}}\right|_{\mathbf{u=0}}
\end{equation}

This linearized system would be a good approximation to the original system in \eqref{eqn:govEqn} for small enough displacements from the linearization point. It is well known that for such a linear system, the system response can be written as a linear combination of constant eigenvectors (also referred to as the eigenmodes or VMs) in the structural dynamics context) which form a basis of $ \mathbb{R}^n $ as follows.
\begin{equation}
\mathbf{u}(t) = \sum_{i=1}^{n} \boldsymbol{\phi}_i \eta_i(t)
\end{equation}
where the eigenmodes $ \boldsymbol{\phi}_i \in \mathbb{R}^n $ are found by the solution of the generalized eigenvalue problem \footnote{Here we neglect the damping contribution in eigenvalue problem to avoid complex eigenvalues and vectors. Note that for damped linear systems with low damping or \textit{modal/Rayleigh}  damping as explained in \cite{MechVibrations}, the eigenvectors for an undamped system are a good approximation for the damped counterpart and still form a good basis for linear modal superposition. Such a damping is very popular in structural dynamics and the theory is illustrated in this context.}
\begin{equation}
\label{EVP}
(\mathbf{K}|_{eq}-\omega_i^2\mathbf{M})\boldsymbol{\phi}_i = \mathbf{0}
\end{equation}
($ \omega_i^2 $ is the eigenvalue or the eigenfrequency squared). This concept of expressing the solution $ \mathbf{u}(t) $ in terms of a basis of eigenvectors is referred to as the principle of \textit{linear modal superposition}. However, if one is considering the slowly varying dynamics of the system, then it can be shown that the response can be very accurately approximated by a few low frequency modes and a modal truncation can be obtained \cite{MechVibrations}.
\begin{equation}
\mathbf{u}(t) \approx \sum_{i=1}^{m} \boldsymbol{\phi}_i \eta_i(t) = \boldsymbol{\Phi}\boldsymbol{\eta}(t),
\end{equation}
where $ \boldsymbol{\Phi}\in \mathbb{R}^{n\times m} $, $ \boldsymbol{\eta}(t) = [\eta_1(t) \ \eta_2(t) \ \cdots \eta_m(t)]^T \in \mathbb{R}^m, m \ll n $. Thus in doing so, we introduce a mapping $ \mathbf{y}:\mathbb{R}^n\longmapsto\mathbb{R}^m $ such that $ \mathbf{u}=\mathbf{u(\boldsymbol{\eta})=}\boldsymbol{\Phi}\boldsymbol{\eta} $. Since $ m \ll n $, this reduces the number of unknowns in the system and an effective ROM is obtained for linear systems. Note that this is equivalent to performing a Galerkin projection as shown in \eqref{eqn:GalerkinProj} for a linearized system, where $ \mathbf{\Phi} $ is the reduction basis and $ \boldsymbol{\eta} $ are the corresponding reduced unknowns.

\subsection{Modal Derivatives}

When the deviation from the linearization point increases, the response of \eqref{Eqn:linEOM} can no longer be considered as a good approximation for the original nonlinear counterpart. One might still think of using the VMs obtained from the linearized model to form a reduction basis for the reduction of the nonlinear set of equations. A basis composed of a few dominant VMs, however, is typically not sufficient for reduction since it does not feature the dominant coupling effects (e.g. membrane-bending), typical of geometrically nonlinear structures.

Earlier work in \cite{IdelCard1,IdelCard2} and recent work in \cite{WWS,WITTEVEEN} discuss the use of the so called Modal Derivatives (MDs) to capture the response of the nonlinear system upon departure from the linear behavior. \review{After replacing $ \mathbf{K}_{eq} $ in \eqref{EVP} with the tangential stiffness matrix $ \mathbf{K} $ and differentiating \eqref{EVP} with respect to the modal amplitude $ \eta_j $ (assuming $ \mathbf{M} $ to be a constant mass matrix), the resulting equation evaluated at equilibrium yields}
\begin{gather}
\label{EVD}
(\mathbf{K}|_{eq}-\omega_i^2|_{eq}\mathbf{M})\left.\parderv{\boldsymbol{\phi}_i}{\eta_j}\right|_{eq} + \left(\left.\parderv{\mathbf{K}}{\eta_j}\right|_{eq}-\left.\parderv{\omega_i^2}{\eta_j}\right|_{eq}\mathbf{M}\right)\boldsymbol{\phi}_i|_{eq} = \mathbf{0},
\end{gather}
where the MD $ \parderv{\boldsymbol{\phi}_i}{q_j} $ denotes the derivative of the $i^{th}$ mode in the $j^{th}$ modal direction. Here the tangent stiffness matrix derivative w.r.t. $ \eta_j $ is obtained by giving the system a displacement in the direction of $ \boldsymbol{\phi}_j|_{eq} $ i.e.\,,
\begin{equation}
\label{eqn:Ksens}
\left.\parderv{\mathbf{K}}{\eta_j}\right|_{eq} = \left.\parderv{\mathbf{K}(\mathbf{u} = \eta_j\boldsymbol{\phi}_j|_{eq}))}{\eta_j}\right|_{\eta_j=0}
\end{equation}

Physically, an MD represents the sensitivity of VM $ \boldsymbol{\phi}_i $ corresponding to a displacement given in the direction of VM $ \boldsymbol{\phi}_j $. As will be shown, these MDs can be used as efficient tools to model the departure from the linear behavior in a nonlinear system.

\begin{framed}
	\begin{rmk}
	Conceptually, VMs are defined only about an equilibrium point and modal amplitudes are not parameters which change the equilibrium points and hence the VMs. Here the spectral expansion of tangent operators leads to this intuitive notion of MDs, which is not linked to free vibration of a nonlinear system.
	\end{rmk}
\end{framed}

\subsection{Calculation of Modal Derivatives}
\label{sec:MDCal}
It is easy to see that  $ \left.\parderv{\boldsymbol{\phi}_i}{q_j}\right|_{eq} $ cannot be trivially obtained from \eqref{EVD} since the coefficient matrix is singular by definition (cf. \eqref{EVP}) . This singularity can be dealt with by imposing a normalization condition for the eigenmodes. Reference \cite{EigenvectorDerivativeThesis} covers an extensive account of different solution techniques and introduces a generalised approach to find eigenvector derivatives for different kinds of normalizations. The popular mass normalization has been adopted here i.e.
\begin{equation}
\label{eq:massnorm}
\boldsymbol{\phi}_i^T \mathbf{M} \boldsymbol{\phi}_i = 1 ~~~~~~~~ \forall i \in \{1,2,\dots,m\}.
\end{equation}
Differentiating the equation above w.r.t. the modal amplitude results in
\begin{equation}
\boldsymbol{\phi}_i^T \mathbf{M} \parderv{\boldsymbol{\phi}_i}{\eta_j} + \boldsymbol{\phi}_i^T \mathbf{M}^T \parderv{\boldsymbol{\phi}_i}{\eta_j} = 0 ~~~~~~~~ \forall i,j \in \{1,2,\dots,m\}.
\end{equation}
Exploiting the symmetry of $ \mathbf{M} $ and subsequent evaluation at the equilibrium position results in the following relation
\begin{equation}
\label{normalization}
\boldsymbol{\phi}_i^T|_{eq} \mathbf{M} \left.\parderv{\boldsymbol{\phi}_i}{\eta_j}\right|_{eq} = 0 ~~~~~~~~ \forall i,j \in \{1,2,\dots,m\}.
\end{equation}
The following \emph{direct} approach to calculate the MDs can then be formulated using \eqref{EVD}, \eqref{normalization}:
\review{
	\begin{equation}
\label{eq:MDs}
\begin{bmatrix}
[\mathbf{K}|_{eq}-\omega_i^2|_{eq}\mathbf{M}]_{n\times n} & -\left[ \mathbf{M}\boldsymbol{\phi}_i|_{eq}\right]_{n \times 1} \\
-\left[ \mathbf{M}\boldsymbol{\phi}_i|_{eq}\right]_{1\times n}^T  & 0_{1\times 1}
\end{bmatrix}
\begin{bmatrix}
\left.\parderv{\boldsymbol{\phi}_i}{\eta_j}\right|_{eq} \\
\left.\parderv{\omega_i^2}{\eta_j}\right|_{eq}
\end{bmatrix}
=
\begin{bmatrix}
-\left.\pardervd{\mathbf{K}}{\eta_j}\right|_{eq}\boldsymbol{\phi}_i|_{eq}\\
0
\end{bmatrix}\,.
\end{equation}}
The above non-singular system can be used to solve for the MDs $ \left.\parderv{\boldsymbol{\phi}_i}{\eta_j}\right|_{eq} $. This approach, however, is not very attractive since it destroys the band structure of the original system. Nonetheless, it is rigorous and accurate, and has been used here for the calculation of MDs. Apart from this direct approach, the pseudo inverse technique (cf. \cite{MechVibrations}) and the Nelson's method \cite{nelson} are some techniques that preserve the band structure of the matrices.\\
\review{Regardless of the method adopted to solve \eqref{EVD}, a high dimensional matrix needs to be factorized for each $\omega_i$. When considered together for a large number of modes, the computational costs of this factorization could be significantly high. Idelsohn and Cardona \cite{IdelCard2} discuss a way to approximate the problem \eqref{EVD} by neglecting the inertial contribution, i.e. by eliminating the terms containing $ \mathbf{M} $ in it to obtain}
\begin{equation}
\label{eqn:SMD}
\mathbf{K}|_{eq}\left.\pardervd{\boldsymbol{\phi}_i}{\eta_j}\right|_{eq}^s = -\left.\pardervd{\mathbf{K}}{\eta_j}\right|_{eq}\boldsymbol{\phi}_i|_{eq}\,.
\end{equation}
We call the MDs calculated in such manner as \emph{Static} MDs (SMDs). \review{It is then easy to see that the computation of SMDs is much easier than solving \eqref{eq:MDs} for every mode, since it involves the factorization of $ \mathbf{K}|_{eq} $ only once. The use of SMDs can be shown to be analogous to Static Condensation \cite{Mignolet} in special cases. This is discussed in more details further in the paper.}  The superscript $ s $ in  $ \left.\parderv{\boldsymbol{\phi}_i}{\eta_j}\right|_{eq}^s $ stands for static.

\begin{thm}[Symmetry of SMDs]
	\label{thm:SMD}
	The Static Modal Derivatives given in \eqref{eqn:SMD} are symmetric, i.e. $ \left.\parderv{\boldsymbol{\phi}_i}{\eta_j}\right|_{eq}^s = \left.\parderv{\boldsymbol{\phi}_j}{\eta_i}\right|_{eq}^s $.
\end{thm}

\textit{Proof}:
It is easy to see that the stiffness matrix derivative given by \eqref{eqn:Ksens} can be written as
\begin{equation}
\left.\parderv{\mathbf{K}}{\eta_j}\right|_{eq} = \left.\parderv{\mathbf{K}(\mathbf{u})}{\mathbf{u}}\right|_{\mathbf{u=0}} \cdot \boldsymbol{\phi}_j|_{eq} = \left.\parderv{^2\mathbf{f}(\mathbf{u})}{\mathbf{u}\partial\mathbf{u}}\right|_{\mathbf{u=0}} \cdot \boldsymbol{\phi}_j|_{eq}\,.
\end{equation}
Substituting this into \eqref{eqn:SMD}, we obtain
\begin{equation}
\mathbf{K}|_{eq}\left.\pardervd{\boldsymbol{\phi}_i}{\eta_j}\right|_{eq}^s = -\left(\left.\parderv{^2\mathbf{f}(\mathbf{u})}{\mathbf{u}\partial\mathbf{u}}\right|_{\mathbf{u=0}} \cdot \boldsymbol{\phi}_j|_{eq} \right) \cdot \boldsymbol{\phi}_i|_{eq}\,.
\end{equation}
But the third order tensor $ \left(\parderv{^2\mathbf{f}(\mathbf{u})}{\mathbf{u}\partial\mathbf{u}}\right) $ which contains the second order partial derivatives, is symmetric by Shwarz' theorem ($ \mathbf{f(u)}\in C^2 (\mathbb{R}^n,\mathbb{R}^n) $) i.e.  $ \left(\parderv{^2\mathbf{f}(\mathbf{u})}{\mathbf{u}\partial\mathbf{u}}\right)_{Iij} = \left(\parderv{^2\mathbf{f}(\mathbf{u})}{\mathbf{u}\partial\mathbf{u}}\right)_{Iji} $. Thus, we get
\begin{align}
\left.\pardervd{\boldsymbol{\phi}_i}{\eta_j}\right|_{eq}^s &= -(\mathbf{K}|_{eq})^{-1}\left[\left(\left.\parderv{^2\mathbf{f}(\mathbf{u})}{\mathbf{u}\partial\mathbf{u}}\right|_{\mathbf{u=0}} \cdot \boldsymbol{\phi}_j|_{eq} \right) \cdot \boldsymbol{\phi}_i|_{eq}\right]\\
&= - (\mathbf{K}|_{eq})^{-1}\left[\left(\left.\parderv{^2\mathbf{f}(\mathbf{u})}{\mathbf{u}\partial\mathbf{u}}\right|_{\mathbf{u=0}} \cdot \boldsymbol{\phi}_i|_{eq} \right) \cdot \boldsymbol{\phi}_j|_{eq} \right]= \left.\pardervd{\boldsymbol{\phi}_j}{\eta_i}\right|_{eq}^s\,.
\end{align}

$ \hfill\blacksquare $
\begin{framed}
	\begin{rmk}[Symmetry of MDs]
	\label{rem:MDs}
	While the SMDs have been shown to be symmetric, such a claim cannot be made for the modal derivatives as given by \eqref{eq:MDs}. These MDs are infact, \textit{not} symmetric in general.
\end{rmk}
\end{framed}

\subsection{(S)MDs in a Reduction Basis}
A linear basis ($ \boldsymbol{\Psi} $) consisting of VMs augmented with these MDs could be used to effectively reduce the nonlinear system \cite{IdelCard1,IdelCard2}.

\begin{equation}
\label{eqn:LinMan}
\boldsymbol{\Psi}= \left[\boldsymbol{\phi}_1|_{eq} ~ \boldsymbol{\phi}_2|_{eq} ~ \dots ~ \boldsymbol{\phi}_m|_{eq} ~ ... ~ \boldsymbol{\theta}_{ij}|_{eq} \dots \right]
\end{equation}

where $ \boldsymbol{\theta}_{ij} = \left.\pardervd{\boldsymbol{\phi}_i}{\eta_j}\right|_{eq} $ or $\left.\pardervd{\boldsymbol{\phi}_i}{\eta_j}\right|_{eq}^s $, are the MDs or the SMDs. Using MDs, one could expect a maximum basis size of $ m^2 $ (if all MDs are linearly independent). Since SMDs are symmetric (cf. Remark \ref{rem:MDs}), a basis $ \boldsymbol{\Psi} \in \mathbb{R}^{n\times M}$ can be obtained using (S)MDs, where $ M = m + \frac{m(m+1)}{2} $ would be its maximum size. Indeed, in both cases inclusion of MDs in the basis is expected to increase the reduced number of unknowns quadratically with the number of VMs ($ m $) in the basis.\\

\begin{framed}
	\review{
		\begin{rmk}[On normalization]
			From the chosen normalization in \eqref{normalization}, it is easy to see that any VM $ \boldsymbol{\phi}_i $ is $ \mathbf{M} $-orthogonal to all corresponding MDs $ \left.\parderv{\boldsymbol{\phi}_i}{\eta_j}\right|_{eq} \forall j\in\{1,\dots,m\}$. However, this is not sufficient to ensure that $ \boldsymbol{\Psi} $ in \eqref{eqn:LinMan} possesses a full column rank, since all vectors in $ \boldsymbol{\Psi} $ may not be mutually orthogonal (or linearly independent). Consequently, the condition number of $\boldsymbol{\Psi}$ might suffer. In the companion paper~\cite{Johannes} it is shown, that the condition number decreases with an increasing number of modes and (S)MDs. Hence, an orthogonalization or a deflation of the linear basis~$\boldsymbol{\Psi}$ is necessary to avoid bad conditioning or even singularities in the reduced model. 
		\end{rmk}	
	}
\end{framed}

\subsection{Optimal (S)MDs basis selection}
\label{Sec:MDS}

As seen above, the MDs capture the essential second order non-linearities of the system. But if all the MDs corresponding to a given set of VMs are used in order to augment the basis, then the size of the basis increases with $ \mathcal{O}(m^2) $. This is undesirable and in practise only a few MDs could be selected to capture the nonlinear response of the system. \review{We propose different heuristic methods to \textit{a priori} select (S)MDs which are expected to produce the highest contribution depending on the applied loading function $ \mathbf{g}(t) $ during time integration.}

\subsubsection{Maximum Modal Interaction (MMI)} Similar to the selection strategy proposed in \cite{MMI}, the basic idea of this method is to calculate the modal interaction between different modes during a linear run and use the SMDs corresponding to the maximum interaction in augmenting the basis. A weighting matrix $ \mathbf{W} $ can be built to rank the SMDs in order of relevance
\begin{equation}
\label{eqn:MMIweightage}
	W_{ij} = \int_{0}^{T}|\eta_i(t)\eta_j(t)|~\mathrm{d}t,
\end{equation}
where $ W_{ij} $ represents the \review{ weight} of the SMD $ \boldsymbol{\theta}_{ij} $ and $ \eta_i(t) $ represents the time varying amplitude of the $ i^{th} $ mode \review{ obtained in response to the applied external loading}, in a linear modal superposition run over time span $[0,T]$. Since this procedure only involves a linear modal superposition run, this weighting matrix is obviously extremely cheap to obtain. By looking at the product of two modal amplitudes, one obtains the interaction between the corresponding modes in the sense that if the relevant weightage becomes high, then the corresponding nonlinearity may get triggered and that SMD becomes important. Furthermore, as the weighting matrix $ \mathbf{W} $ is symmetric, this technique would be suitable for ranking only SMDs (and not MDs), since they are also symmetric. Using it rank to MDs would imply equality between $ \left.\pardervd{\boldsymbol{\phi}_j}{\eta_i}\right|_{eq} $ and $ \left.\pardervd{\boldsymbol{\phi}_i}{\eta_j}\right|_{eq} $,  which is general does not hold, as stated in \eqref{rem:sym}.

\subsubsection{Modal Virtual Work (MVW)}
Here, the basic idea of assigning weights is to compute the virtual work done by the nonlinear elastic forces arising from one mode upon another mode.
\review{First, the maximum modal amplitudes for each of the $ m $ VMs in response to the applied external forcing are computed with a linear modal superposition run. They are multiplied with the corresponding modes to obtain $m$ modal displacement vectors scaled with the maximum amplitude. Then, the nonlinear internal force corresponding to these displacement fields is computed, and projected onto each mode to obtain the resultant virtual-work. Finally, the magnitude of this work is collected as the MD-weight in the matrix $ \mathbf{W} \in \mathbb{R}^{m\times m} $. This can be mathematically written as follows:
\begin{align}
	t_{i}^{max} &= \arg\max_{t\in T}|\eta_i(t))|,\\
	W_{ij} &= |\boldsymbol{\phi}_j^T\mathbf{f} (\eta_i(t_{i}^{max})\boldsymbol{\phi}_i  )| \qquad (\text{no summation}).
\end{align}}
Physically, $ W_{ij} $ would also represent the interaction between modes $ \boldsymbol{\phi}_i $ and $ \boldsymbol{\phi}_j $, thereby establishing importance of the corresponding MD. It is easy to see that $ \mathbf{W} $ would not be symmetric in general (it would though be always diagonal when the internal force $ \mathbf{f(u)} $ is linear). Due to this asymmetry, this makes the MVW a suitable technique for ranking MDs obtained from \eqref{eq:MDs}, which are not symmetric in general.

It should be noted that the (S)MDs are only "ranked" using the weighing matrices described above. While the contribution of a low ranked MD is intuitively expected to be less than that of a high ranked one during a reduced nonlinear run, the weights cannot be taken as a quantitative measure of their relative contribution since they are obtained from a linear analysis. \review{Furthermore, different normalizations of the VMs can lead to potentially different weights. In all the presented examples, the VMs are mass-normalized according to \eqref{eq:massnorm}.}

\section{Quadratic Manifold}
\label{Sec:QM}
As explained before, the linear modal superposition using a few VMs is a good technique to obtain the reduced solution of a linear system. However, when the nonlinearities become significant, the modal basis can be augmented with MDs to effectively capture the response. This quickly increases the size of the basis as more VMs are used and defeats the purpose of reduction. Though an MD selection alleviates this problem to some extent, this selection is not robust for different loads. If we assume that the linearised modal subspace which was good for capturing small displacement smoothly persists for displacements in the nonlinear range but this subspace deforms into an analytic manifold around the linearisation point, then, interestingly, it can be shown that the MDs capture the second order components of this analytic modal manifold. To this effect,  we aim to introduce a nonlinear (quadratic) mapping for our purpose of reduction. Such a mapping can be written as
\begin{equation}
\label{eqn:mapping}
\mathbf{u} \approx \mathbf{\Gamma(q)} := \boldsymbol{\Phi}\cdot\mathbf{q} + \dfrac{1}{2}\left( \boldsymbol{\Omega}\cdot \mathbf{q}\right) \cdot \mathbf{q},
\end{equation}
where $ \mathbf{q}\in\mathbb{R}^m, m\ll n $ are the reduced unknowns, $ \mathbf{\Phi}\in \mathbb{R}^{n\times m} $, and $ \mathbf{\Omega} \in \mathbb{R}^{n\times m\times m} $ is a third order tensor. \review{ The mapping~\eqref{eqn:mapping} can be written using the Einstein summation convention in the indicial notation as
\begin{equation}
\label{eqn:mappingein}
\Gamma_I  = \Phi_{Ii}q_j + \frac{1}{2}\Omega_{Iij}q_iq_j\,\quad I \in \{1,\dots,n\},\quad i,j\in \{1,\dots,m\}.
\end{equation}
}
\begin{framed}
	\begin{rmk} [on symmetry of $ \boldsymbol{\Omega} $]
		\label{rem:sym}
It should be noted that this mapping is independent of the anti-symmetric part of $ \boldsymbol{\Omega} $. Indeed, $ \boldsymbol{\Omega} $ can be split into its symmetric and antisymmetric parts as follows.
\begin{equation*}
\boldsymbol{\Omega} = \boldsymbol{\Theta} + \boldsymbol{\Lambda},
\end{equation*}
where 
\begin{align}\label{eqn:theta}
 \Theta_{IJK}  &:= \frac{1}{2}(\Omega_{IJK} + \Omega_{IKJ}), \\ 
 \Lambda_{IJK}  &:= \frac{1}{2}(\Omega_{IJK} - \Omega_{IKJ}),
\end{align}
$ \forall I \in \{1,\dots, n\},~J,K \{1,\dots,m\}$. Thus, $ \boldsymbol{\Theta} $ and $ \boldsymbol{\Lambda} $ are the symmetric and anti-symmetric parts of $ \boldsymbol{\Omega} $, respectively. It is then easy to see in the indicial notation that
\begin{align*}
\Gamma_I  &= \Phi_{Ii}q_j + \frac{1}{2}\Omega_{Iij}q_iq_j \\
&=\Phi_{Ii}q_j +  \frac{1}{2}\Theta_{Iij}q_iq_j + \frac{1}{2}\Lambda_{Iij}q_iq_j  	 \\
&= \Phi_{Ii}q_j + \frac{1}{2}\Theta_{Iij}q_iq_j + \frac{1}{4}(\Omega_{Iij}q_iq_j - \Omega_{Iji}q_iq_j)\\
&\review{= \Phi_{Ii}q_j + \frac{1}{2}\Theta_{Iij}q_iq_j + \frac{1}{4}(\Omega_{Iij}q_iq_j - \Omega_{Iij}q_jq_i)}\\
&= \Phi_{Ii}q_j + \frac{1}{2}\Theta_{Iij}q_iq_j ,
\end{align*}
\end{rmk}
is independent of $ \boldsymbol{\Lambda} $.
\end{framed}

Taking the mapping in \eqref{eqn:mapping} to be a nonlinear extension of linear modal superposition, the first order derivative evaluated at equilibrium would correspond to the set of VMs at equilibrium such that

\begin{gather}
\label{Eqn:VM}
\left.\parderv{\mathbf{\Gamma}}{{q}_j}\right|_{\mathbf{q=0}} = \boldsymbol{\phi}_j|_{eq}.
\end{gather}
Thus $ \mathbf{\Phi} $ would be a matrix containing $ m $ VMs (essentially the same VMs which are good for capturing the linearized system response). Intuitively this means that the tangent space of the manifold $ \mathbf{\Gamma} $ at the equilibrium point is the subspace spanned by the $ m $ VMs. In \cite{Johannes}, we show that this can be generalized to any linear subspace, and not just the subspace spanned by VMs at equilibrium.

The second order derivatives evaluated at equilibrium would then be
\begin{gather}
\label{Eqn:MD}
\left. \frac{\partial^2 \mathbf{\Gamma}}{\partial \mathbf{q}\partial \mathbf{q}}\right|_{\mathbf{q=0}} =  \mathbf{\Theta} ,
\end{gather}
where $ \mathbf{\Theta} $ is the symmetric part of $ \mathbf{\Omega}$ as defined above. This second order derivative gives information about how the tangent space of the manifold $ \mathbf{\Gamma} $ changes on departure from the equilibrium.

We would now like to use the (S)MDs (as given by \eqref{EVD},\eqref{eqn:SMD}) in choosing the components for $ \mathbf{\Omega} $ such that
\begin{equation}
\mathbf{\Omega}_{:ij} = \left.\parderv{\boldsymbol{\phi}_i}{q_j}\right|_{eq}^s \text{ or } \left.\parderv{\boldsymbol{\phi}_i}{q_j}\right|_{eq},
\end{equation}
where $ \mathbf{\Omega}_{:ij}  = \{\mathbf{v}\in \mathbb{R}^n: v_I = \Omega_{Iij}   \}$. We attempt to justify this choice for the quadratic component in the following manner.

If $ \mathbf{\Omega} $ is chosen to consist of the SMDs, then due to Remark~\ref{thm:SMD}, we would obtain that $ \left. \frac{\partial^2 \mathbf{\Gamma}}{\partial q_i\partial q_j}\right|_{\mathbf{q=0}} = \mathbf{\Theta}_{:ij}=  \left.\parderv{\boldsymbol{\phi}_i}{\eta_j}\right|_{eq}^s $\footnote{This is contrary to the suggestions in \cite{IdelCard1} where the author motivates the modal derivatives using a Taylor expansion which leads to an inconsistent definition of the second order components of \eqref{eqn:mapping} by a factor of $ \frac{1}{2} $. This of course would not pose a problem if the MDs are used as independent components in a linear basis (which was the focus of that work). But when MDs are quadratically enslaved to the VM amplitudes (as is done here) then this factor plays a crucial role in the mapping. }.
The resulting tangent space of the manifold would then be given by
\begin{equation}
\label{eqn:tangentspace}
\parderv{\mathbf{\Gamma}}{q_i} = \boldsymbol{\phi}_i|_{eq} + \sum_{j} \left.\parderv{\boldsymbol{\phi}_i}{q_j}\right|_{eq}^s q_j.
\end{equation}
Since the MDs intuitively represent how VMs change upon departure from equilibrium, \eqref{eqn:tangentspace} shows that the tangent space is being corrected using SMDs upon departure from equilibrium. The choice $ \mathbf{\Omega}_{:ij} = \left.\parderv{\boldsymbol{\phi}_i}{q_j}\right|_{eq}^s $ is thus intuitively justified.

A similar argument would hold for the choice $ \mathbf{\Omega}_{:ij} = \left.\parderv{\boldsymbol{\phi}_i}{q_j}\right|_{eq} $, except that one would then obtain $ \left. \frac{\partial^2 \mathbf{\Gamma}}{\partial q_i\partial q_j}\right|_{\mathbf{q=0}} =\frac{1}{2}\left(  \left.\parderv{\boldsymbol{\phi}_i}{\eta_j}\right|_{eq} +  \left.\parderv{\boldsymbol{\phi}_j}{\eta_i}\right|_{eq} \right) $, since the MDs are not symmetric in general (cf. Remark~\ref{rem:MDs}). This is not an issue since due to Remark~\ref{rem:sym}, the mapping is only dependent on the symmetric part of $ \mathbf{\Omega} $.

Thus, with the help of such a quadratic mapping (using (S)MDs), one can extend the classical notion of linear modal truncation to nonlinear systems while remarkably preserving the reduced number of unknowns.

For the purpose of further discussion, we treat $ \mathbf{\Gamma(q)} $ as a general nonlinear mapping as given by \eqref{eqn:mapping}.

The velocity and acceleration are then expressed as functions of the modal coordinates $\mathbf{q}$ as:
\begin{align}
\dot{\mathbf{\Gamma}} = \mathbf{P_\Gamma}\cdot\dot{\mathbf{q}} &= \boldsymbol{\Phi}\cdot\dot{\mathbf{q}} + (\boldsymbol{\Theta}\cdot\dot{\mathbf{q}})\cdot \mathbf{q} ,
\end{align}
\begin{equation}
\ddot{\mathbf{\Gamma}} = \mathbf{P_\Gamma}\cdot\ddot{\mathbf{q}} + \left[\parderv{\mathbf{P_\Gamma}}{\mathbf{q}}\cdot\dot{\mathbf{q}}\right] \cdot\dot{\mathbf{q}} =\boldsymbol{\Phi}\cdot\ddot{\mathbf{q}} + (\boldsymbol{\Theta}\cdot\ddot{\mathbf{q}}) \cdot \mathbf{q} + (\boldsymbol{\Theta}\cdot\dot{\mathbf{q}})\cdot\dot{\mathbf{q}},
\end{equation}
where
\begin{equation}
\label{eqn:PGamma}
\underbrace{\mathbf{P_\Gamma}}_{\in \mathbb{R}^{n\times m}} = \dfrac{\partial \mathbf{\Gamma(q)}}{\partial \mathbf{q}} = \boldsymbol{\Phi} + \boldsymbol{\Theta}\cdot\mathbf{q}.
\end{equation}
\review{The equations above write in index notation:
\begin{align}
(P_\Gamma)_{IJ} &= \dfrac{\partial \Gamma_I}{\partial q_J} = \Phi_{IJ} + \frac{1}{2}\Omega_{IJj}q_j + \frac{1}{2}\Omega_{IjJ}q_j = \Phi_{IJ} + \Theta_{IJj}q_j,\quad I \in \{1,\dots,n\},\quad J,j\in \{1,\dots,m\},\\
\dot{\Gamma}_I &= (P_\Gamma)_{Ii}\dot{q}_i = \Phi_{Ii}\dot{q}_i + \Theta_{Iij}\dot{q}_iq_j,\quad I \in \{1,\dots,n\},\quad i,j\in \{1,\dots,m\},\\
\ddot{\Gamma}_I &= (P_\Gamma)_{Ii}\ddot{q}_i + \parderv{(P_\Gamma)_{Ii}}{q_j}\dot{q_i}\dot{q}_j =\Phi_{Ii}\ddot{q}_i + \Theta_{Iij}\ddot{q}_iq_j + \Theta_{Iij}\dot{q}_i\dot{q}_j,\quad I \in \{1,\dots,n\},\quad i,j\in \{1,\dots,m\}.
\end{align}}

The above calculated expressions can be substituted into \eqref{eqn:RedEqm} to obtain the reduced order model in $ m $ unknowns i.e. $ \mathbf{q} \in \mathbb{R}^m $ as follows:
\review{ 
\begin{gather}
 \label{QMreducedEqn}
 \widetilde{\mathbf{M}}\ddot{\mathbf{q}} + \tilde{\mathbf{p}} + \widetilde{\mathbf{C}}\dot{\mathbf{q}} + \tilde{\mathbf{f}} = \tilde{\mathbf{g}}\,,
 \end{gather}
where $ \widetilde{\mathbf{M}} =\mathbf{P_\Gamma}^T \mathbf{M}\mathbf{P_\Gamma}  $ and $ \widetilde{\mathbf{C}} =\mathbf{P_\Gamma}^T \mathbf{C}\mathbf{P_\Gamma}  $ are the configuration-dependent reduced mass and damping matrix respectively;  $ \tilde{\mathbf{f}} = \mathbf{P_\Gamma}^T \mathbf{f(\Gamma(q))} $ is  the reduced internal force vector; $ \tilde{\mathbf{g}} =\mathbf{P_\Gamma}^T \mathbf{g}(t)  $ is the configuration-dependent, reduced external load vector; and $ \tilde{\mathbf{p}} = \mathbf{P_\Gamma}^T(\mathbf{M}(\boldsymbol{\Theta}\cdot\dot{\mathbf{q}})\cdot\dot{\mathbf{q}}) $ is a convective term, quadratic in generalized velocities, which is similar to the convective terms that appear in the multibody dynamical systems in rotating frames. For the sake of completeness, we have included the steps to obtain a ROM using quadratic manifold and to perform time integration in Algorithm \ref{alg:QM}}.\\

\begin{framed}
	\begin{rmk} [Reduction Error]
		\label{rem:reductionerror}
		The quadratic manifold~(\ref{eqn:mapping}) is a special instance of the more general subspace defined by~(\ref{eqn:LinMan}), i.e. where the amplitudes of the (S)MDs are taken as predefined functions of the amplitudes of the VMs. Hence it is obvious that the approximation resulting from \eqref{eqn:mapping} is never better (and in general worse)  than the approximation obtained with~(\ref{eqn:LinMan}) (in the sense of Galerkin).
	\end{rmk}
\end{framed}

\review{\paragraph{Comparison of the quadratic manifold with static condensation:}It is interesting to compare the proposed reduction method using the Quadratic Manifold with the idea of \textit{static condensation}, whereby the axial modes of the structure are statically condensed out of the system leaving only the bending degrees of freedom in the ROM (see \cite{Mignolet} for a review). The reduction mapping obtained using the quadratic manifold (when SMDs are used) is in fact equivalent to the one obtained from static condensation for the special case of a flat and isotropic structure. This is shown with the help of an example in Appendix \ref{sec:app}. In the case of a more general layout (such as the example that will be presented in this paper), one may not be able to distinguish between transverse and axial DOFs to perform a static condensation approximation. However, for slender structures, this dichotomy is still present in modal coordinates, as low-frequency modes are indeed bending dominated. The quadratic manifold is applicable to such structures. Another distinction from the static condensation approximation lies in the fact that the ROM is obtained after projection of governing equations onto a configuration dependent tangent space, giving rise to quadratic generalized velocity terms and configuration-dependent mass matrix.  Though the reduction mapping is the same in some special cases (as discussed above), the ROM from static condensation approximation misses these terms. The static-condensation-like approaches do have an advantage of being non-intrusive in nature, whereby commercial FE packages can be treated as a black box. Nonetheless, it is worth mentioning that the essential components of a quadratic manifold reduction, namely, the (S)MDs can also be computed in a non-intrusive manner using finite difference schemes from commercial packages.}

\begin{algorithm}
	\caption{Quadratic Manifold Reduction
		\label{alg:QM}}
	\begin{algorithmic}[1]
\review{		\Statex \underline{\textit{Offline} Stage}
		\Statex
		\Require{VMs : $\mathbf{\Phi} = [ \boldsymbol{\phi}_1,\dots, \boldsymbol{\phi}_m] \in \mathbb{R}^{n\times m}$ }
		\Ensure{ Quadratic manifold and tangent space mapping  }
		\Statex
		\State $ \mathbf{\Omega}, \mathbf{\Theta} \gets $ all empty 3-tensor $ \in \mathbb{R}^{n\times m \times m} $ 
		\For{$j \gets 1 \textrm{ to } m$}
		\State Calculate $ \left.\parderv{\mathbf{K}}{\eta_j}\right|_{eq} $ using \eqref{eqn:Ksens}
		\For{$i \gets 1 \textrm{ to } m$}
		\State $ \boldsymbol{\theta}_{ij} \gets \left.\parderv{\boldsymbol{\phi}_i}{\eta_j}\right|_{eq}^s$ or $ \left.\parderv{\boldsymbol{\phi}_i}{\eta_j}\right|_{eq} $ \Comment{Use SMDs from \eqref{eqn:SMD} or MDs from \eqref{eq:MDs} in the quadratic part} 
		\State $ \mathbf{\Omega}(\texttt{:,i,j}) \gets \boldsymbol{\theta}_{ij} $
		\EndFor
		\EndFor
		\State $ \mathbf{\Theta} \gets \texttt{SYMMETRIZE}(\mathbf{\Omega})$ \Comment{\texttt{SYMMETRIZE} returns the symmetric part of a 3-tensor $ \mathbf{\Omega} $ as given by \eqref{eqn:theta}}  
		\State Build \texttt{QM}:	\Comment{Build function \texttt{QM} with argument $ \mathbf{q} $,  which returns $ \mathbf{\Gamma(q)} $ from \eqref{eq:qm1} and $ \mathbf{P_{\Gamma}} $ from \eqref{eqn:PGamma}}
		\Statex $ [\mathbf{\Gamma(q)}, \mathbf{P_{\Gamma}(q)}] \gets \texttt{QM}(\mathbf{q})$
		\Statex
		\Statex \underline{\textit{Online} Stage:} using implicit Newmark time integration (\cite{Newmark}, see \cite{MechVibrations} for a review)
		\Statex
		\Require Initial conditions for \eqref{QMreducedEqn}: $ \mathbf{q}_0, \dot{\mathbf{q}}_0 \in \mathbb{R}^m$;  Newmark time integration parameters: $ \alpha, \beta, \gamma $; time step: $ h $; residual error tolerance for iteration convergence: $ \epsilon $.
		\Ensure{ Reduced-order model using QM  }
		\Statex
		\State $ [\mathbf{u}_0, \mathbf{P_{\Gamma}}] \gets \texttt{QM}(\mathbf{q}_0)$, \Comment{Initialization}
		\State $ \widetilde{\mathbf{M}} \gets \mathbf{P_\Gamma}^T \mathbf{M}\mathbf{P_\Gamma} $, $ \widetilde{\mathbf{C}} \gets \mathbf{P_\Gamma}^T \mathbf{C}\mathbf{P_\Gamma}  $, $ \tilde{\mathbf{f}} \gets \mathbf{P_\Gamma}^T \mathbf{f(u}_0) $, $ \tilde{\mathbf{g}} \gets\mathbf{P_\Gamma}^T \mathbf{g}(t_0)  $, $ \tilde{\mathbf{p}} \gets \mathbf{P_\Gamma}^T(\mathbf{M}(\boldsymbol{\Theta}\cdot\dot{\mathbf{q}}_0)\cdot\dot{\mathbf{q}}_0) $
		\State $ \ddot{\mathbf{q}}_0 = \widetilde{\mathbf{M}}^{-1}[\tilde{\mathbf{g}} - \tilde{\mathbf{f}} - \widetilde{\mathbf{C}}\dot{\mathbf{q}}_0 - \tilde{\mathbf{p}}  ] $ 
		\Statex $ k\gets0 $ 
		\While{$ t < t_{max} $} \Comment{Time-marching loop}
		\State $ k \gets k + 1$
		\State $ t_{k+1}  \gets t_k + h$ \Comment{Time increment}
		\State $ \dot{\mathbf{q}}_{k+1} \gets \dot{\mathbf{q}}_{k} + (1-\gamma)h\ddot{\mathbf{q}}_k $ \Comment{Prediction}
		\State $ {\mathbf{q}}_{k+1} \gets \mathbf{q}_k + h\dot{\mathbf{q}}_{k} + (0.5-\beta)h^2\ddot{\mathbf{q}}_k $
		\State $ \ddot{\mathbf{q}}_k \gets \mathbf{0} $
		\While{\textbf{true}}	\Comment{Newton-Raphson loop}
			\State $ [\mathbf{u}_{k+1}, \mathbf{P_{\Gamma}}] \gets \texttt{QM}(\mathbf{q}_{k+1})$, \Comment{Full solution using QM}
			\State $ \widetilde{\mathbf{M}} \gets \mathbf{P_\Gamma}^T \mathbf{M}\mathbf{P_\Gamma} $, $ \widetilde{\mathbf{C}} \gets\mathbf{P_\Gamma}^T \mathbf{C}\mathbf{P_\Gamma}  $, $ \tilde{\mathbf{f}} \gets \mathbf{P_\Gamma}^T \mathbf{f(u}_{k+1}) $, 
			\Statex \hspace{1cm} $ \tilde{\mathbf{g}} \gets \mathbf{P_\Gamma}^T \mathbf{g}(t_{k+1})  $, $ \tilde{\mathbf{p}} \gets \mathbf{P_\Gamma}^T(\mathbf{M}(\boldsymbol{\Theta}\cdot\dot{\mathbf{q}}_{k+1})\cdot\dot{\mathbf{q}}_{k+1}) $
			\State $\mathbf{r}_{k+1} \gets \widetilde{\mathbf{M}}\ddot{\mathbf{q}}_{k+1} + \tilde{\mathbf{p}} + \widetilde{\mathbf{C}}\dot{\mathbf{q}}_{k+1} + \tilde{\mathbf{f}} - \tilde{\mathbf{g}}  $	\Comment{Residual evaluation}
			\If{$ \|\mathbf{r}_{k+1}\| <\epsilon\|\tilde{\mathbf{f}}\| $} \Comment{Convergence criterion}
			\State \textbf{break}
			\EndIf
			\State $ \mathbf{S} \gets \mathbf{P}_{\Gamma}^T\mathbf{K}(\mathbf{u}_{k+1})\mathbf{P}_{\Gamma} + \frac{\gamma}{\beta h}\widetilde{\mathbf{C}} + \frac{1}{\beta h^2} \widetilde{\mathbf{M}}$ \Comment{(Approximate) Jacobian evaluation}
			\State $ \Delta \mathbf{q} \gets -\mathbf{S}^{-1}\mathbf{r}_{k+1} $ \Comment{Calculation of correction}
			\State $ \mathbf{q}_{k+1} \gets \mathbf{q}_{k+1} + \Delta\mathbf{q}  $ \Comment{Correction}
			\State $ \dot{\mathbf{q}}_{k+1} \gets \dot{\mathbf{q}}_{k+1} + \frac{\gamma}{\beta h}\Delta \mathbf{q} $
			\State $ \ddot{\mathbf{q}}_{k+1} \gets \ddot{\mathbf{q}}_{k+1} + \frac{1}{\beta h^2}\Delta \mathbf{q} $

			\EndWhile
		\EndWhile
}		
	\end{algorithmic}
\end{algorithm}


\clearpage
\section{Applications and Results}
\label{chap:Results}
The proposed reduction techniques are tested and compared on examples. Two models are considered, each being a thin-walled structure with different levels of complexity. Both structures are modelled using triangular shell elements featuring 6 degrees of freedom (DOFs) per node or $ 18 $ DOFs per element. Rayleigh damping is used as structural damping in all models \footnote{A modal damping assumption is used to create a so called \textit{diagonal damping} matrix using weighted sum of Mass and stiffness matrices ($ \mathbf{M} $ and $ \mathbf{K} $ respectively). A modal damping factor of 0.4\% for the first two modes is used to determine the weights (see \cite{MechVibrations} for details about this implementation). This low value is realistic and is chosen to make sure the VMs of undamped system can be used for the reduction.}.

For a general nonlinear system, the evaluation of the tangent stiffness and internal forces required during formation of the Jacobian and the residual respectively, is done by element level assembly during each iteration. This is an expensive online cost apart from the linear system solution. The linear system solution cost is mitigated by projection onto a ROB. But as the system becomes larger, the mapping, nonlinearity evaluation and projection become dominant in taking the CPU time during the time integration. An effective way to deal with this is the evaluation of nonlinearities offline using tensors (see for instance \cite{Mignolet}) or to use hyper-reduction, thereby making time integration independent of the system size. However, as discussed in the introduction of the paper, we perform time integration in the presented examples without such hyper-reduction, as the focus of this paper is on investigating the appropriateness and effectiveness of MDs and quadratic manifolds to approximate the response of non-linear structures. For this reason, we do not report computational speedup but only the reduction in problem size.

For each of the models, the accuracy of the results has been compared to the corresponding full nonlinear solutions. In this context, a global relative error measure shall be used, defined as
\begin{equation}
\label{eqn:GRE}
GRE_{M} = \dfrac{\sqrt{\sum\limits_{t\in S} (\mathbf{u}(t) - \tilde{\mathbf{u}}(t)  )^T \mathbf{M} (\mathbf{u}(t) - \tilde{\mathbf{u}}(t)  ) }}{\sqrt{\sum\limits_{t\in S} \mathbf{u}(t)^T \mathbf{M} \mathbf{u}(t) }} \times 100 \%,
\end{equation}
where $ \mathbf{u}(t) \in \mathbb{R}^{n} $ is the vector of generalised displacements at the time $ t $ obtained from the full nonlinear solution,  $ \tilde{\mathbf{u}}(t) \in \mathbb{R}^{n} $ is the solution based on the reduced model, and $ S $ is the set of time instants at which the error is recorded. The mass matrix $ \mathbf{M} $ provides a relevant normalisation for the generalised displacements, which could be a combination of physical displacements and rotations, as is the case in the shell models shown here.

\subsection{Flat Structure}
\label{sec:Flat}
\begin{figure}[h!]
	\center
	\begin{subfigure}{0.49\linewidth}
		\centering
		\includegraphics[width=\linewidth]{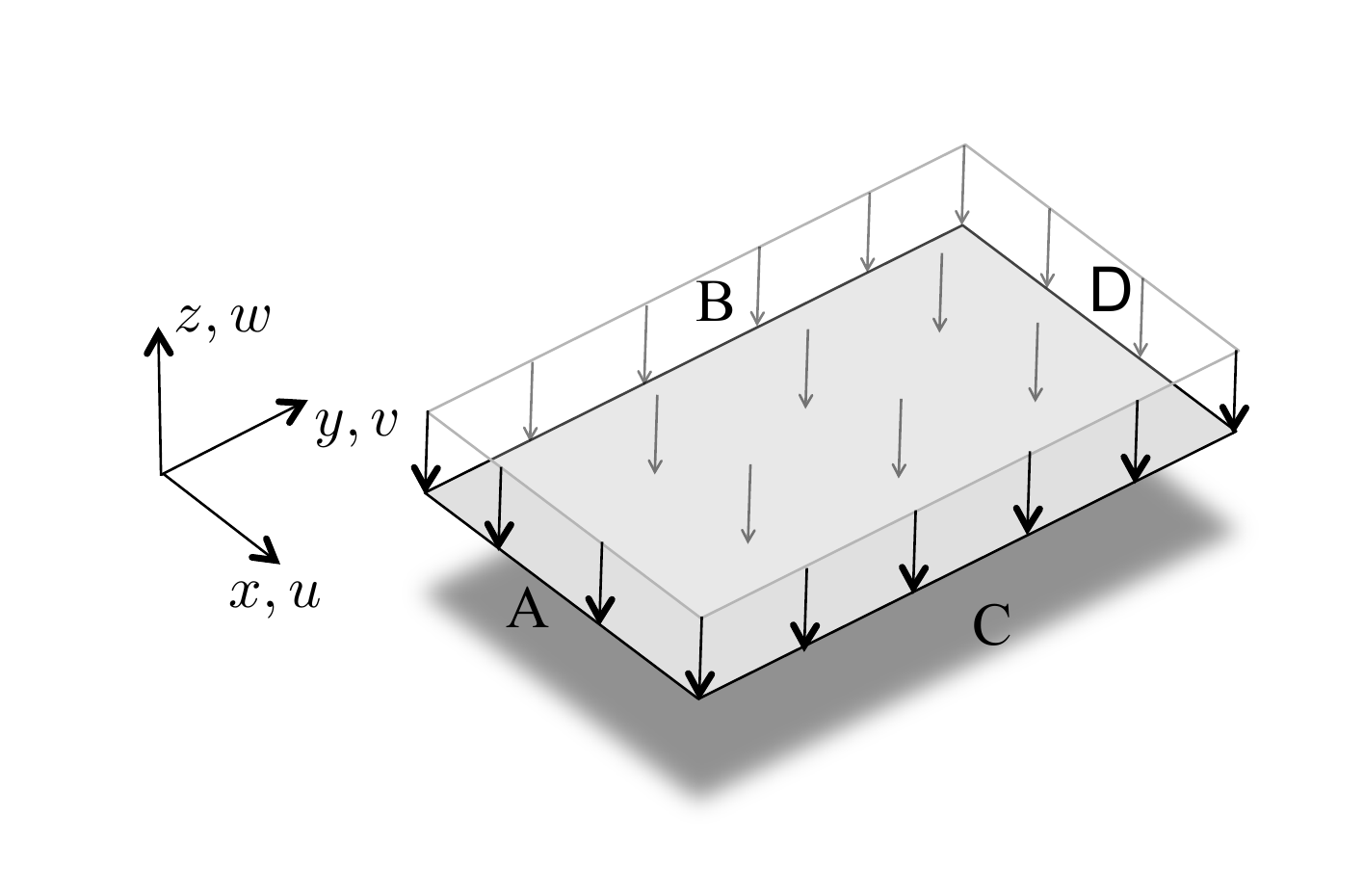}
		\captionsetup{justification=centering}
		\caption{\scriptsize Structure sketch }
		\label{fig:flat_plate_sketch}
	\end{subfigure}
	\begin{subfigure}{0.49\linewidth}
		\centering
		\includegraphics[width=\linewidth]{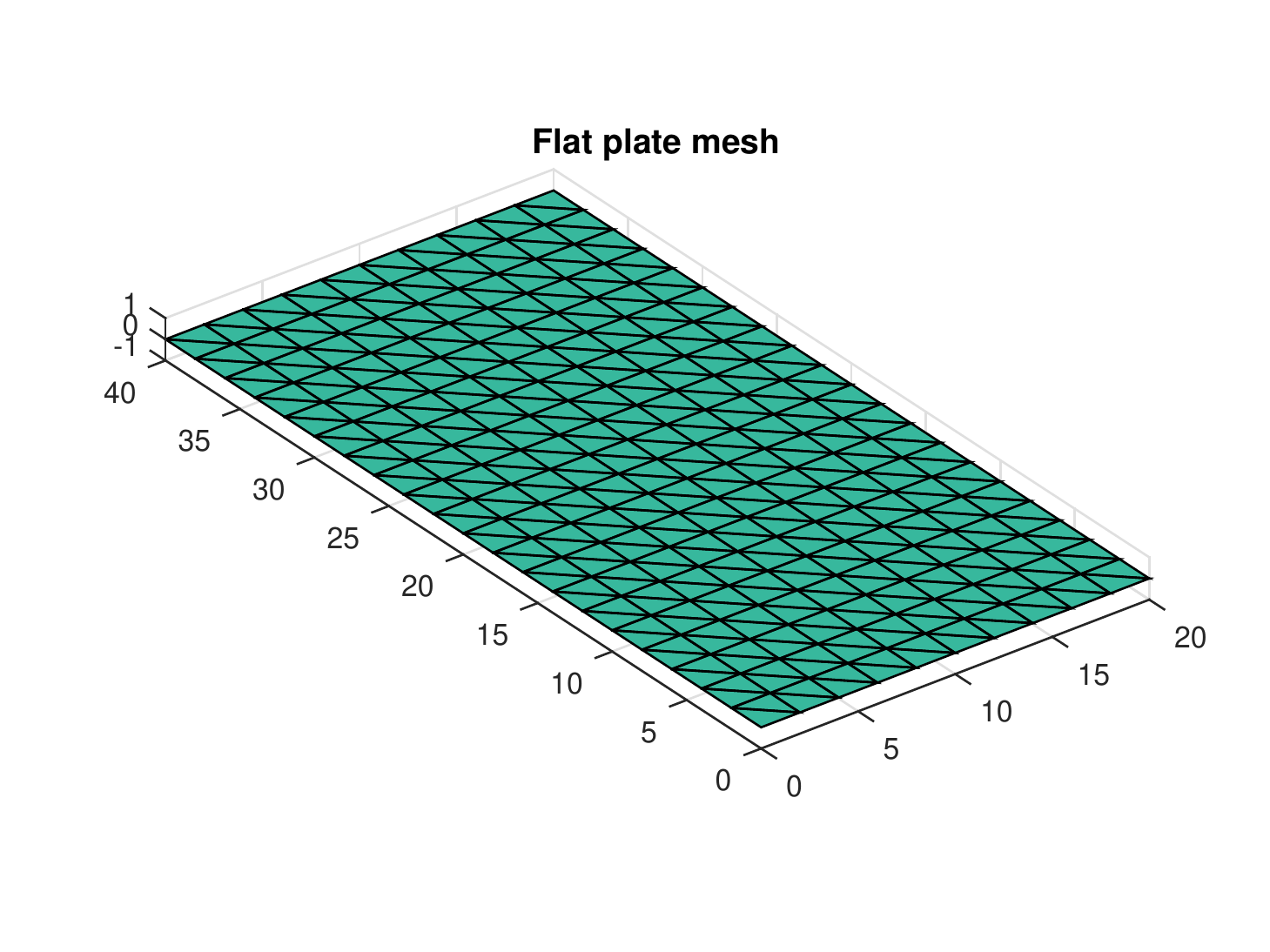}
		\captionsetup{justification=centering}
		\caption{\scriptsize mesh containing 1386 DOFs}
		\label{fig:flatplatemesh}
	\end{subfigure}
	\captionsetup{justification=justified,
		singlelinecheck=false
	}
	\caption{{\bf Model - I: }Flat plate example - simply supported on sides $ A $ and $ D $. The plate is $L=40$ mm long, $H=20$ mm wide, $t=0.8$ mm thick. The Young Modulus is $E=70$ GPa, the Poisson's ratio is $\nu=0.33$, and the density is $\rho=2700$ Kg/m$^{3}$. For simply supported sides, we have $u=v=w=0$. A uniform pressure is applied on the plate, according to the time history $p(t)=P[\sin(\omega t) +\sin(\pi\omega t)]$, where $P = 50 $ N/mm$^2$ and $\omega=2.097\times 10^4$ rad/s. }
	\label{fig:FlatModel}
\end{figure}

A flat plate simply supported on two opposite sides is considered. The Model (henceforth referred to as Model-I) sketch and parameters are shown in Figure~\ref{fig:FlatModel}.

\begin{figure}[h!]
	\center
	\begin{subfigure}{0.45\linewidth}
		\centering
		\includegraphics[width=\linewidth]{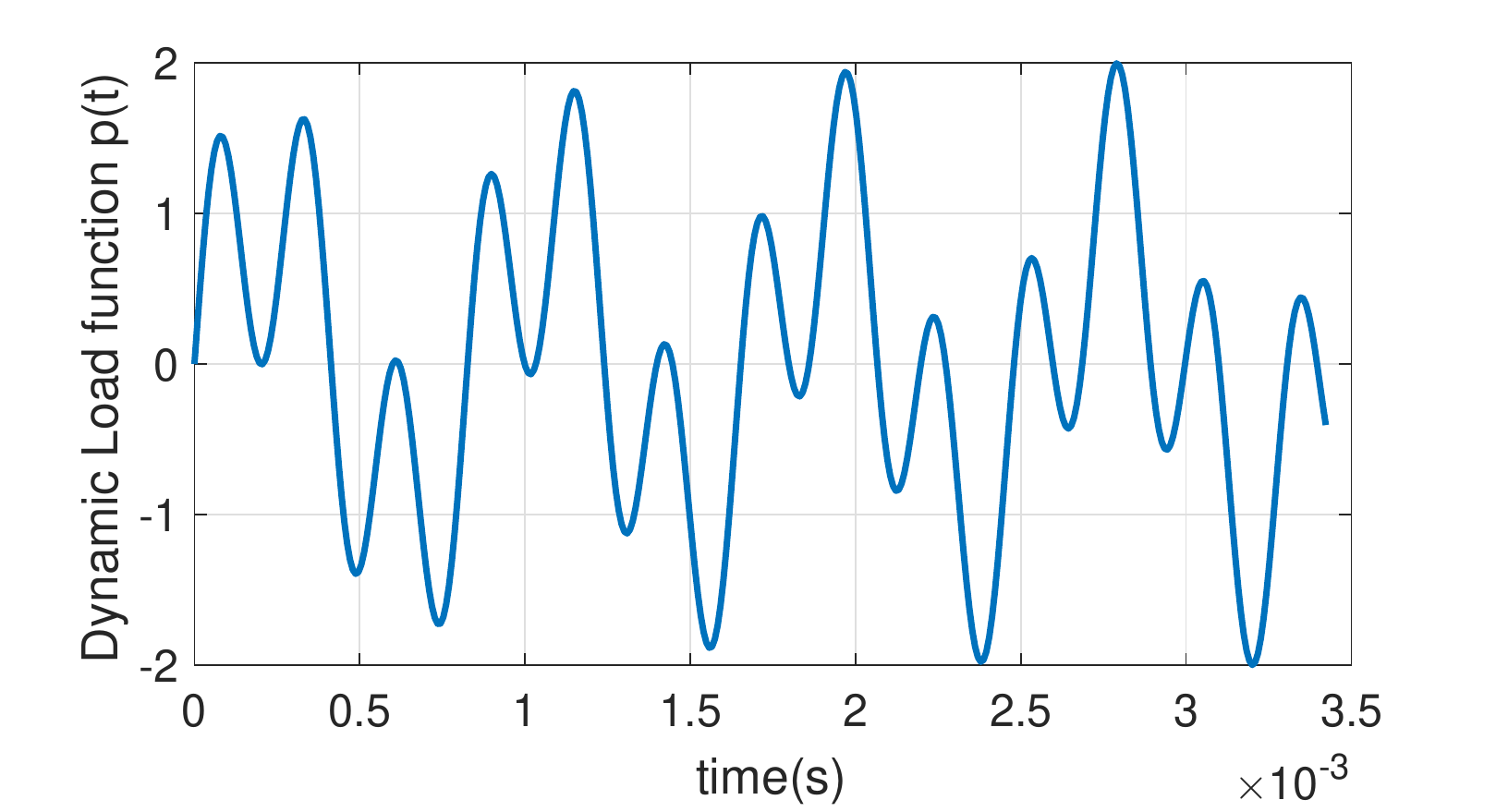}
		\captionsetup{justification=centering}
		\caption{}
		\label{fig:load}
	\end{subfigure}
	\begin{subfigure}{0.54\linewidth}
		\centering
		\includegraphics[width=\linewidth]{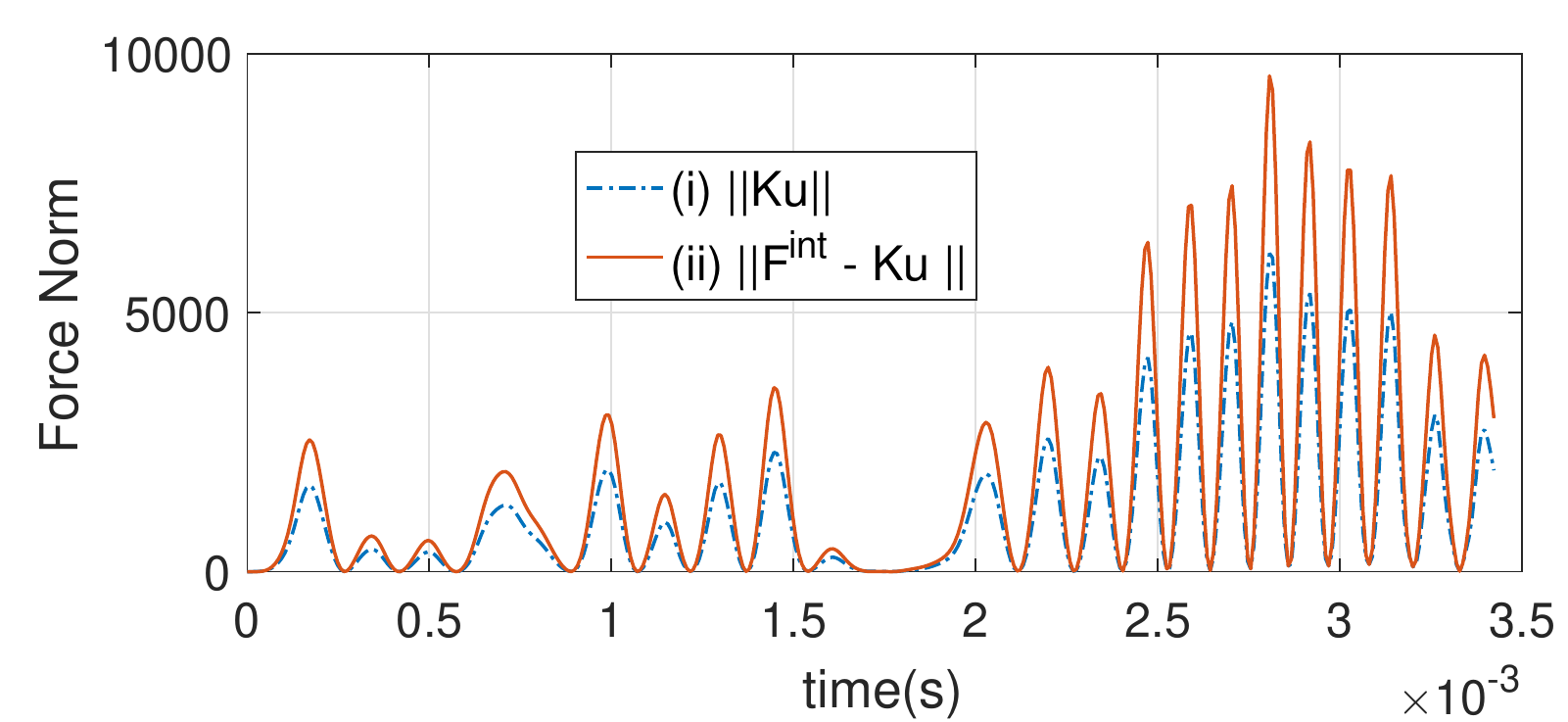}
		\captionsetup{justification=centering}
		\caption{}
		\label{fig:response}
	\end{subfigure}
		\captionsetup{justification=justified,
			singlelinecheck=false
		}
	\caption{ (a) Dynamic load function for a quasi-periodic loading (cf.~\eqref{eqn:loadfactor}). (b) The comparison of the norm of the linear and the nonlinear internal forces during the full nonlinear solution for a flat plate simply supported on two opposite sides (cf. Figure~\ref{fig:FlatModel}).}
	\label{fig:loadresponse}
\end{figure}

 A uniform pressure distribution is chosen to act normal to the plate surface as the external load. A time varying amplitude (load function) is used given by

\begin{gather}
	\mathbf{g}(t) = p(t)\mathbf{l}\\
	\label{eqn:loadfactor} p(t) = p_0[\sin(\omega t) + \sin(\pi\omega t)]
\end{gather}
where $ \mathbf{l} $ is a constant load vector corresponding to a uniform pressure distribution of 1 Pa. Here, $ p(t) $ can be termed as the \textit{dynamic load function} which determines the time-dependency of the external load. Here the results are shown for a quasi periodic choice for $ p(t) $, $ \omega $ is a typical loading frequency chosen as the first eigenfrequency of the linearized system (cf. \eqref{eqn:loadfactor}). The amplitude of loading is kept large enough to trigger significant nonlinear behaviour. The magnitude of $ p_0 $ is chosen such that the magnitude of  the linear and nonlinear internal forces are of the same order (cf. Figure~\ref{fig:response}). This is in agreement with the domain of applicability of von K\'{a}rm\'{a}n kinematics here adopted.

The \textit{full} nonlinear solution was computed by updating the Jacobian of the residual at each Newton-Raphson (N-R) iteration (within every time step). No reduction is involved here and thus the linearised system solve of full size is a costly procedure. This also involves the element level calculation and assembly of tangent-stiffness matrix at each N-R iteration which further adds to the \textit{online} cost (at least for large systems). Different levels of mesh refinement were considered to reach an optimum number of degrees of freedom in terms of accuracy. The resulting mesh, containing 1386 DOFs  and $ n_e = 400 $ elements, is shown in Figure~\ref{fig:flatplatemesh}. This nonlinear solution is used as a reference for comparing the various reduction techniques. For the shown time span, the solution is computed over $ n_h = 400 $ time steps of equal size for all the techniques.

\FloatBarrier
\subsubsection{Linear Manifold}
\begin{figure}[h!]
	\centering
	\begin{subfigure}{0.32\linewidth}
		\centering
		\includegraphics[width=\linewidth]{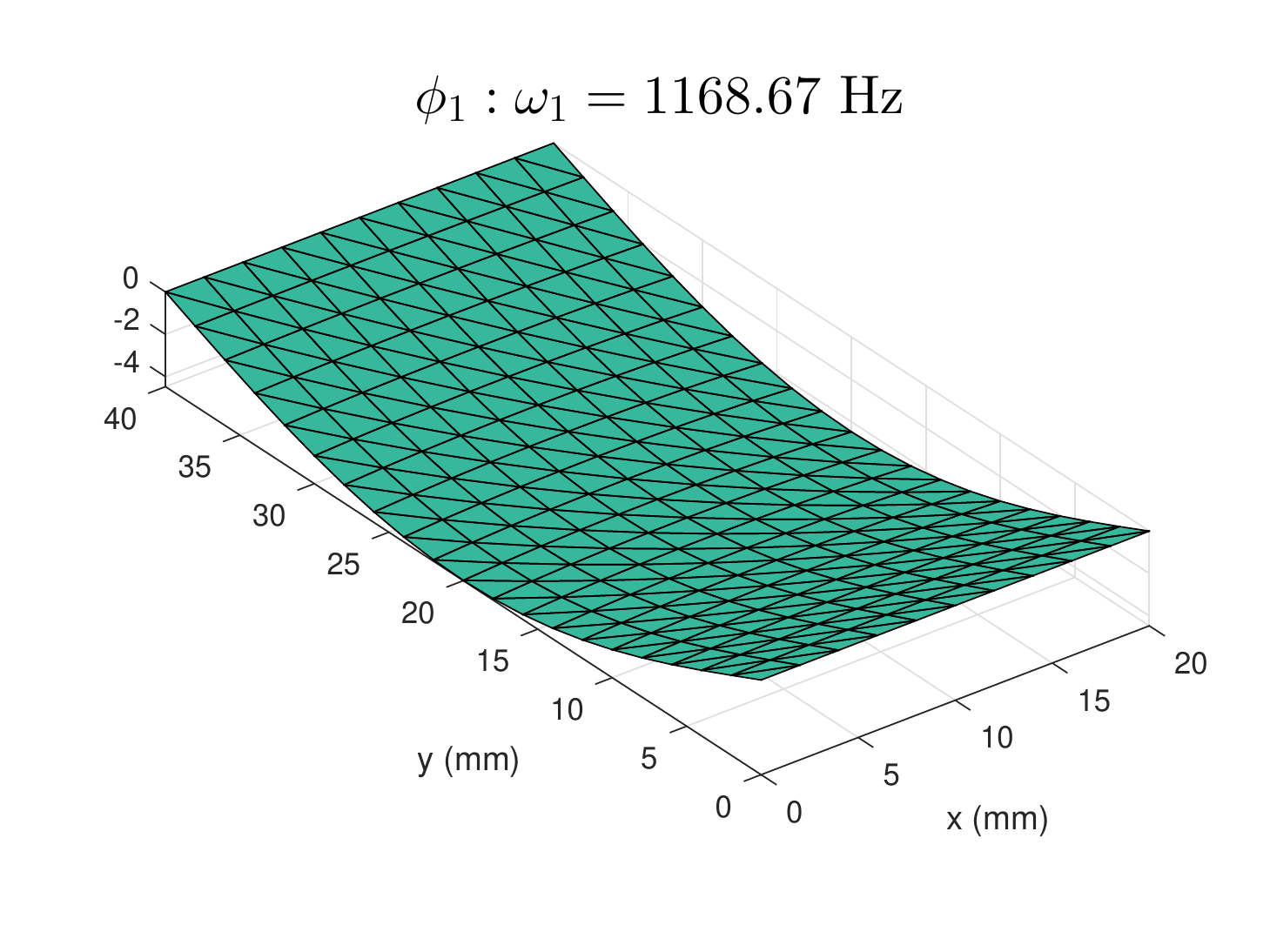}
		\captionsetup{justification=centering}
		\caption{ \footnotesize }
		\label{fig:VM1}
	\end{subfigure}
	\begin{subfigure}{0.32\linewidth}
		\centering
		\includegraphics[width=\linewidth]{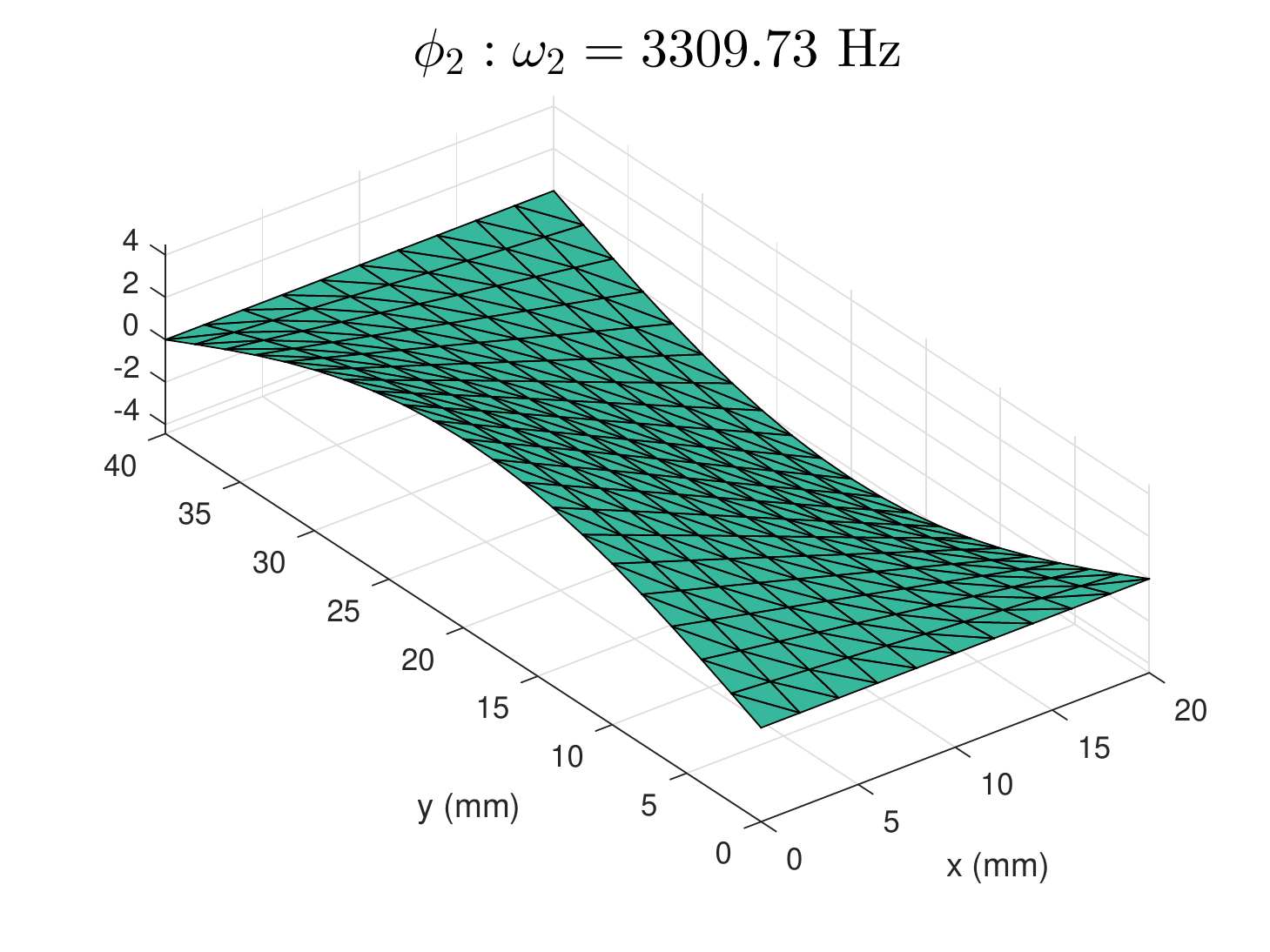}
		\captionsetup{justification=centering}
		\caption{\footnotesize }
		\label{fig:VM2}
	\end{subfigure}
	\begin{subfigure}{0.32\linewidth}
		\centering
		\includegraphics[width=\linewidth]{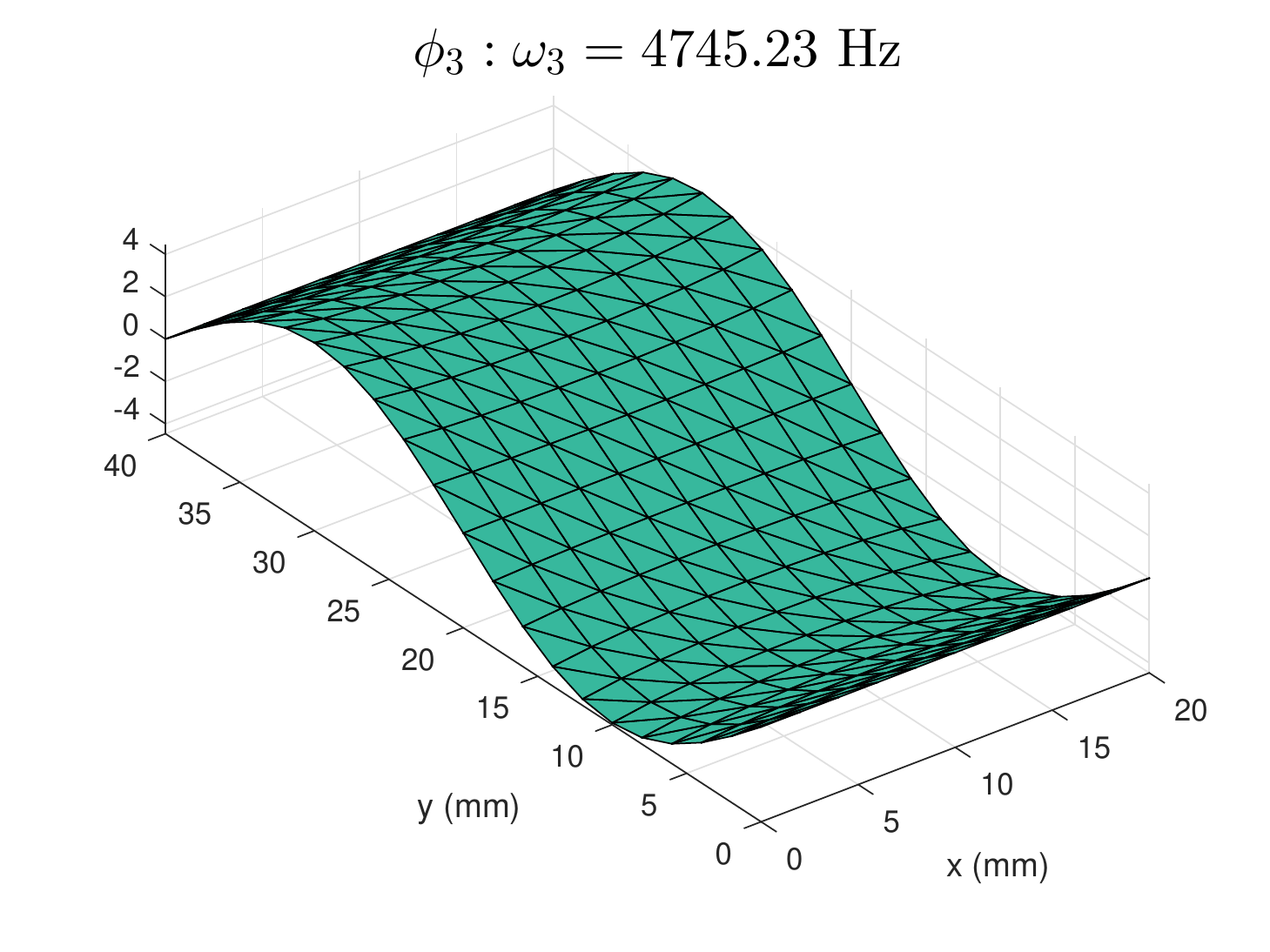}
		\captionsetup{justification=centering}
		\caption{\footnotesize }
		\label{fig:VM3}
	\end{subfigure}
	\begin{subfigure}{0.32\linewidth}
		\centering
		\includegraphics[width=\linewidth]{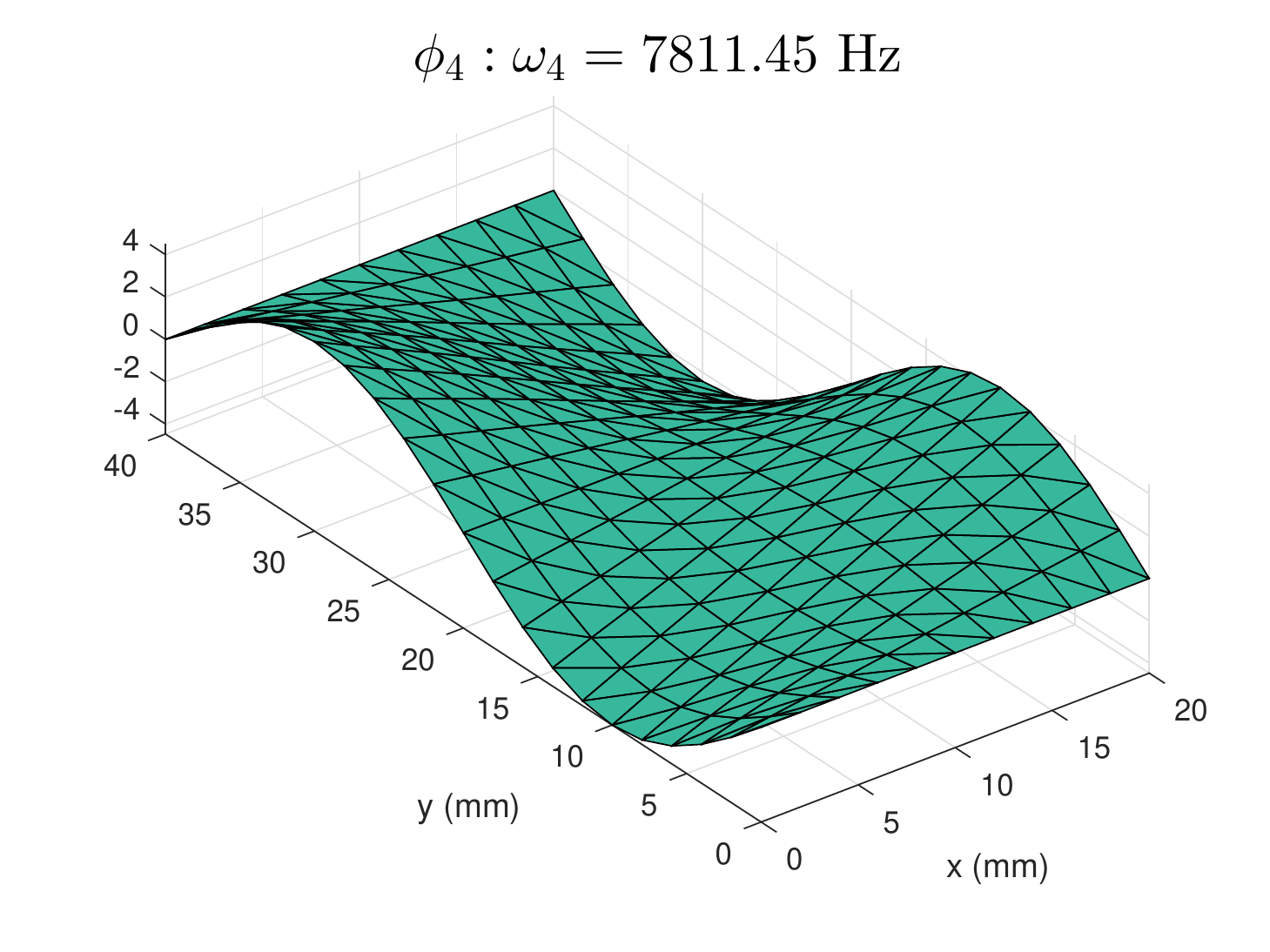}
		\captionsetup{justification=centering}
		\caption{\footnotesize }
		\label{fig:VM4}
	\end{subfigure}
	\begin{subfigure}{0.32\linewidth}
		\centering
		\includegraphics[width=\linewidth]{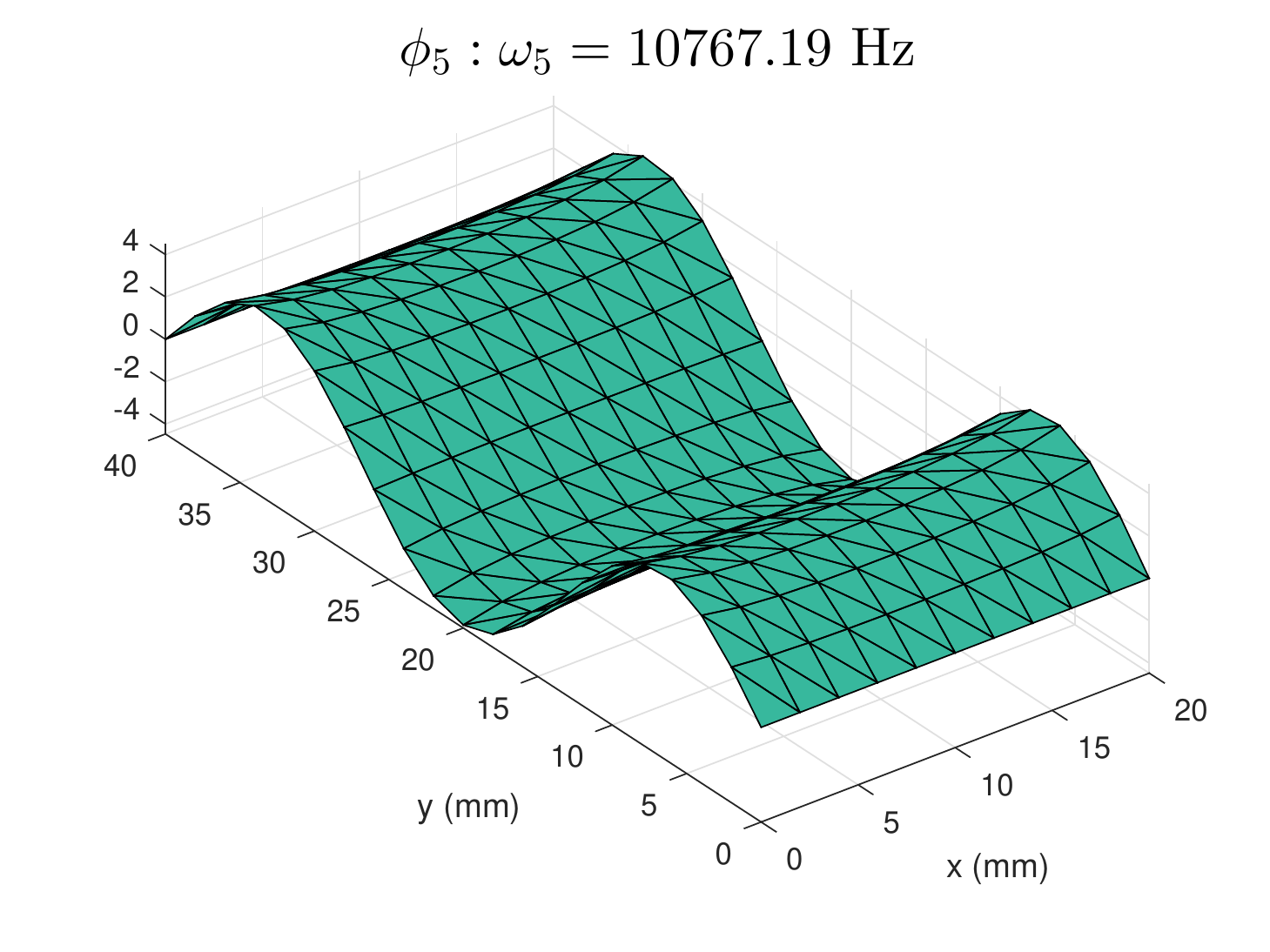}
		\captionsetup{justification=centering}
		\caption{\footnotesize }
		\label{fig:VM5}
	\end{subfigure}
	\caption{{\bf Model - I: } Mode shapes and frequencies for first 5 vibration modes of Model-I. }
	\label{fig:VMs}
\end{figure}

\begin{figure}[h!]
	\centering
	\begin{subfigure}{0.32\linewidth}
		\centering
		\includegraphics[width=\linewidth]{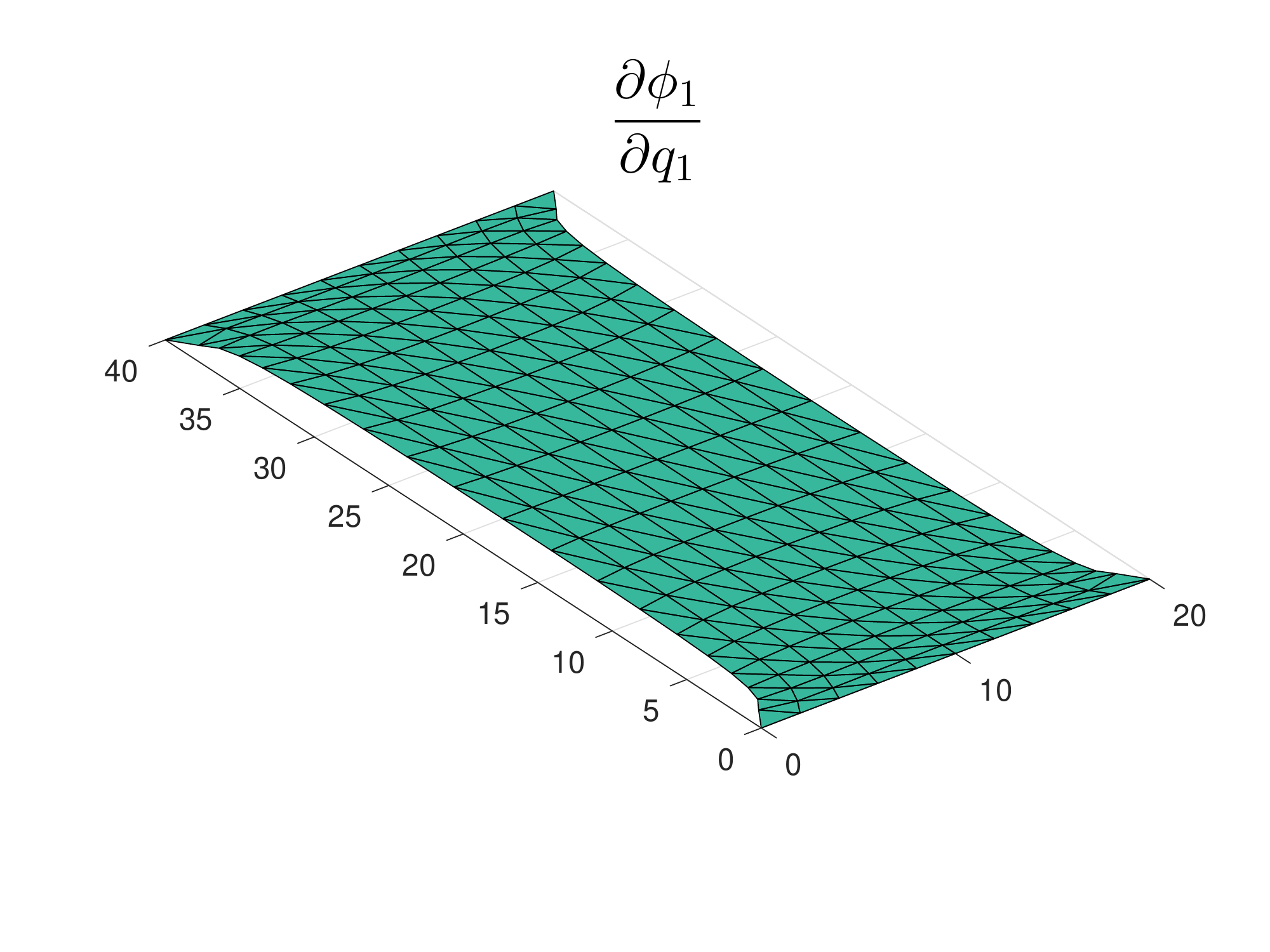}
	\end{subfigure}
	\begin{subfigure}{0.32\linewidth}
		\centering
		\includegraphics[width=\linewidth]{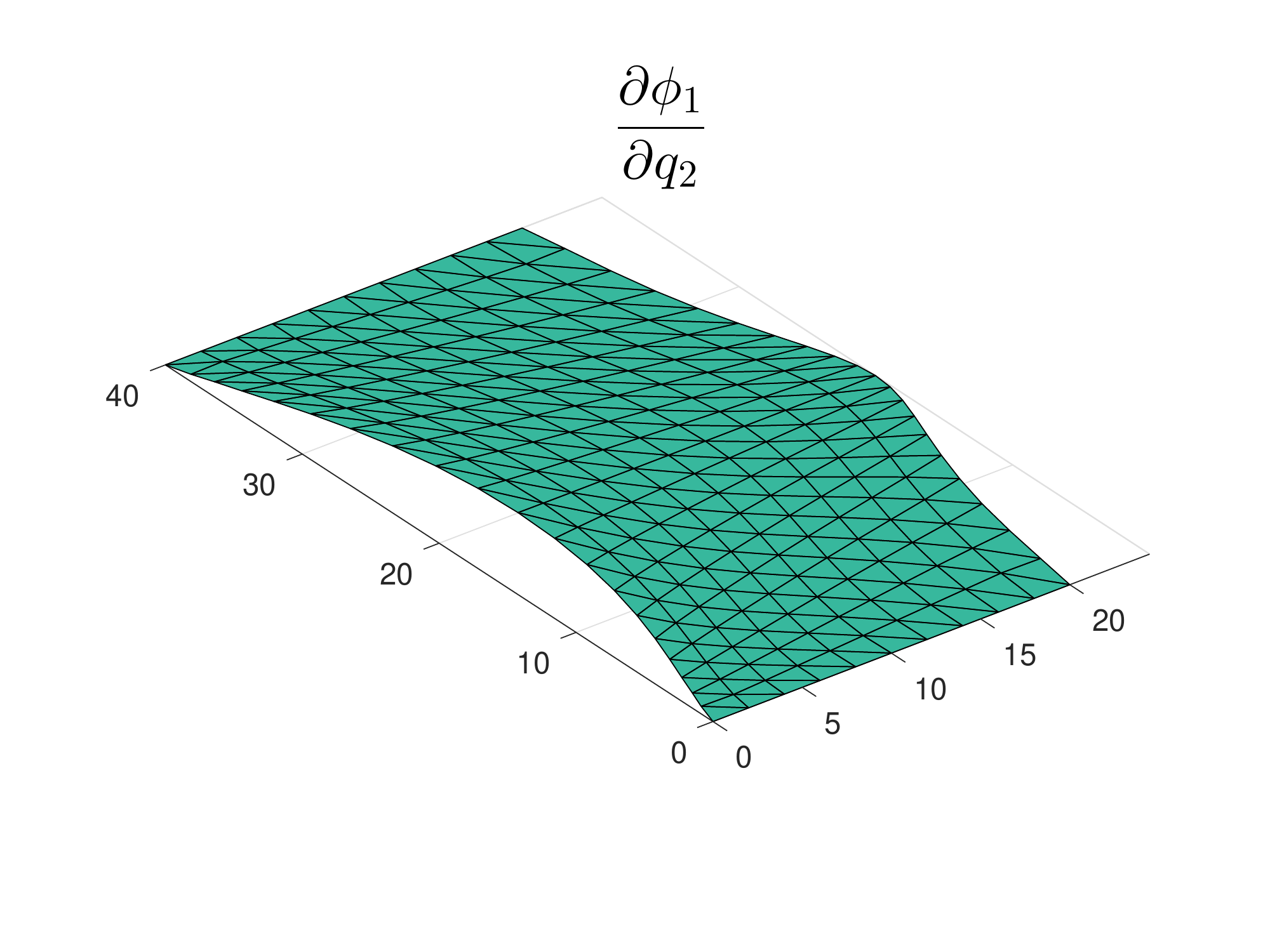}
	\end{subfigure}
	\begin{subfigure}{0.32\linewidth}
		\centering
		\includegraphics[width=\linewidth]{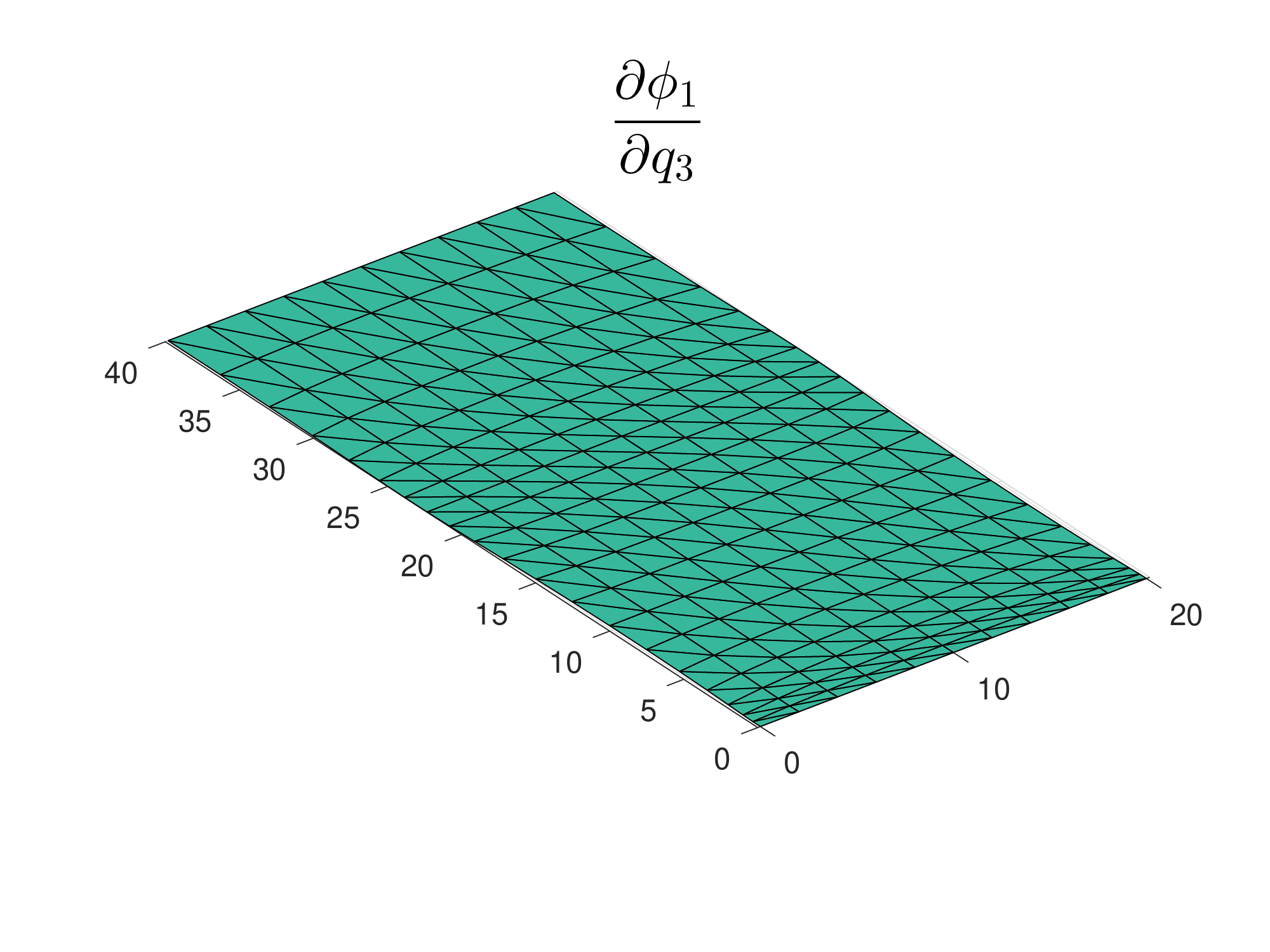}
	\end{subfigure}
	\begin{subfigure}{0.32\linewidth}
		\centering
		\includegraphics[width=\linewidth]{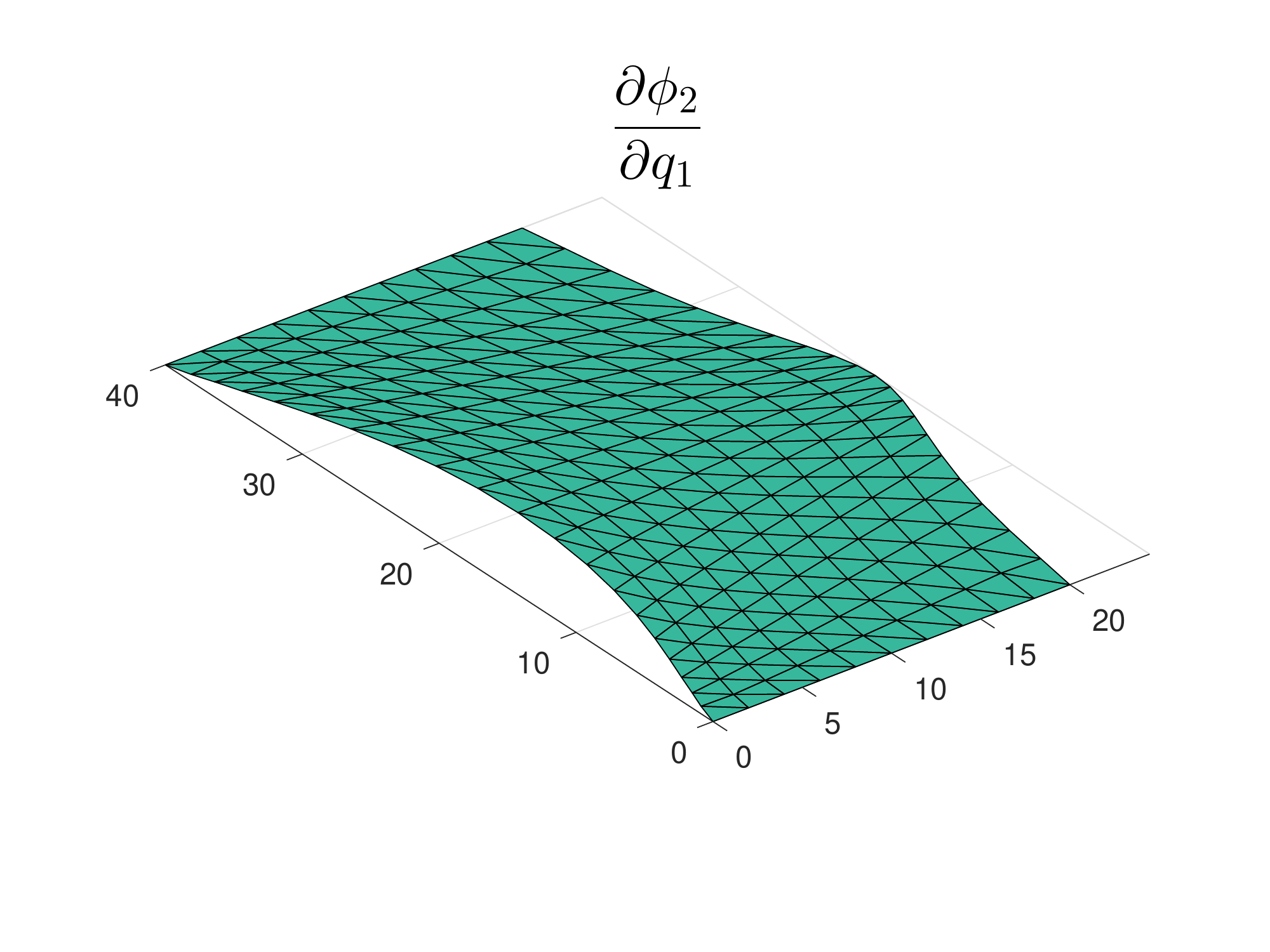}
	\end{subfigure}
	\begin{subfigure}{0.32\linewidth}
		\centering
		\includegraphics[width=\linewidth]{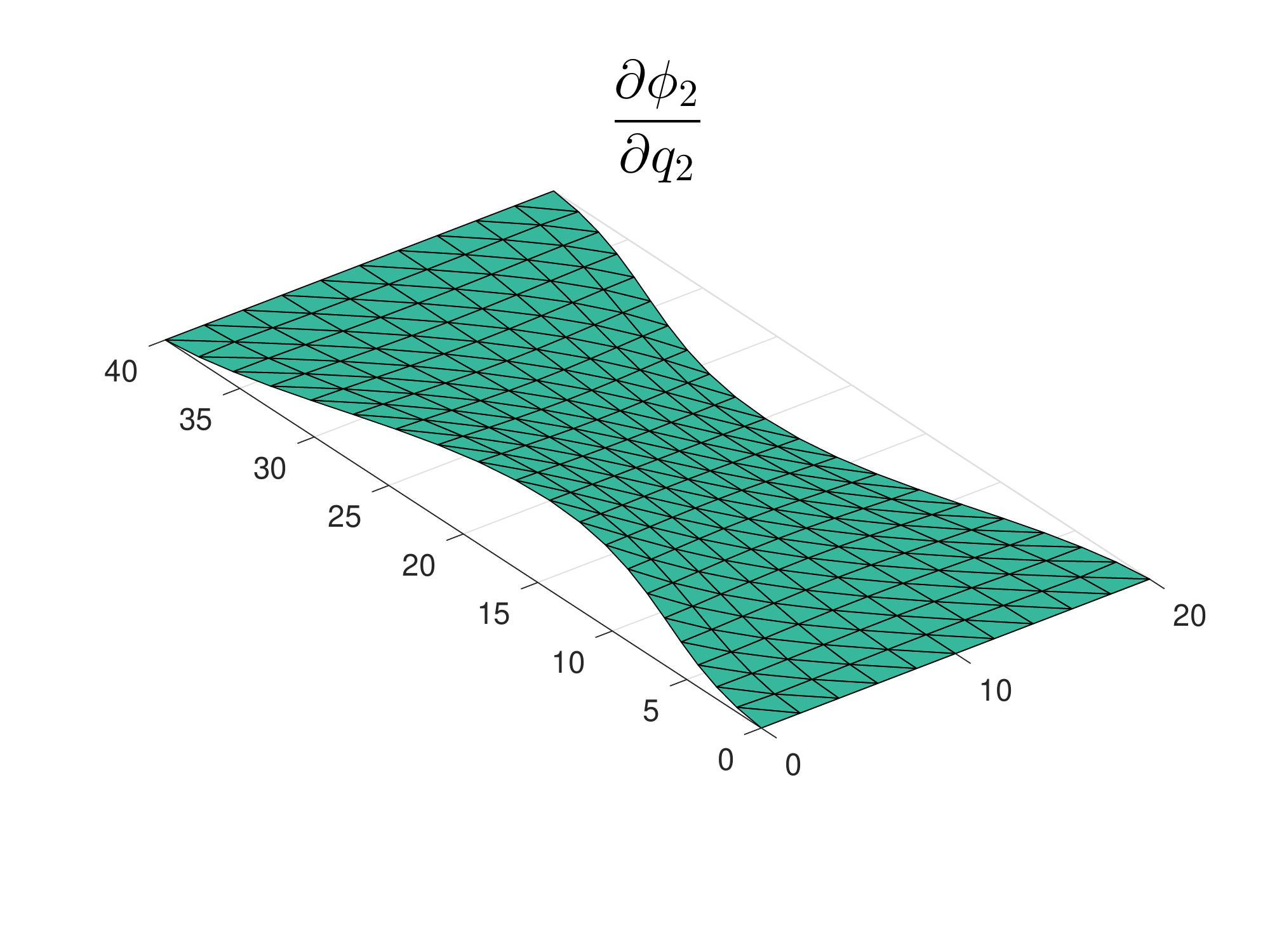}
	\end{subfigure}
	\begin{subfigure}{0.32\linewidth}
		\centering
		\includegraphics[width=\linewidth]{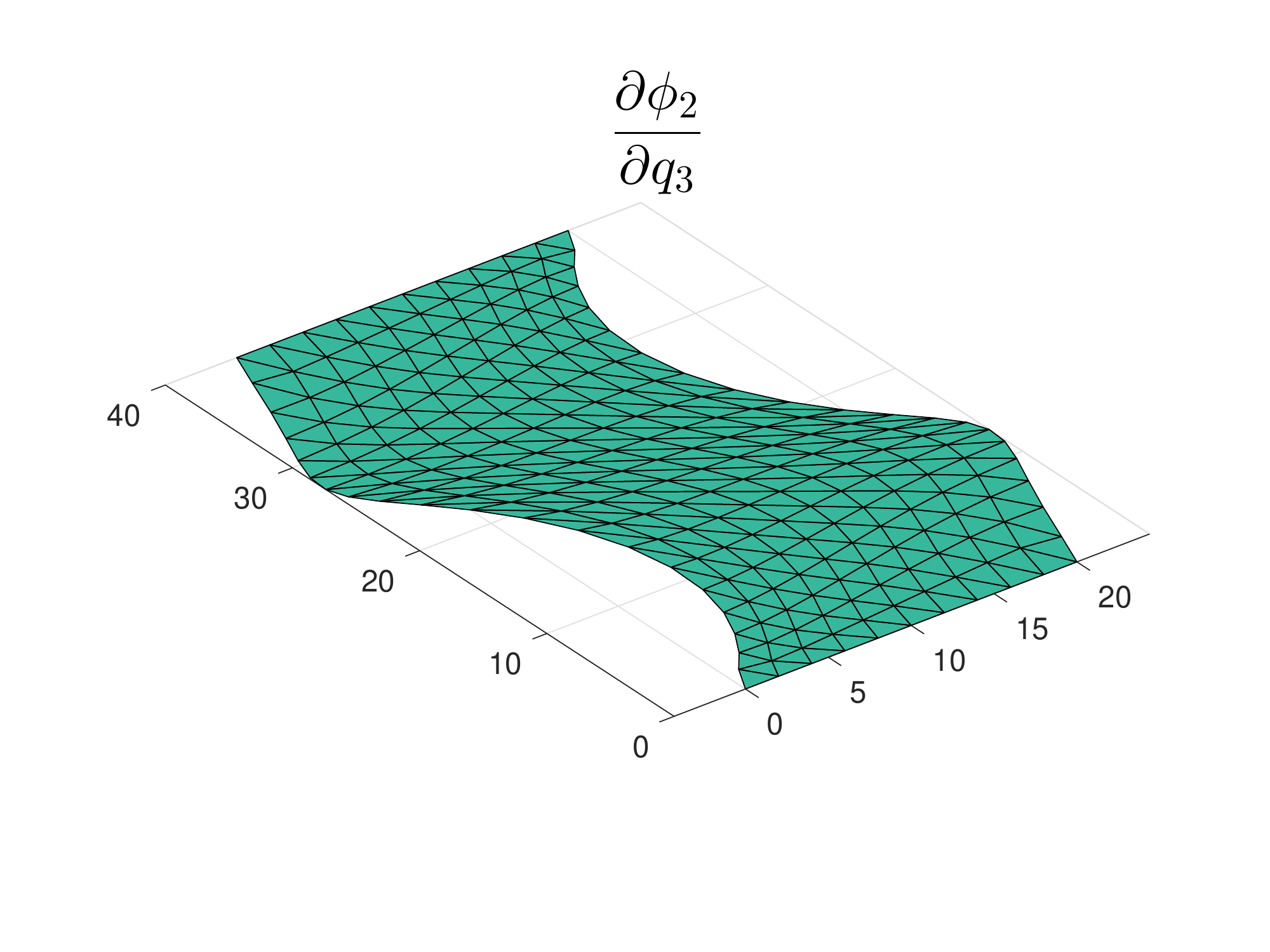}
	\end{subfigure}
	\begin{subfigure}{0.32\linewidth}
		\centering
		\includegraphics[width=\linewidth]{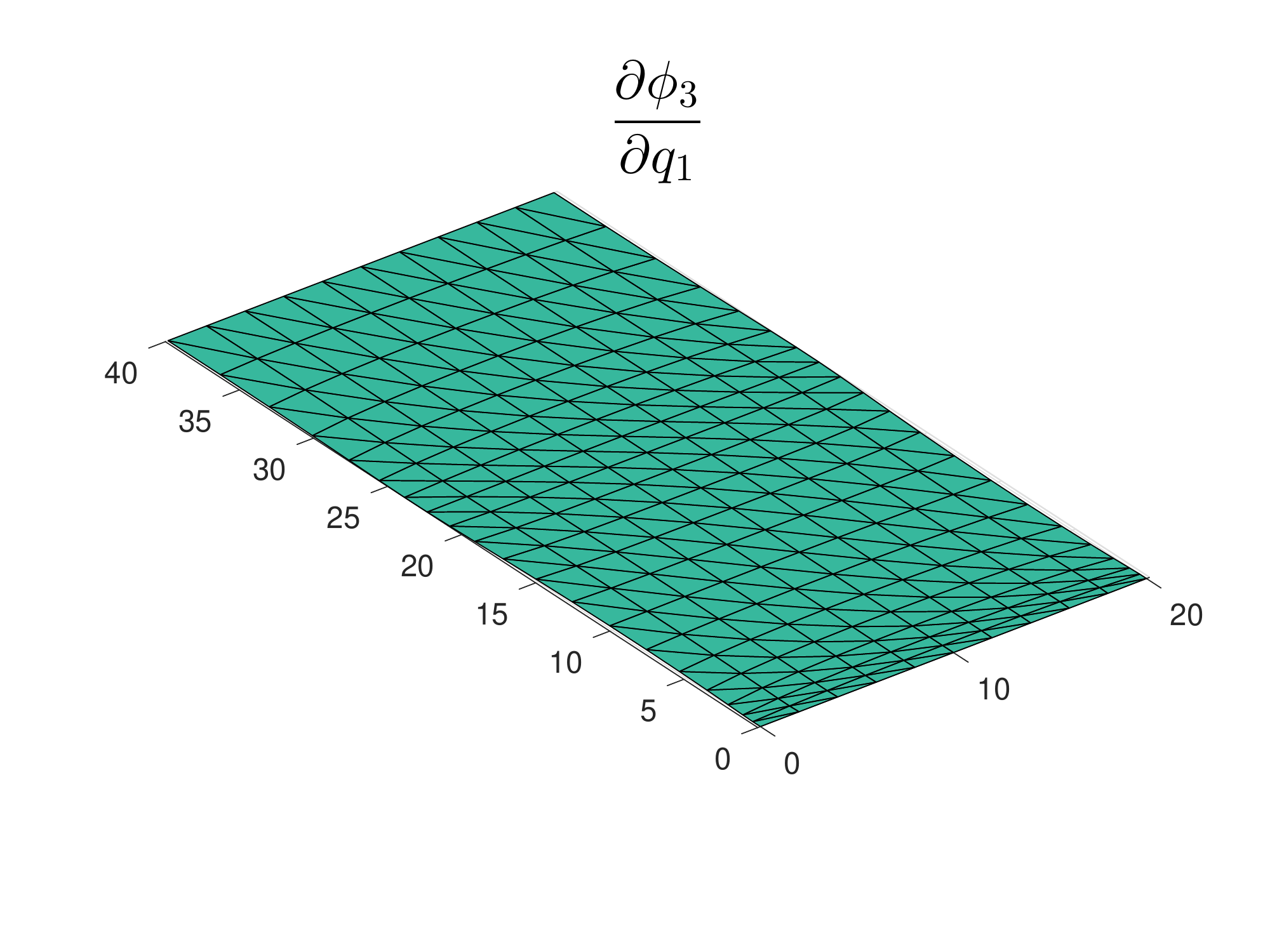}
	\end{subfigure}
	\begin{subfigure}{0.32\linewidth}
		\centering
		\includegraphics[width=\linewidth]{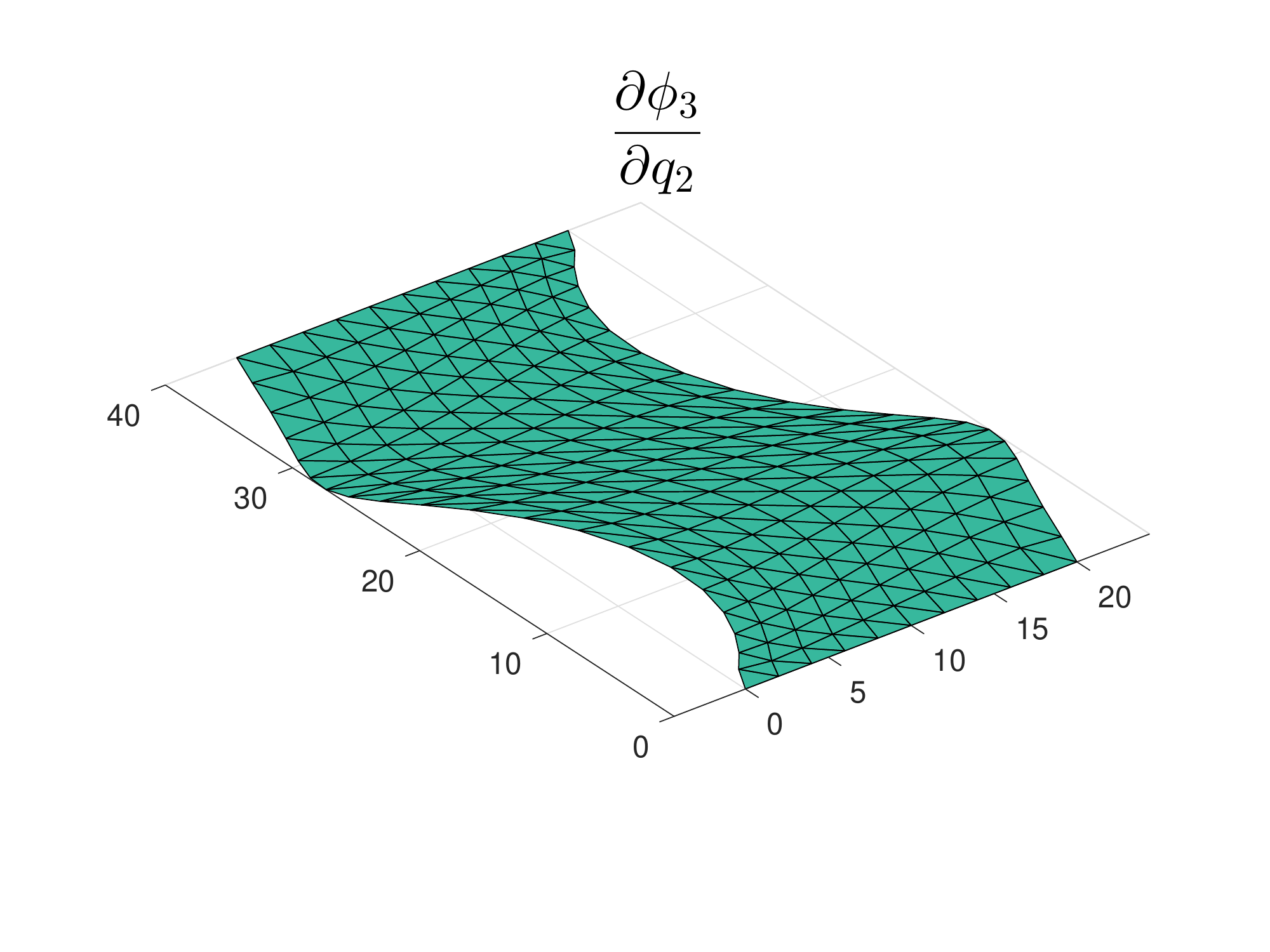}
	\end{subfigure}
	\begin{subfigure}{0.32\linewidth}
		\centering
		\includegraphics[width=\linewidth]{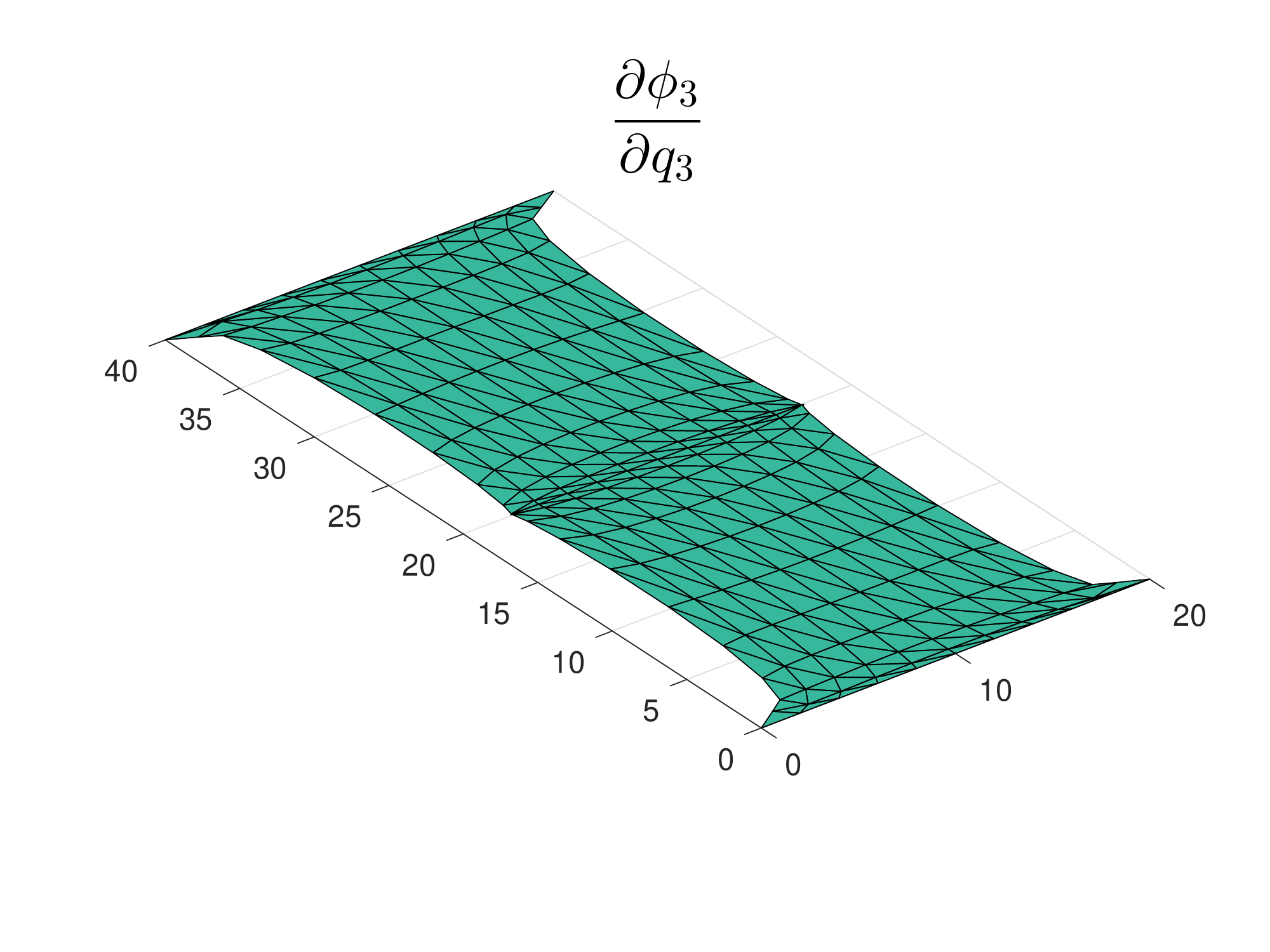}
	\end{subfigure}
	\caption{\review{(S)MDs corresponding to first 3 VMs of Model-1. The (S)MDs for a flat plate contain only in-plane displacement contributions. They capture the inherent membrane effects associated to the bending modes of the flat structure (cf. first three VMs in Figure~\ref{fig:VMs}). Physically, an MD represents the sensitivity of VM $ \boldsymbol{\phi}_i $ corresponding to a displacement given in the direction of VM $ \boldsymbol{\phi}_j $. For example, if the structure is given a displacement in the direction of the first VM, then we expect in-plane stretching effects due to geometric nonlinearities as given by $ \parderv{\boldsymbol{\phi}_1}{q_1} $}.}
	\label{fig:MDs}
\end{figure}

\FloatBarrier
Model-I is a rather simplistic example and linear modal superposition using the first and the fifth VMs (corresponding to the first and second bending mode of the plate) is able to accurately reproduce the linear solution \footnote{ The intermediate modes (i.e. the 2$ ^{\mathrm{nd}} $, 3$ ^{\mathrm{rd}} $ and the 4$ ^{\mathrm{th}} $ modes) do not contribute towards the linear response because these shapes are antisymmetric, see Figure~\ref{fig:VMs}, and are not excited by the uniform loading case being considered here.}. However, a basis containing just these modes, is not good enough for capturing the nonlinear response. \review{ The Linear Manifold (LM) reduction involves the use of a VMs basis augmented by all or some of the (S)MDs, as explained in Section~\ref{Sec:LM}. The (S)MDs corresponding to the first 3 bending VMs are shown in Figure~\ref{fig:MDs}. This figure shows that the MDs inherently capture the bending-stretching coupling associated to geometrical nonlinearities in the structure.} Using 2 VMs (first and fifth), 4(3) (S)MDs can be obtained, to constitute a basis of 5(6) vectors. This reduced basis was used for integration and the results (Figure~\ref{fig:FlatResults},Table~\ref{table:GREModel1}) are quite accurate.

\begin{figure}[h!]
	\center
	\begin{subfigure}{0.4\linewidth}
		\centering
		\includegraphics[width=\linewidth]{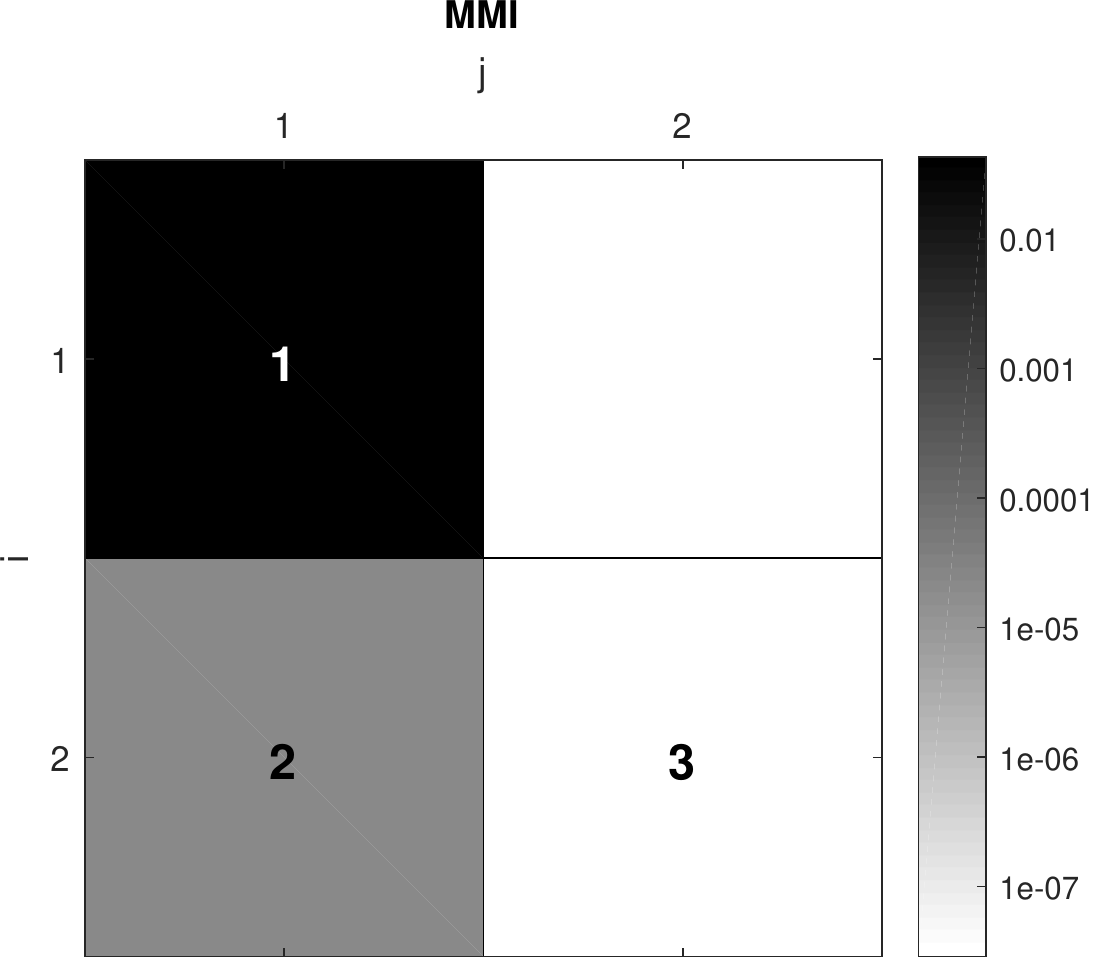}
		\captionsetup{justification=centering}
		\caption{ \footnotesize }
		\label{fig:FlatMMIWeightage}
	\end{subfigure}
	\begin{subfigure}{0.4\linewidth}
		\centering
		\includegraphics[width=\linewidth]{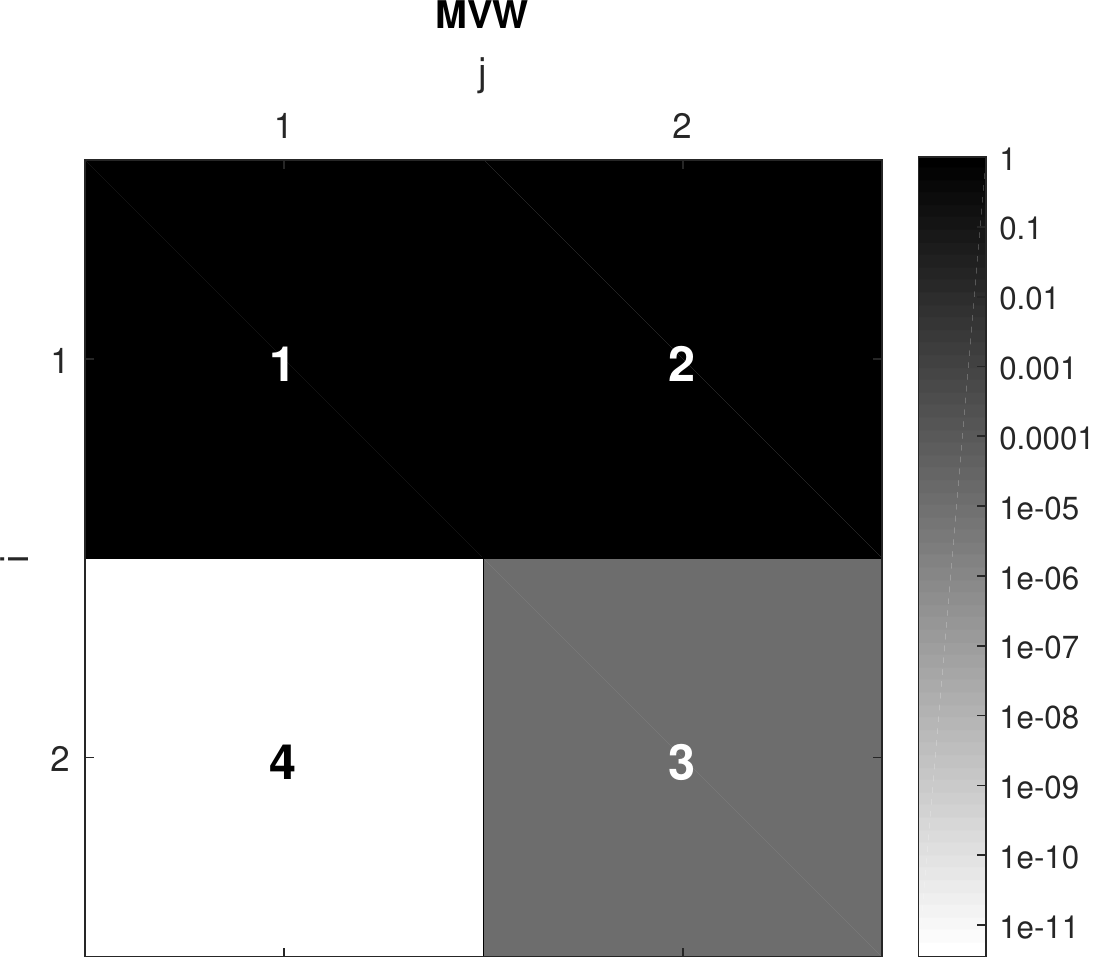}
		\captionsetup{justification=centering}
		\caption{\footnotesize }
		\label{fig:FlatMVWWeightage}
	\end{subfigure}
		\captionsetup{justification=justified,
			singlelinecheck=false
		}
	\caption{ The ranking and the weights obtained for all (S)MDs using 2 VMs in Model-I. (a) shows the colour intensity plot for weights found using Maximum Modal Interaction (MMI) technique. Cell (i,j) contains the normalized weight $ W_{ij} $ for $ SMD_{ij} $ based on \eqref{eqn:MMIweightage} (Note: part above diagonal not shown due to symmetry),  (b) shows the color plot for ranking found using Modal Virtual Work (MVW) technique. Cell (i,j) contains weight for $ MD_{ij} $. \\ }
	\label{fig:FlatMDS}
\end{figure}

\paragraph{MD Selection}
As noted earlier, the reduction basis size grows with complexity $ \mathcal{O}(m^2) $ for $ m $ VMs and all corresponding (S)MDs in the basis. However, only a few of these MDs might be important for capturing the nonlinear behaviour. Section~\ref{Sec:MDS} describes techniques to rank the (S)MDs to be considered in the basis. The ranking and the weights obtained for all (S)MDs using the corresponding techniques are shown in Figure~\ref{fig:FlatMDS}.
As a rule of thumb, a total of $ n_{MD} = m $ MDs has been chosen in this work for a LM basis containing $ m $ VMs, thus making the basis size $ 2m $.  This is done for a fair comparison and keeping the basis size linear with $ m $.

Incidentally, the SMDs and MDs turn out to be identical for this example. Thus using a basis of size 4 (2 VMs + 2 selected (S)MDs), both techniques give results with same accuracy(Table~\ref{table:GREModel1}) since the (S)MDs selected are the identical.
\FloatBarrier
\subsubsection{Quadratic Manifold}
\label{sec:FlatQM}
\begin{figure}[h!]
	\center
	\begin{subfigure}{0.49\linewidth}
		\centering
		\includegraphics[width=\linewidth]{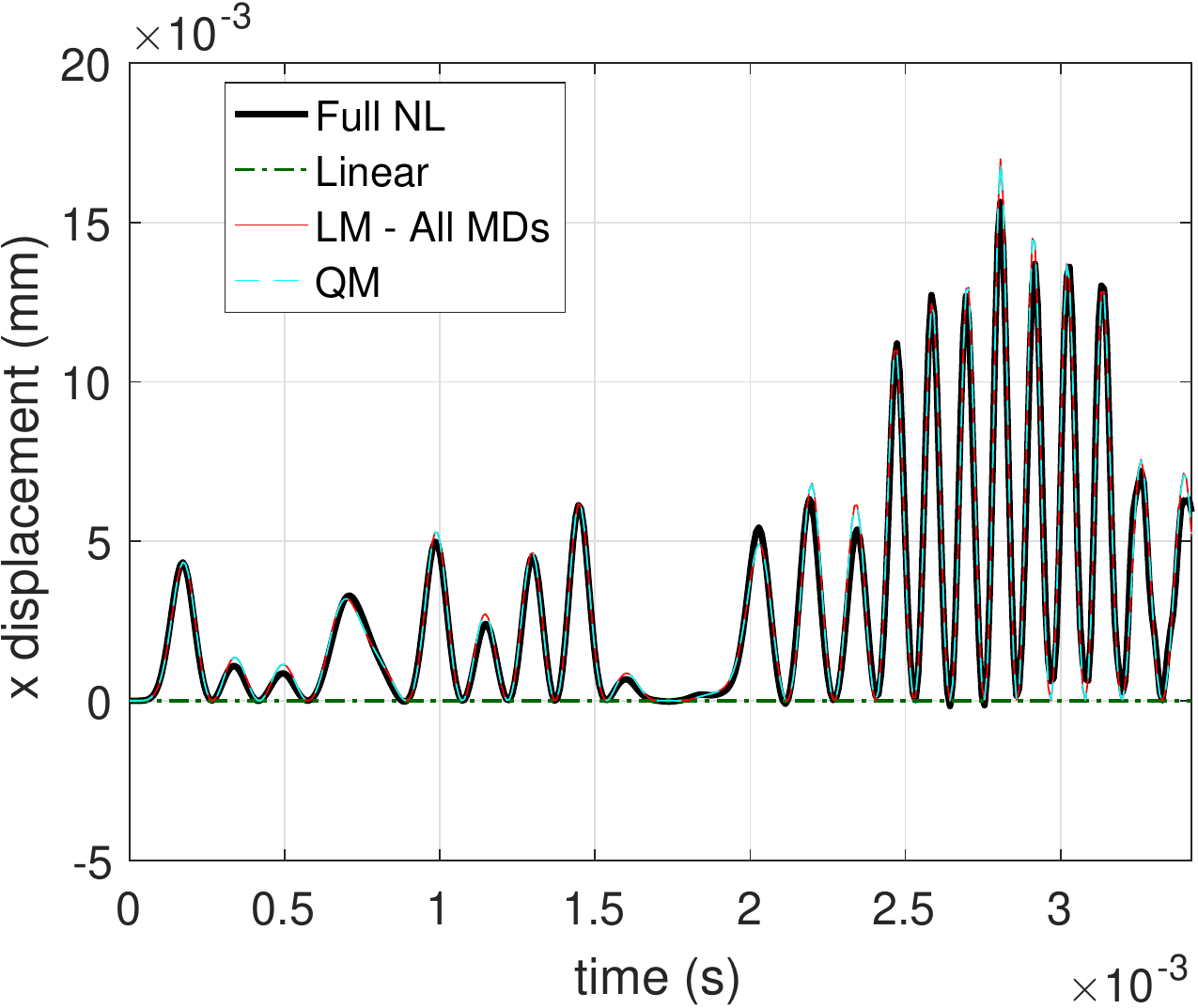}
		\caption{x-Displacement at tip node-1}
	\end{subfigure}
	\begin{subfigure}{0.49\linewidth}
		\centering
		\includegraphics[width=\linewidth]{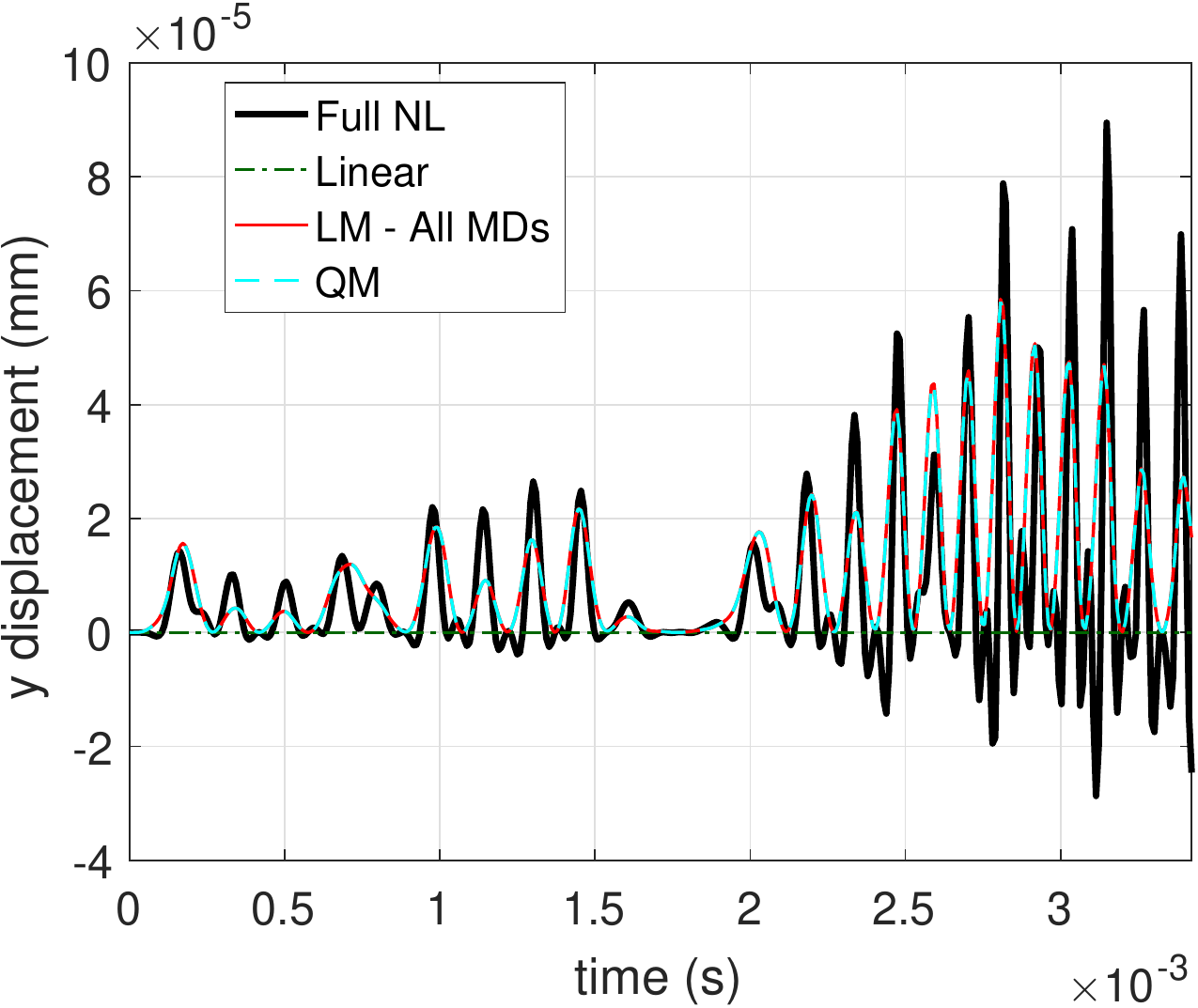}
		\caption{y-Displacement at tip node-1}
	\end{subfigure}
	\begin{subfigure}{0.49\linewidth}
		\centering
		\includegraphics[width=\linewidth]{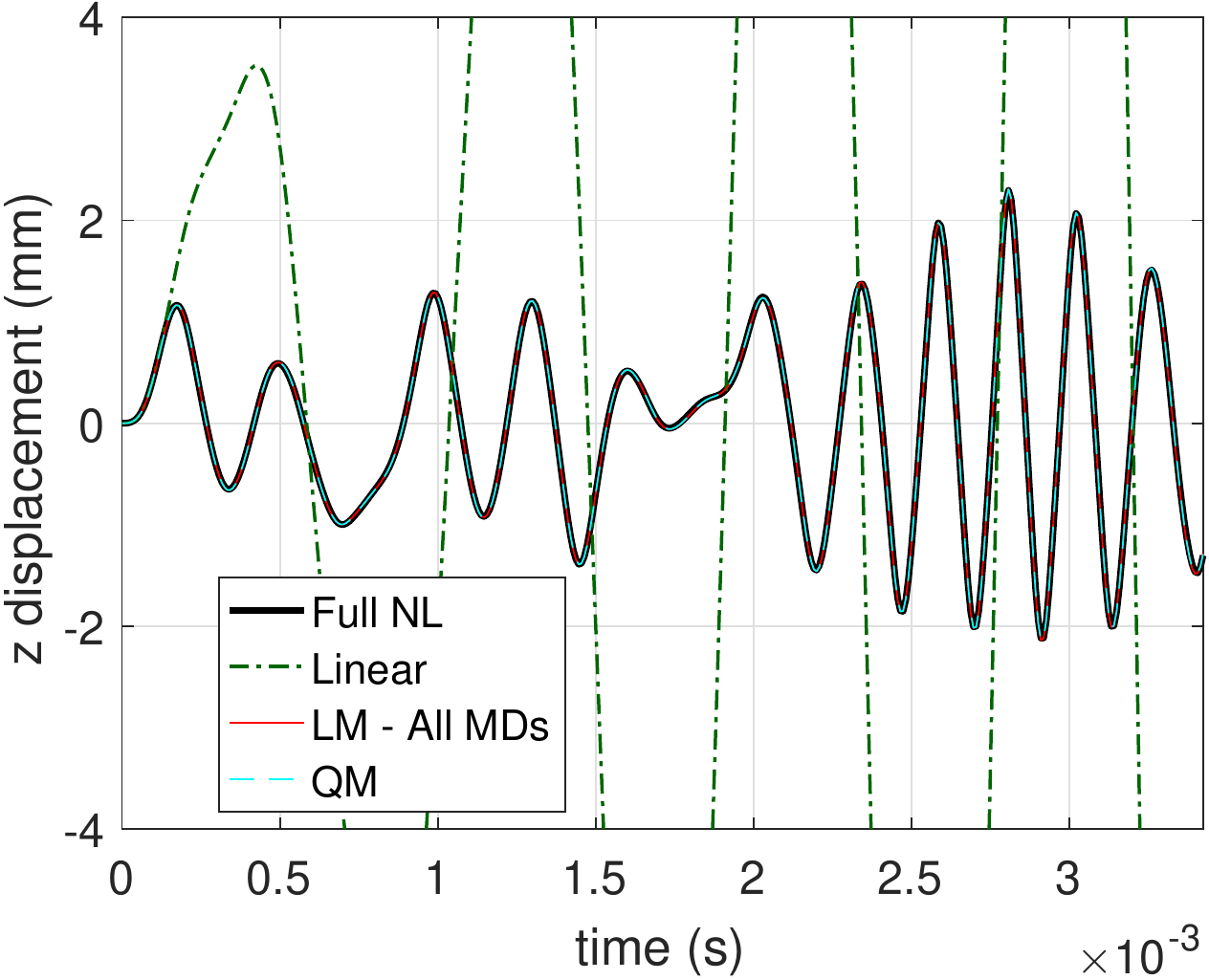}
		\caption{z-Displacement at tip node-1}
	\end{subfigure}
	\captionsetup{justification=justified,
		singlelinecheck=false
	}
	\caption{ Time history of solution (displacement) in x,y,z direction at a node situated at the centroid of the Flat structure (cf. Figure~\ref{fig:FlatModel}) is shown. Full nonlinear solution, Linearized system solution, reduced solution on the Linear Manifold (using All MDs) and Quadratic Manifold are shown for comparison. The remarkable difference in Linearised and Full nonlinear solution shows the system being in the nonlinear range of operation. System is successfully reduced on the Linear and Quadratic Manifold with practically identical accuracy.}
	\label{fig:FlatResults}
\end{figure}

A quadratic manifold is constructed using the first and the fifth VMs. The size of the reduced system is therefore $ m = 2 $ instead of $m=5$ as in the linear Manifold consisting of all MDs. As shown in Table~\ref{table:GREModel1} and Figure~\ref{fig:FlatResults}, the QM is able to reach similar accuracy for the response in this case.

\begin{table}[h!]
	\centering
	\caption{Global Relative Error in Flat structure model (Model-I) for different reduction techniques.}
	\label{table:GREModel1}
	\begin{tabular}{m{6cm} | c c}
		{\bf Reduction Technique} 	& 	$ GRE_M  $ (\%) 	& 	\# unknowns	\\
		\midrule

		LM (All (S)MDs)	&	2.29	&	5	\\
		LM-Selected SMDs (MMI)	& 2.32 &	4	\\
		LM-Selected MDs  (MVW)	& 2.32 &	4	\\
		\midrule
		Quadratic Manifold -(S)MDs	&	1.91 &	2	\\
		\midrule
		POD 	&	14.83	&	5
	\end{tabular}
\end{table}

For the sake of comparison, a POD basis was created containing the same number of vectors (5 in case of Model-I) as the LM basis (containing all (S)MDs) and it was observed that 5 POD modes do not provide a comparable accuracy as reported in Table~\ref{table:GREModel1}. It is interesting to note that the $ GRE_M $ for Quadratic Manifold reduction is lower than that for the Linear Manifold reduction using MDs. The quadratic manifold cannot give a better result as the LM (with the same MDs), as specified earlier in the Remark~\ref{rem:reductionerror}. This, howerer, holds for the error in the residual in the sense of Galerkin projection at each time step, which is not what is measured by the $ GRE_M $ here adopted.
\FloatBarrier
\subsection{NACA airfoil wing structure}
\label{sec:Real}
In the context of a more realistic application, a thin-walled wing structure is considered. The mesh for this structural model (referred to as Model-II hereafter) contains a realistically high number of DOFs so that the accuracy and reduction in problem size can be compared and appreciated.
\begin{figure}[h!]
	\centering
	\begin{subfigure}{0.49\linewidth}
		\centering
		\includegraphics[width=\linewidth]{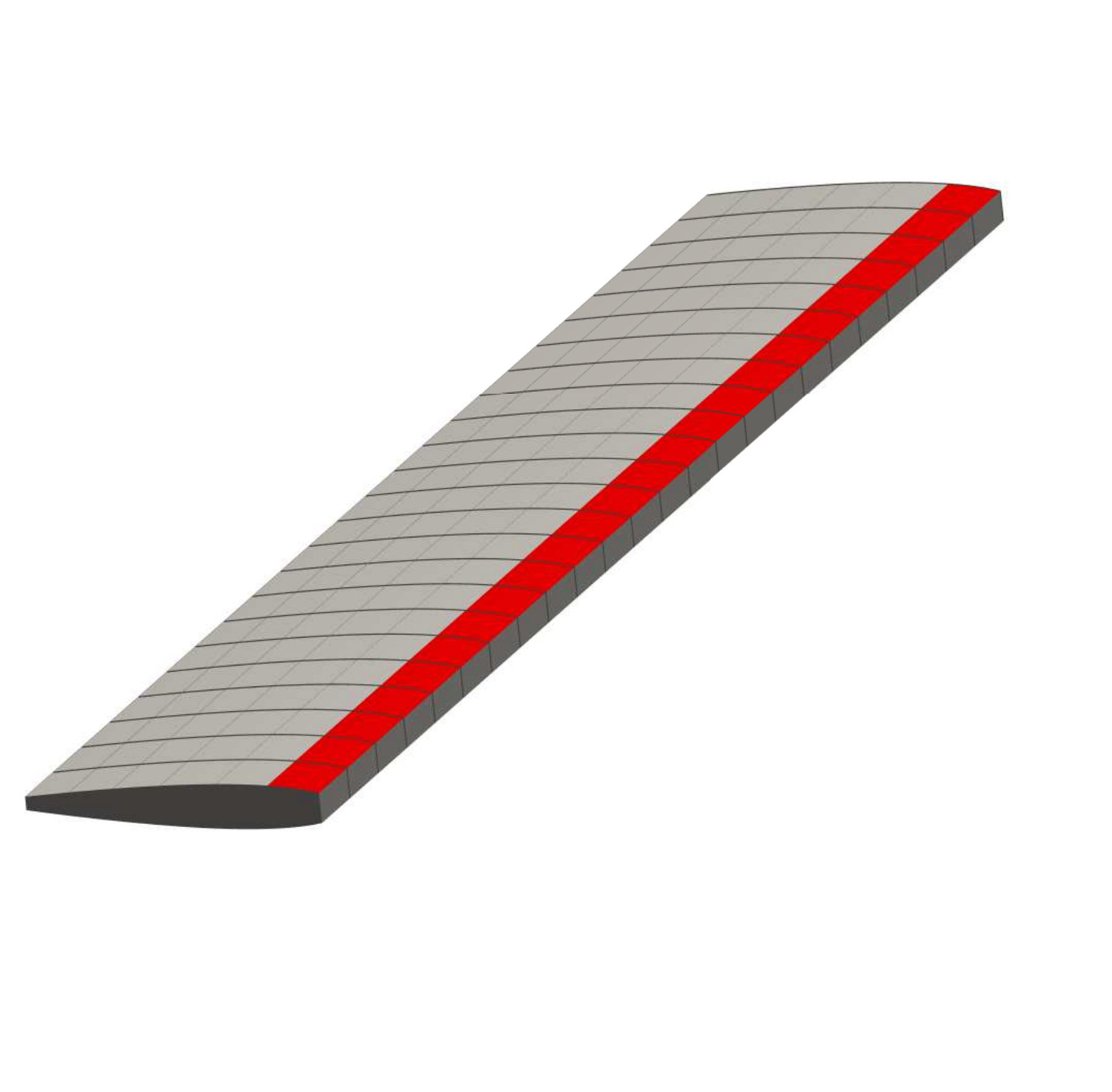}
		\captionsetup{justification=centering}
		\caption{ \footnotesize }
		\label{fig:wing1}
	\end{subfigure}
	\begin{subfigure}{0.49\linewidth}
		\centering
		\includegraphics[width=\linewidth]{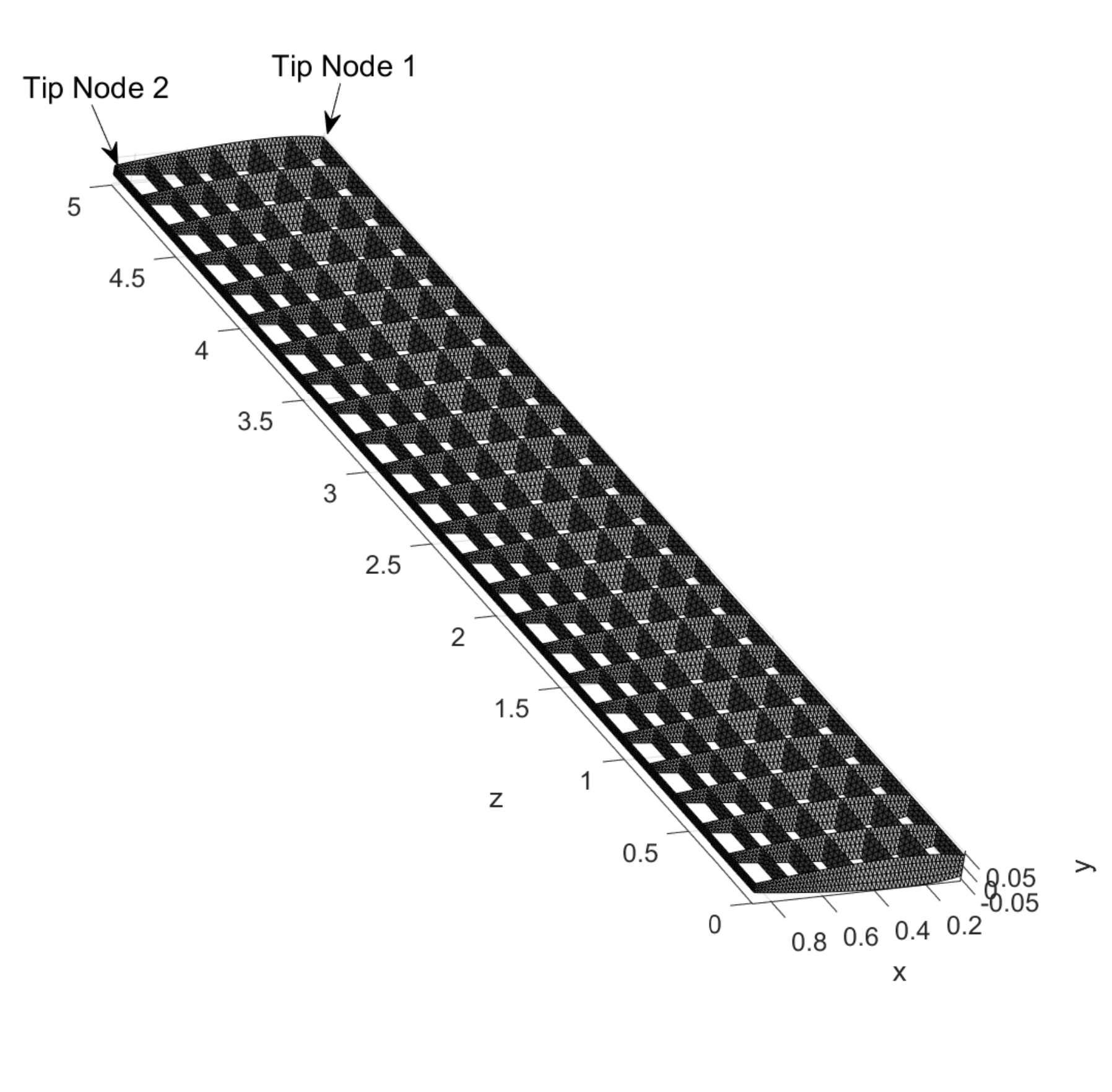}
		\captionsetup{justification=centering}
		\caption{\footnotesize }
		\label{fig:wing2}
	\end{subfigure}
		\captionsetup{justification=justified,
			singlelinecheck=false
		}
	\caption{{\bf Model - II: }A wing structure with NACA 0012 airfoil (length(L) = 5 m, Width(W) $ \approx $ 0.9 m, Height(H) = 0.1 m ) stiffened with ribs along the longitudinal and lateral direction. The Young Modulus is $E=70$ GPa, the Poisson's ratio is $\nu=0.33$, and the density is $\rho=2700$ Kg/m$^{3}$. The wing is cantilevered at one end. Uniform pressure is applied on the highlighted area (shown in (a)). The structure is meshed with triangular flat shell elements with 6 DOFs per node and each with a thickness of 1.5 mm. The mesh contains $ n = 135770 $ DOFs, $ n_{el} = 49968 $ elements. For illustration purposes, the skin panels are removed and mesh is shown in (b).}
	\label{fig:Wing}
\end{figure}

\begin{figure}[h!]
		\center
		\begin{subfigure}{0.46\linewidth}
			\centering
			\includegraphics[width=\linewidth]{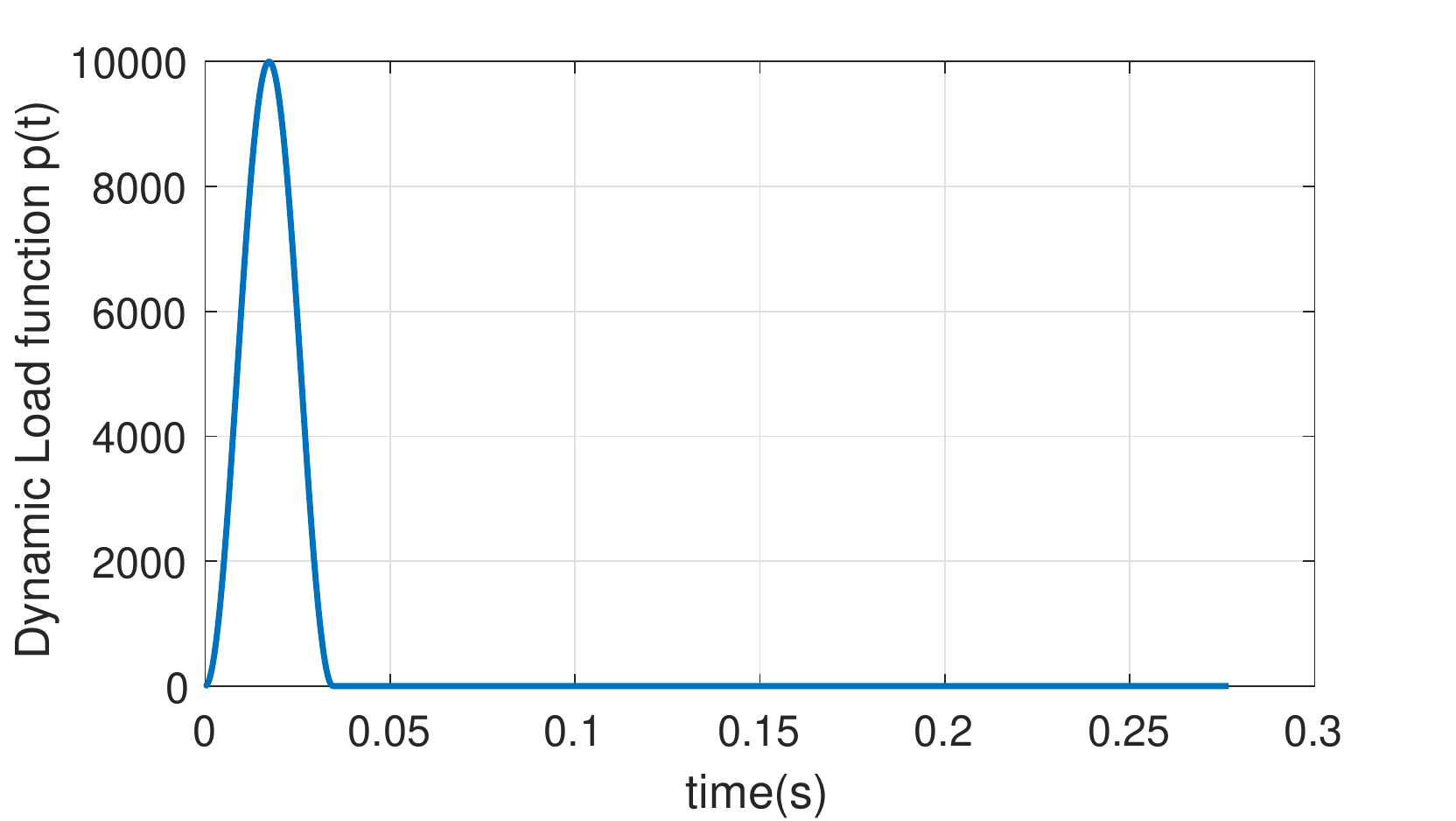}
			\captionsetup{justification=centering}
			\caption{}
			\label{fig:Realload2}
		\end{subfigure}
		\begin{subfigure}{0.53\linewidth}
			\centering
			\includegraphics[width=\linewidth]{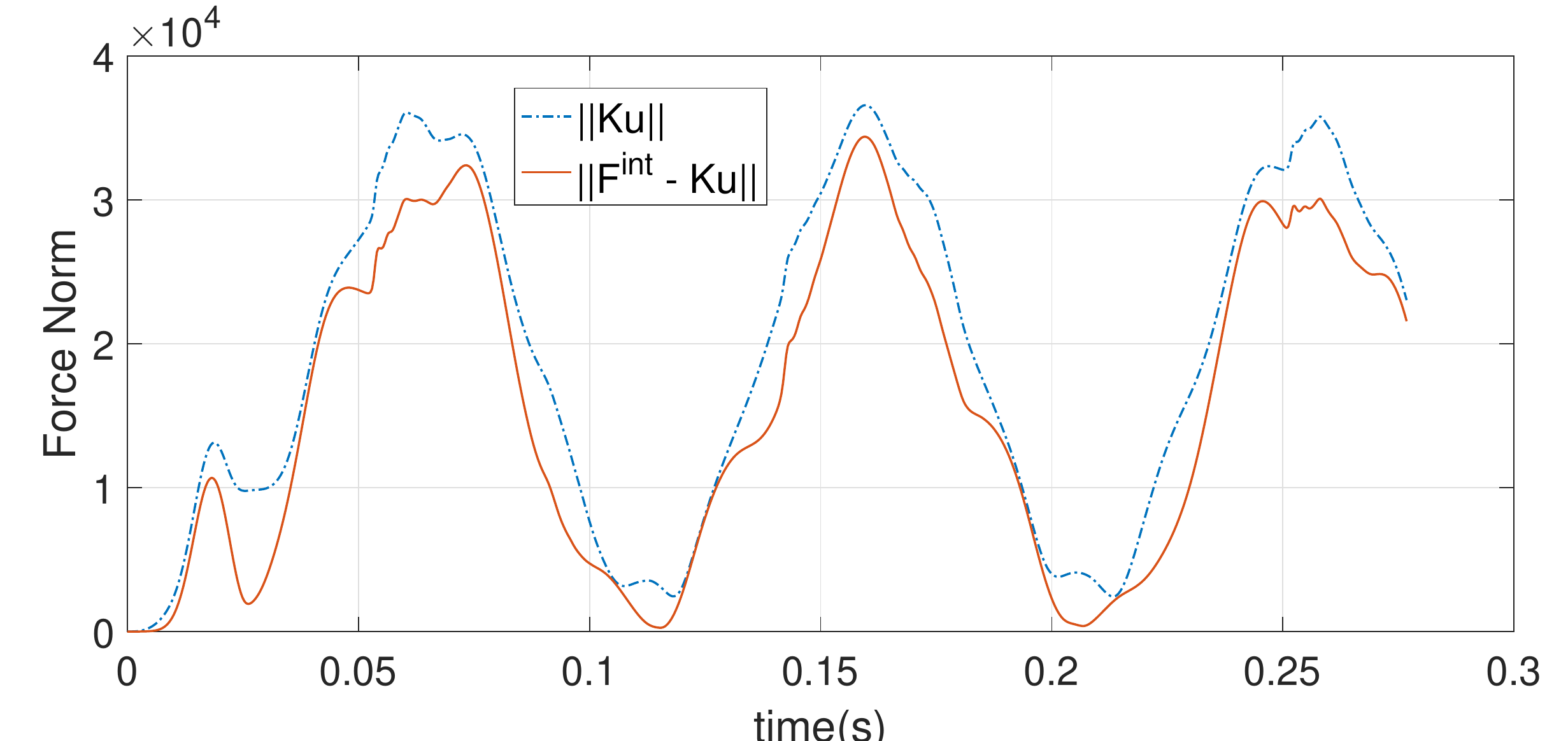}
			\captionsetup{justification=centering}
			\caption{}
			\label{fig:RealRefNL2}
		\end{subfigure}
			\captionsetup{justification=justified,
				singlelinecheck=false
			}
		\caption{ (a) Dynamic load function for pulse loading (cf.~\eqref{eqn:pulseload}). (b) The comparison of the norm of the linear and the nonlinear internal force during a full nonlinear solution for model-II (cf. Figure~\ref{fig:Wing}).}
		\label{fig:Realloadresponse2}
	\end{figure}

For illustration purposes, the results for a low frequency pulse load are shown here. A spatially uniform pressure load is applied locally on the structure skin at an area highlighted in Figure \ref{fig:wing1}. The pressure load takes the shape of a pulse in time as shown in Figure \ref{fig:Realload2}. The dynamic load function is given as
	\begin{equation}
	\label{eqn:pulseload}
	p(t) = A\sin^2(\omega t)\left[ H(t) - H\left( \frac{\pi}{\omega} - t\right) \right] ,
	\end{equation}
	where $ H(t) $ is the heaviside function and $ \omega $ chosen as the average of the first and second natural frequency of vibration. Again the load amplitude is chosen so that the linear and nonlinear internal forces have magnitudes of similar order (see Figure~\ref{fig:RealRefNL2}).
The displacement response against time are shown for two nodes on the tip of the structure in Figure \ref{fig:wing2}.

\begin{figure}[h!]
	\center
	\begin{subfigure}{0.49\linewidth}
		\centering
		\includegraphics[width=\linewidth]{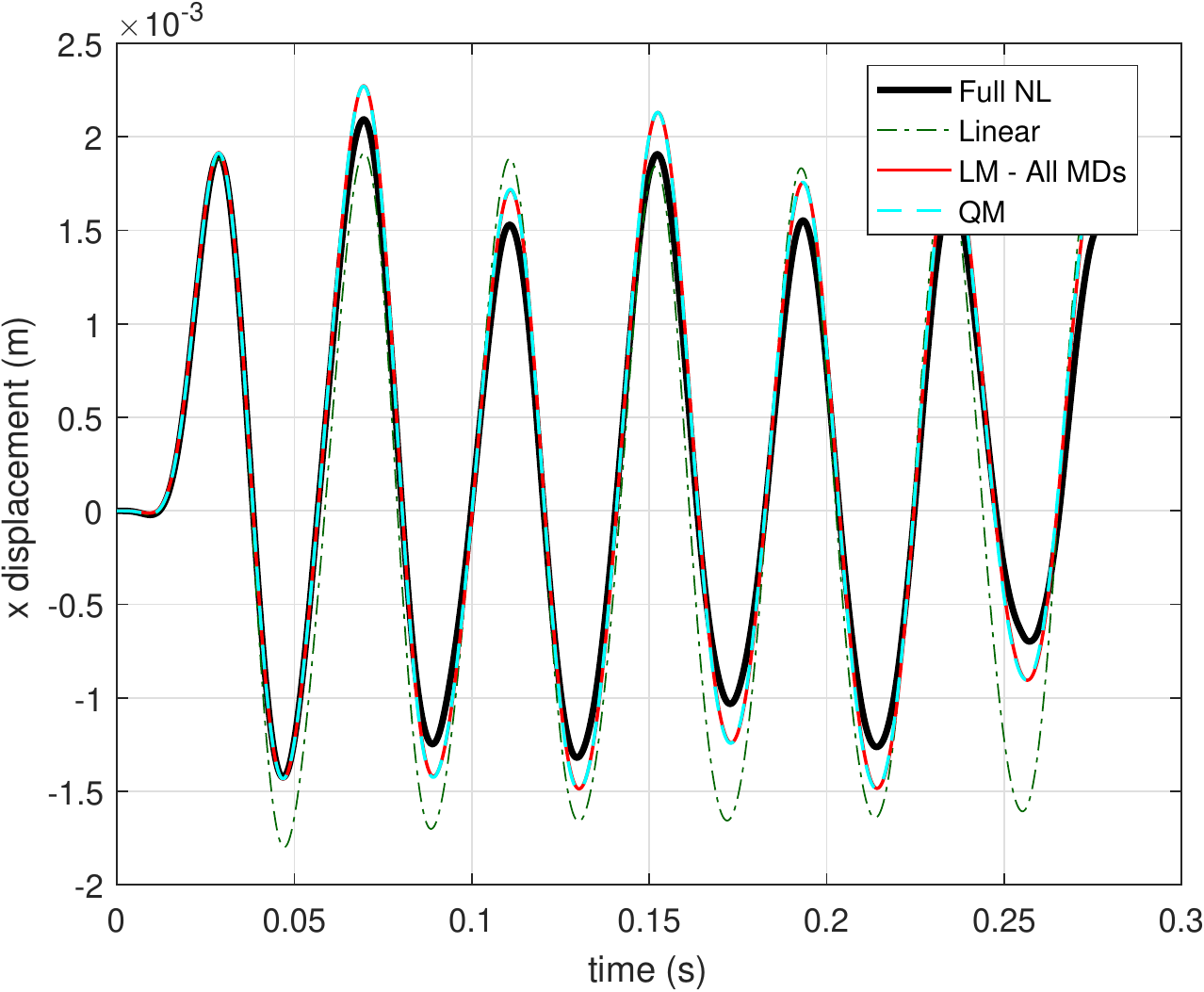}
		\caption{x-Displacement at tip node-1}
		\end{subfigure}
		\begin{subfigure}{0.49\linewidth}
			\centering
			\includegraphics[width=\linewidth]{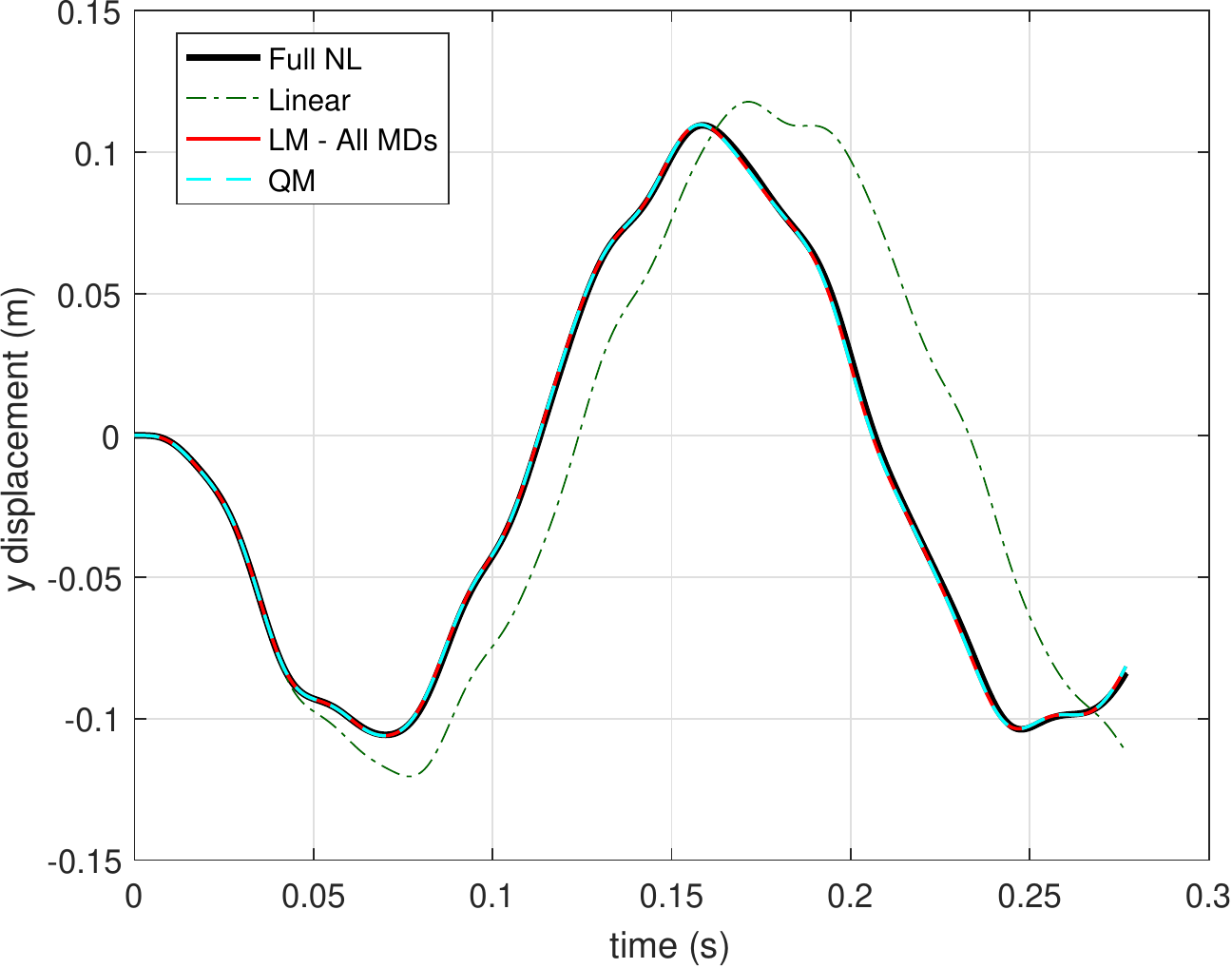}
			\caption{y-Displacement at tip node-1}
		\end{subfigure}
		\begin{subfigure}{0.49\linewidth}
			\centering
			\includegraphics[width=\linewidth]{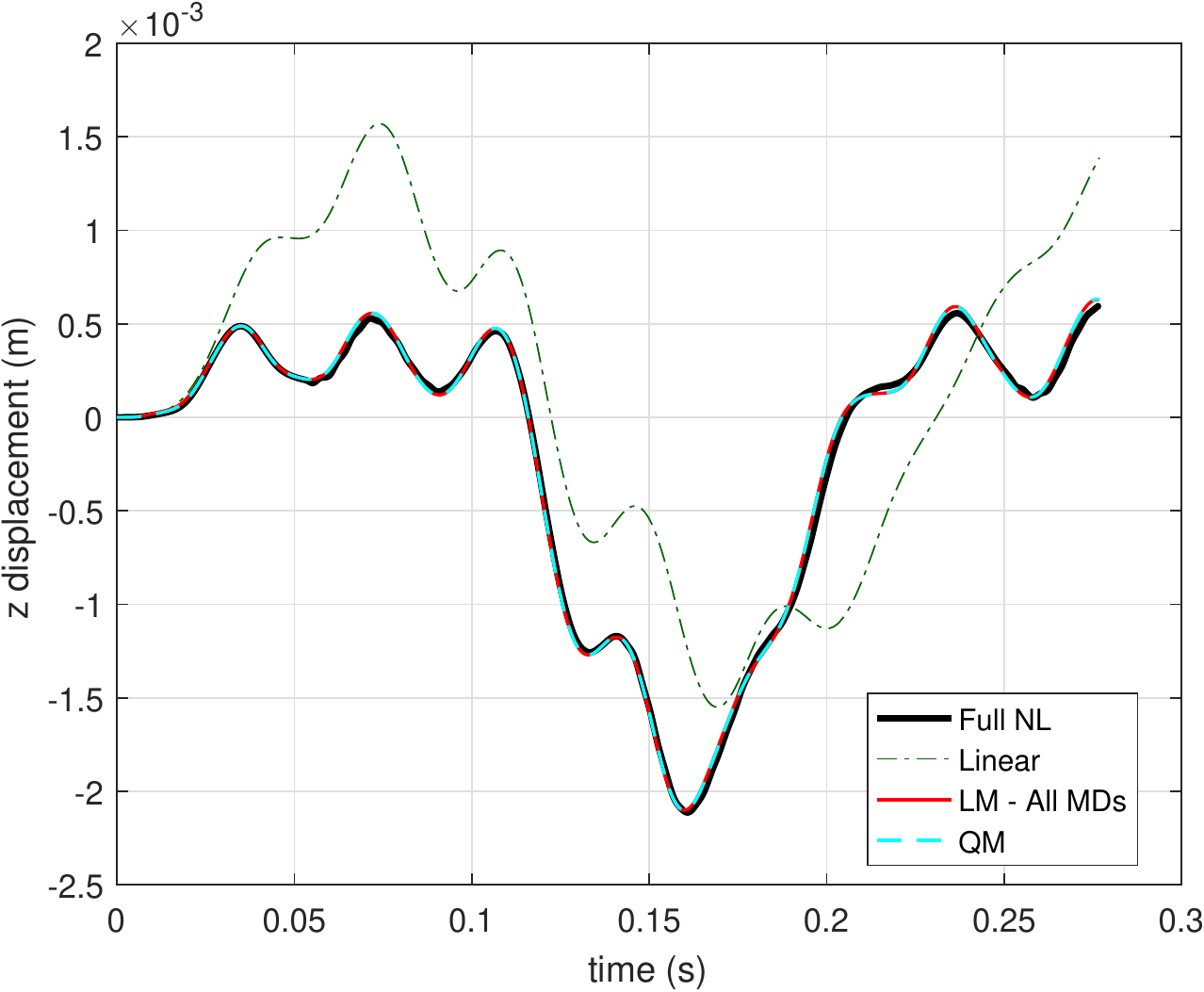}
			\caption{z-Displacement at tip node-1}
		\end{subfigure}
		\captionsetup{justification=justified,
			singlelinecheck=false
			}
		\caption{Time history of solution (displacement) in x,y,z direction at the tip-node 1 of Model-II (see Figure~\ref{fig:Wing}) is shown. Full nonlinear solution, Linearized system solution, reduced solution using the Linear Manifold (using All MDs) and Quadratic Manifold are shown for comparison. The remarkable difference in Linearised and Full nonlinear solution shows the system being in the nonlinear range of operation. System is successfully reduced with the Linear and Quadratic Manifold with practically identical accuracy.}
		\label{fig:Node1}
		\end{figure}

		\begin{figure}[h!]
			\center
			\begin{subfigure}{0.49\linewidth}
				\centering
				\includegraphics[width=\linewidth]{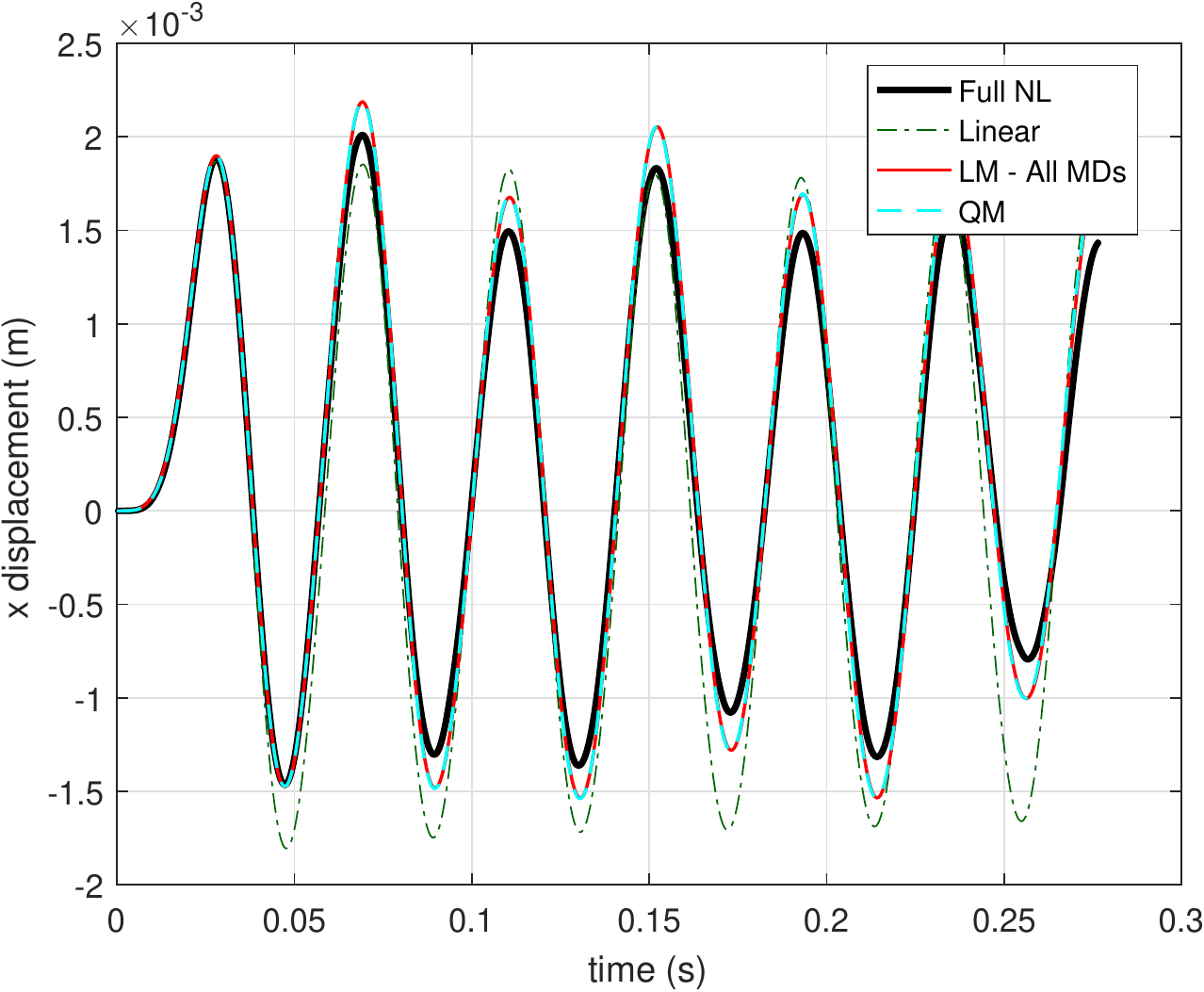}
				\caption{x-Displacement at tip node-2}
			\end{subfigure}
			\begin{subfigure}{0.49\linewidth}
				\centering
				\includegraphics[width=\linewidth]{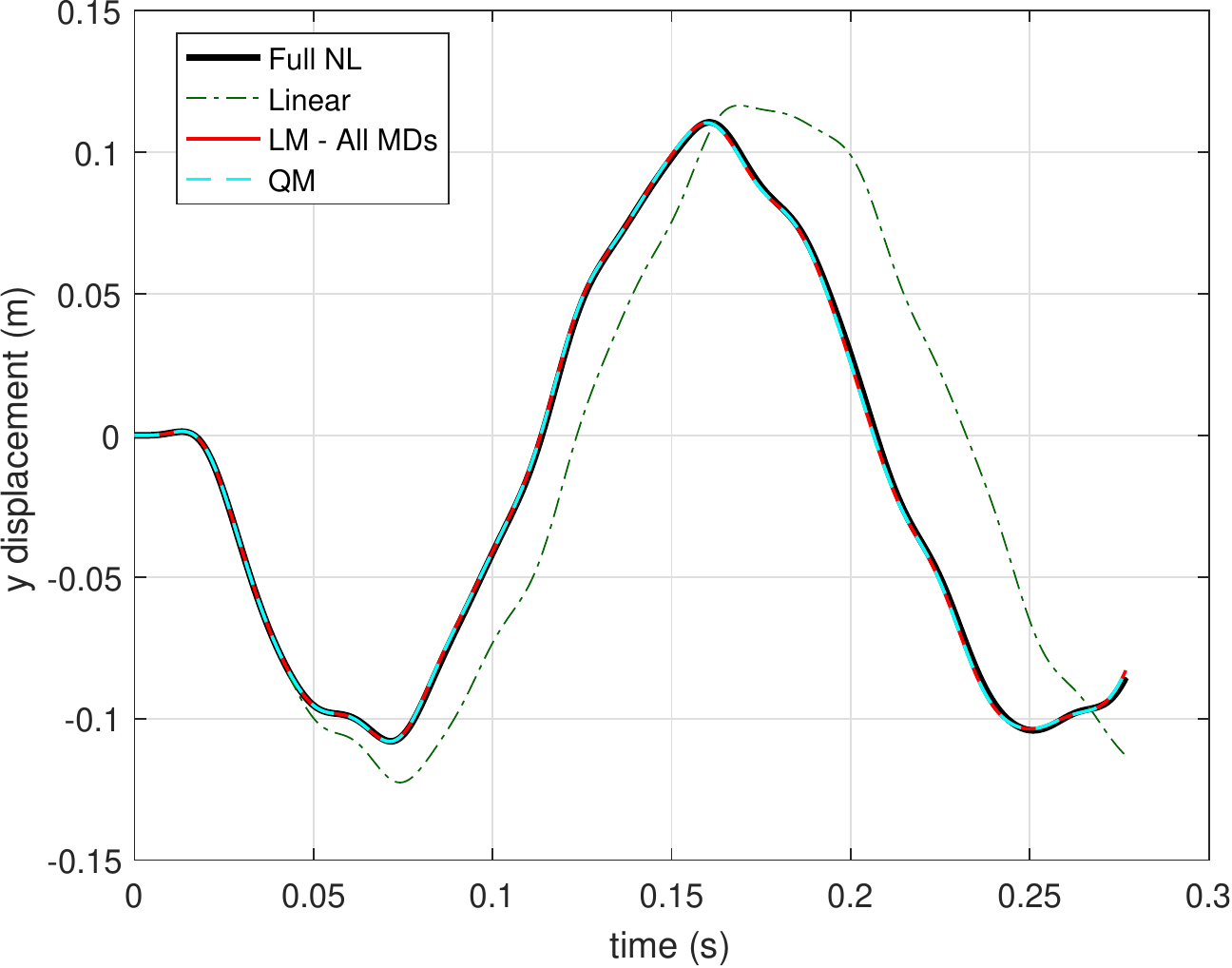}
				\caption{y-Displacement at tip node-2}
			\end{subfigure}
			\begin{subfigure}{0.49\linewidth}
				\centering
				\includegraphics[width=\linewidth]{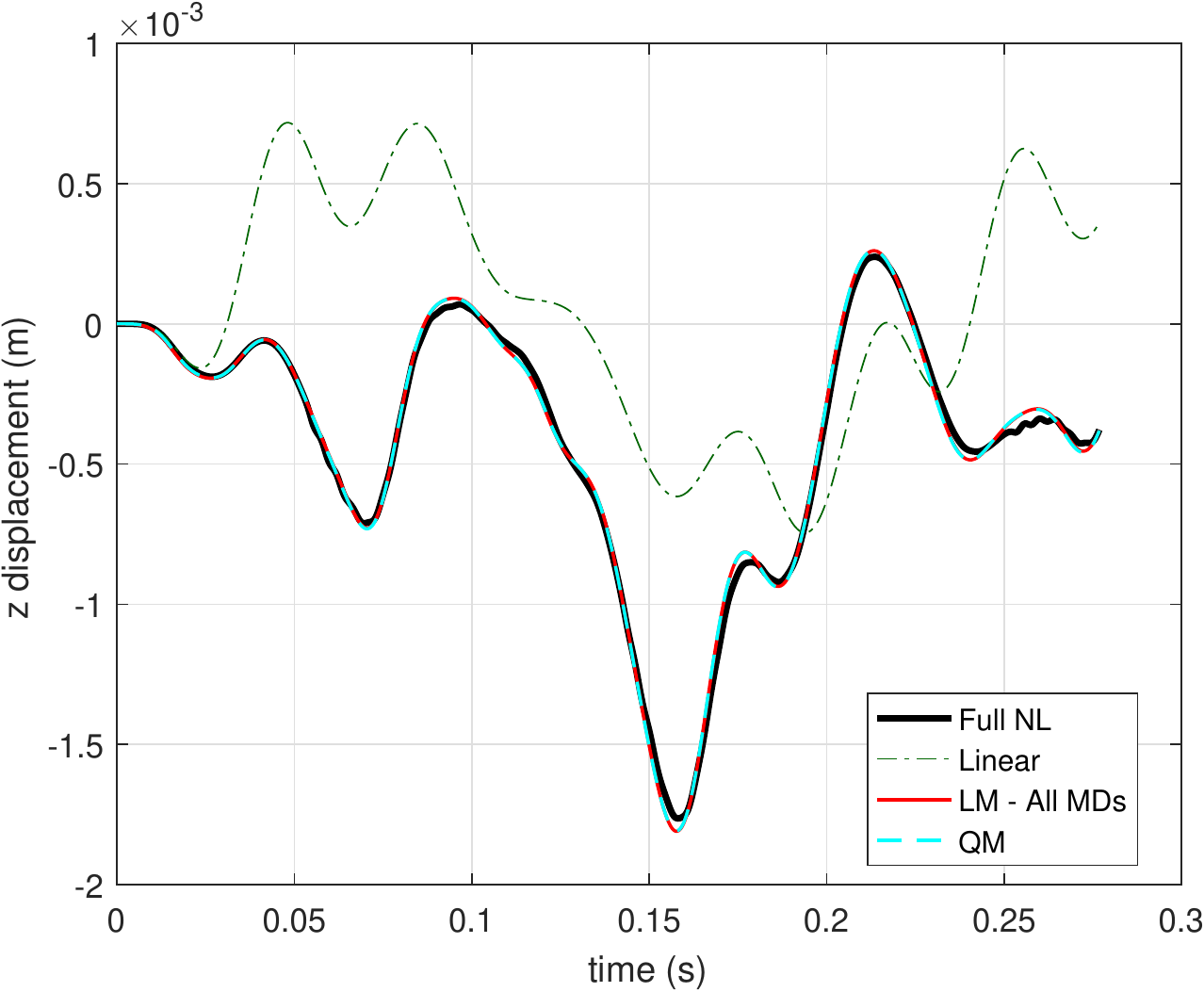}
				\caption{z-Displacement at tip node-2}
			\end{subfigure}
			\captionsetup{justification=justified,
				singlelinecheck=false
				}
			\caption{ Time history of solution (displacement) in x,y,z direction at the tip-node 2 of Model-II (see Figure~\ref{fig:Wing}) is shown. Full nonlinear solution, Linearized system solution, reduced solution over the Linear Manifold (using All MDs) and Quadratic Manifold are shown for comparison. The remarkable difference in Linearised and Full nonlinear solution shows the system being in the nonlinear range of operation. System is successfully reduced over the Linear and Quadratic Manifold with practically identical accuracy.}
			\label{fig:Node2}
			\end{figure}

	\FloatBarrier
	\subsubsection{Linear Manifold}
\begin{figure}[h!]
		\centering
		\begin{subfigure}{0.48\linewidth}
			\centering
			\includegraphics[width=0.8\linewidth]{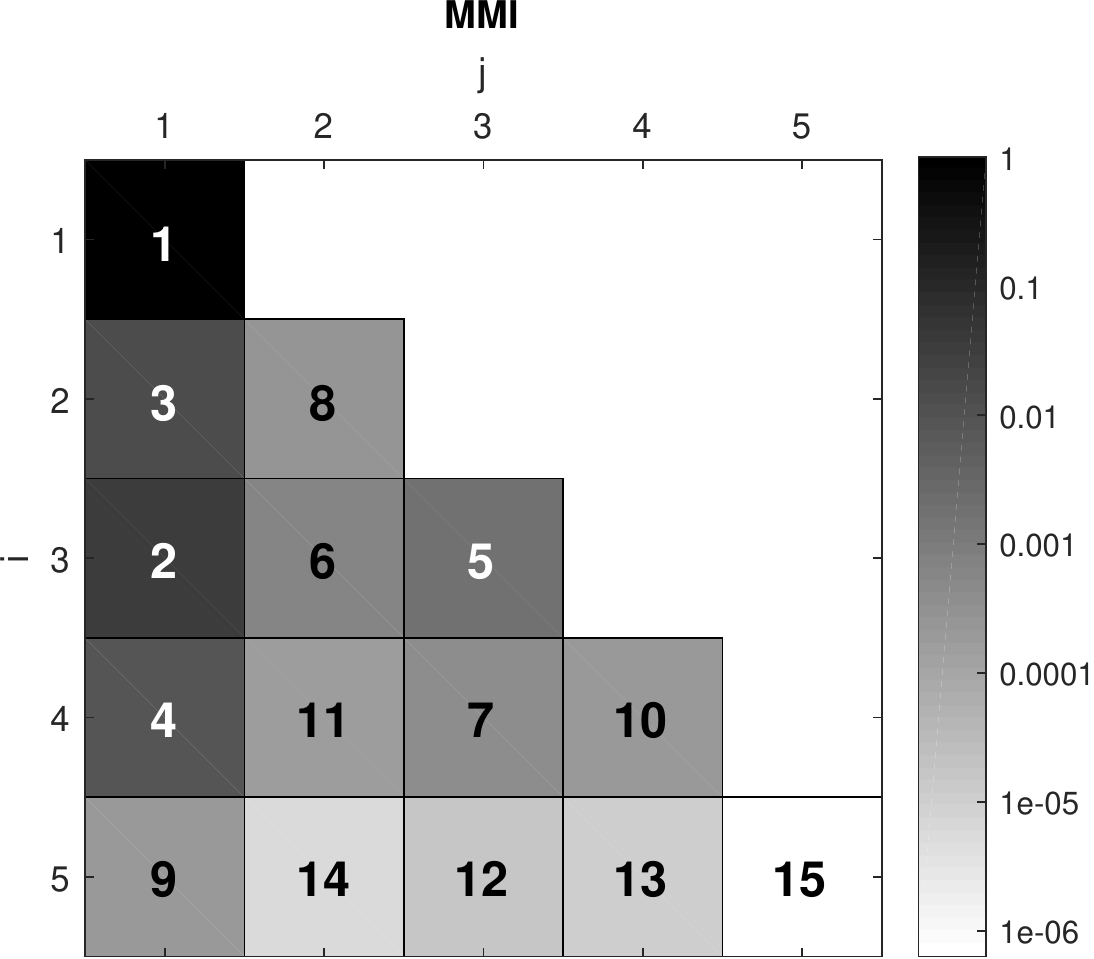}
			\captionsetup{justification=centering}
			\caption{ \footnotesize }
			\label{fig:RealMMIWeightage}
		\end{subfigure}
		\begin{subfigure}{0.48\linewidth}
			\centering
			\includegraphics[width=0.8\linewidth]{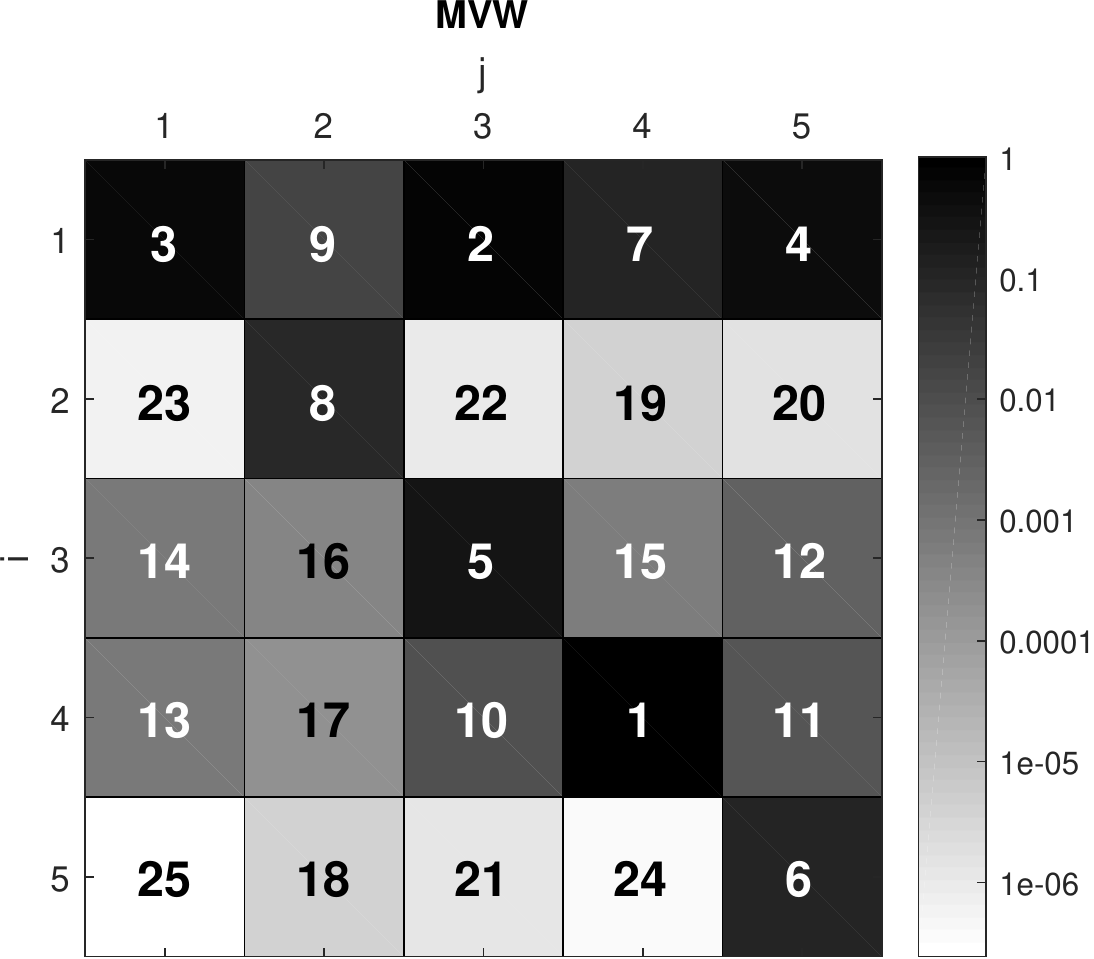}
			\captionsetup{justification=centering}
			\caption{\footnotesize }
			\label{fig:RealMVWWeightage}
		\end{subfigure}\\
			\captionsetup{justification=justified,
				singlelinecheck=false
			}
		\caption{ The ranking and the weights obtained for all (15)25 (S)MDs using 5 VMs in Model-II. (a) shows the colour intensity plot for weights found using Maximum Modal Interaction (MMI) technique, cell (i, j) contains rank (number) and weight (colour) for $ MD_{ij} $. (b) shows the colour intensity plot for ranking found using Modal Virtual Work (MVW) technique, Cell (i,j) contains rank (number) and weight (colour) for $ MD_{ij} $. (Note: part above diagonal not shown due to symmetry)\\ }
		\label{fig:RealMDS}
	\end{figure}

	\FloatBarrier
	A linear manifold is constructed with the first 5 VMs and corresponding (S)MDs. A maximum of  (15)25 (S)MDs can be obtained from 5 VMs, which makes the ROB size (20)30 if all the (S)MDs are considered. However, as discussed in Section~\ref{Sec:MDS}, the (S)MD selection techniques (MMI fo SMDs \& MVW for MDs) are also tested by selecting 5 (S)MDs instead of 15 (ranking shown in Figure~\ref{fig:RealMDS}). It can be seen that both the methods are able to reproduce the nonlinear response with good accuracy. The results are shown in Figures~\ref{fig:Node1} and~\ref{fig:Node2}, and Table~\ref{Tab:GREReal2}. \review{Note that the LM composed of all SMDs has 20 unknowns whereas the LM composed of all MDs has 30 modes, since the SMDs are symmetric and the MDs are not (cf. Theorem 1 and Remark 2). Based on $ m=5 $ VMs in this case, there are $ m(m+1)/2 = 15 $ SMDs, and  $ m^2 = 25 $ MDs.} 
\FloatBarrier
	\subsubsection{Quadratic Manifold}
	A quadratic manifold is used with first $ 5 $ modes reducing the number of unknowns $ m $ to 5 instead of 20 as in case of LM. As shown in Figures~\ref{fig:Node1} and~\ref{fig:Node2}, and Table~\ref{Tab:GREReal2}, the QM is able to provide same accuracy as the LM using quarter the number of unknowns in case of SMDs (six times less in case of MDs). The quadratic manifold was constructed using MDs as well as SMDs and both show very similar accuracy (cf. Table~\ref{Tab:GREReal2}).  \\
	\begin{table}[h!]
		\centering
		\caption{Global Relative Error in Model-II for different reduction techniques.}
		\label{Tab:GREReal2}
		\begin{tabular}{m{5cm} | c |c c}
			\toprule[1.5pt]
			{\bf Reduction Technique} 	& 	\# unknowns  	& 	$ GRE_M  $ (\%)	\\
			\midrule

			LM - MDs (All)	&	30 &	1.60 \\
			LM - Selected MDs (MVW)	&	10 &	2.06	\\
			LM - SMDs (All) & 20 	& 1.65 \\
			LM - Selected SMDs (MMI)	&	10 &	1.84	\\
			\midrule
			Quadratic Manifold (MDs)	&	5 &	1.66	\\
			Quadratic Manifold (SMDs)	&	5 &	1.65	\\
			\midrule
			POD 	&	5	&	4.90\\
			POD     &   20 	& 	0.31
		\end{tabular}
	\end{table}
	\\
It is interesting to see that a POD based approach with 5 basis vectors performs worse than the QM in this case. Indeed the performance improves significantly if more POD modes are included in the basis (cf. Table~\ref{Tab:GREReal2}).
\FloatBarrier

\section{Conclusions}
\label{chap:conlusions}
In this work, we consider projection based Model Order Reduction (MOR) techniques in the context of thin walled structural dynamics characterized by Von K\'{a}rm\'{a}n kinematics. The main focus of the research is on MOR methods which essentially do not require a full non-linear solution run to construct the reduced-order model. Though the use of Modal Derivatives (MDs) for reduction addresses this need, it quickly becomes inhibitive due to growing number of unknowns with the number of Vibration Modes (VMs). The proposed quadratic mapping using MDs avoids this undesirable growth with a negligible loss of accuracy. Some conclusions are as follows.

\emph{Linear Manifold vs. Quadratic Manifold:} 
\begin{itemize}
	\item \textbf{Accuracy:} It is easy to see that the amplitudes connected to the (S)MDs in the linear manifold are unknowns in the corresponding reduced equations and thus are free to choose any value, where as in the quadratic manifold they are inherently constrained by the amplitudes of VMs.  Indeed this makes the former more accurate in the Galerkin sense (cf. Remark~\ref{rem:reductionerror}), but results on tested examples show that the solution accuracy is very similar in both cases. However, it is anticipated that the quadratic enslavement of the MDs to the VM amplitudes would not, in general, hold for an arbitrary structural system. In the experience of the authors, this approach provides excellent accuracy for systems characterized by slow dynamics which is dominated by a few, separated VMs, while the (fast) nonlinear coupling effects are merely triggered quasi-statically by the slow dynamics. Systems featuring beam-like behavior (as the wing box structure shown) are excellent candidates for the advocated approach. Likewise, as the nonlinearities in the system rise, it is expected that there would be a range where (S)MDs could still be good candidates for reduction without obeying a quadratic enslavement to the VM amplitudes. Further discussion on this will be presented in the companion paper \cite{Johannes}.

	\item \textbf{Speed:} Even after selection of important (S)MDs, it is easy to see that the number of unknowns is greater in the Linear Manifold than the Quadratic Manifold. Thus, the reduced system is smaller in case, when a Quadratic Manifold is used. Since the evaluation of nonlinearity and projection are the bottlenecks, the reduction in problem size would not be tantamount to the computational speed up in the current implementation. A hyper-reduction technique or a tensor based approach (in case of polynomial nonlinearities) would be required in order to obtain much appreciable computational speedups. These aspects are currently being investigated by the authors.
	\end{itemize}

\emph{MD Selection:} The search for a selection criteria of (S)MDs to reduce the Linear manifold basis size resulted in two possible candidates (MMI for SMDs and MVW for MDs cf. Section~\ref{Sec:MDS}). Both of them give results with very similar accuracy when all (S)MDs are used in the reduction basis. 
It should be noted that both selection techniques use "weights" to rank the MDs in the order of relative importance, but these techniques did not indicate how many of these ranked (S)MDs should be used to accurately reproduce the solution obtained using all the (S)MDs. Thus, there is a need for further work along this course. As a rule of thumb, a total of $ n_{MD} = m $ (S)MDs was chosen in this work in a linear manifold basis containing $ m $ VMs. This was done for fair comparison and for keeping the basis size linear with $ m $. 

 \review{ In this work, we do not specifically address problems characterized by local or global structural instabilities. However, a quadratic manifold approach was also adopted for this class of problems, and shown to be able to reproduce static complex post-buckling behavior even leading to structural instability  \cite{PaoloThesis}. Furthermore, for transient analysis, it has been shown that the inclusion of modes and their derivatives at different linearization points could capture buckling phenomena \cite{Tiso2011}. The reduced model using a quadratic manifold, as proposed here, considers the quadratic manifold as a local extension around the equilibrium point only. Thus, its applicability to systems with mild noninearities is justified.} The presented examples have given a good overview of the relative behaviour and accuracy of the results. General claims about accuracy of these techniques require more in-depth research by taking an abstract and analytical approach.

\appendix
\review{
\section{Comparison of Quadratic Manifold with the Static Condensation approach}
\label{sec:app}
We consider a 2-DOF dynamical system in variables $ w,v $ given as
\begin{align}
\label{eq:transverse}
m_1\ddot{w} + c_1\dot{w} + k_1 w + avw + bw^3 &= g(t),\\ 
\label{eq:axial}
m_2\ddot{v} + c_2\dot{v} + k_2 v + c w^2  &= 0.
\end{align} 
The tangent stiffness to this system can be written as 
\begin{equation}\label{key}
\mathbf{K}(\mathbf{u}) = \begin{bmatrix}
k_1 + 3bw^2	& aw\\
2cw & k_2
\end{bmatrix}, 
\end{equation}
with $ \mathbf{u} \in \mathbb{R}^2 $ denoting the full vector of unknowns $ \begin{bmatrix}
w\\ v 
\end{bmatrix} $. The above system can be considered as a 2-DOF-variant of the FE discretized equations of the von karman beam (see e.g. \cite{Rutzmoser2014} ), where the solution variables $ w(t), v(t) \in \mathbb{R} $ are analogous to the transverse and axial displacements of the beam, respectively, $ m_1,m_2,k_1,k_2,a,b,c \in \mathbb{R} $ are physical parameters, and $ g(t) $ correspond to the externally applied load in the transverse direction.

The Static Condensation (SC) approach applied to this example implies that the membrane variables, being stiff, are not dynamically excited and statically follow the load. This leads to the quadratic enslavement of the axial displacement variable $ v $ in \eqref{eq:axial} to the transverse displacement $ w $  as
\begin{equation}\label{}
 v = k_2^{-1}(- cw^2),   
\end{equation}
which can be substituted into \eqref{eq:transverse} to obtain a single DOF reduced-order model in $ w $ as
\begin{equation}\label{eq:rom}
m_1\ddot{w} + c_1\dot{w} + k_1 w - ak_2^{-1}cw^3 + bw^3 = g(t).
\end{equation}
Thus, the SC approach has mapped the full system of two unknowns into the single unknown $ w $ as
\begin{equation}\label{eq:sc}
\mathbf{u}\approx\boldsymbol{\Gamma}_{SC} = \begin{bmatrix}
w \\ -k_2^{-1}cw^2
\end{bmatrix}.
\end{equation}

After linearizing the system \eqref{eq:transverse}-\eqref{eq:axial} around the equilibrium position ($ w=v=0 $), it is easily seen that the two VMs of the system are given by $ \boldsymbol{\phi}_1=\begin{bmatrix}
1 \\ 0 
\end{bmatrix} $ and $ \boldsymbol{\phi}_2=\begin{bmatrix}
0 \\ 1 
\end{bmatrix} $. Thus, the system is in modal coordinates, with the modal unknowns $ q_1=w $ and $ q_2 = v $. We consider a quadratic manifold constructed using the first VM and its corresponding SMD (cf. \eqref{eqn:SMD}) given by
\begin{equation}\label{eq:1}
\begin{bmatrix}
k_1 & 0 \\
0  & k_2
\end{bmatrix} \left.\pardervd{\boldsymbol{\phi}_1}{q_1}\right|_{eq}^s = -\left.\parderv{\mathbf{K}}{q_1}\right|_{eq}\boldsymbol{\phi}_1,
\end{equation}
where the tangent stiffness sensitivity can be calculated as
\begin{equation}\label{eq:2}
\left.\parderv{\mathbf{K}}{q_1}\right|_{eq} = \left.\parderv{\mathbf{K}(\mathbf{u}=q_1\boldsymbol{\phi}_1)}{q_1}\right|_{eq} = \begin{bmatrix}
0 & a \\
2c & 0
\end{bmatrix}\,.
\end{equation}
From \eqref{eq:1} and \eqref{eq:2}, we get
\begin{equation}
 \left.\pardervd{\boldsymbol{\phi}_1}{q_1}\right|_{eq}^s =- \begin{bmatrix}
0\\
k_2^{-1}(2c)
\end{bmatrix}\,.
\end{equation}
The single mode quadratic manifold using the SMD is then given by
\begin{equation}\label{eq:qm1}
\mathbf{\Gamma}_{QM} = \boldsymbol{\phi}_1q_1 + \dfrac{1}{2} \left.\pardervd{\boldsymbol{\phi}_1}{q_1}\right|_{eq}^s q_1^2 = \underbrace{\begin{bmatrix}
1\\0
\end{bmatrix} w }_{\text{bending contribution}}+ \underbrace{\begin{bmatrix}
0\\-k_2^{-1} c
\end{bmatrix} w^2}_{\text{membrane contribution}} = \begin{bmatrix}
w\\-k_2^{-1} cw^2
\end{bmatrix} .
\end{equation}
The final expression in \eqref{eq:qm1}, along with \eqref{eq:sc}, shows that for this example, the reduction mapping constructed using the quadratic manifold is exactly the same as that obtained using the static condensation approach. The reduced-order model obtained using the QM can be written using \eqref{QMreducedEqn} as
\begin{equation}\label{eqn:Qrom}
m_1\ddot{w} + \underbrace{4k_2^{-2}c^2m_2(w^2\ddot{w} + w\dot{w}^2)}_{\text{extra inertial terms}} +  c_1\dot{w} + \underbrace{4k_2^{-2}c^2c_2w^2\dot{w}}_{\text{extra damping terms}} + k_1 w - ak_2^{-1}cw^3 + bw^3 = g(t)\,.
\end{equation}
Though the reduction mapping is the same, it can be seen that the two ROMs in \eqref{eq:rom} and \eqref{eqn:Qrom} are different. In particular, the ROM \eqref{eqn:Qrom} created using QM  differs only by inclusion of extra inertial and damping contributions for the statically condensed axial variables. 
This is due to the nonlinear mapping and projection of full equations on to the tangent space in case of the QM-based ROM.  For the ease of readability, we demonstrated a 2-DOF example, but it is easy to see that these conclusions hold for multi-dimensional analogues of \eqref{eq:transverse}-\eqref{eq:axial} as well.
}
\paragraph{Acknowledgements}
The authors are thankful to the anonymous reviewers of this work for their valuable suggestions. The first and the second authors acknowledge the support of the Air Force Office of Scientific Research, Air Force Material Command, USAF under Award No.FA9550-16-1-0096.

\bibliography{report}
\end{document}